\documentclass[12pt,preprint]{aastex}
\usepackage{natbib}
\usepackage{apjfonts}

\usepackage{amsmath}

\begin{document}
\title{Dark Matter in the Milky Way's Dwarf Spheroidal Satellites}
\shorttitle{Dark Matter in the Milky Way's Dwarf Spheroidal Satellites}
\author{Matthew G. Walker\altaffilmark{1,2}}
\altaffiltext{1}{Harvard-Smithsonian Center for Astrophysics, 60 Garden St., Cambridge, MA 02138}
\altaffiltext{2}{Hubble Fellow}

\begin{abstract} 
The Milky Way's dwarf spheroidal satellites include the nearest, smallest and least luminous galaxies known.  They also exhibit the largest discrepancies between dynamical and luminous masses.  This article reviews the development of empirical constraints on the structure and kinematics of dSph stellar populations and discusses how this phenomenology translates into constraints on the amount and distribution of dark matter within dSphs.  Some implications for cosmology and the particle nature of dark matter are discussed, and some topics/questions for future study are identified.
\end{abstract}

\section{Introduction}
\label{sec:intro}

Physics assigns to gravity the responsibility of forming structure on scales ranging from terrestrial to cosmological.  An apparent threshold arises somewhere between scales characteristic of star clusters and those characteristic of galaxies.  The internal dynamics of gravitationally bound structures smaller than a few parsecs, of which globular clusters are the largest examples, are reasonably well described in terms of standard gravity (i.e., Einstein's general theory and/or its Newtonian approximation) sourced by known substances.  The internal dynamics of gravitationally bound structures larger than a few tens of parsecs, of which dwarf spheroidal (dSph) galaxies are the smallest examples, are not.  

The ubiquity of dark matter on galactic and larger scales signifies new physics.  Either there exists an otherwise unknown substance that contributes to dynamical mass but not to baryonic mass, or the standard dynamical framework requires modification (or both).  The `substance' hypothesis is not falsifiable, but in principle it can be confirmed with the detection of non-gravitational interactions involving dark matter particles.  In any case, dSphs provide the most extreme examples of dark matter phenomenology, with dynamical mass-to-light ratios $M/L_V\ga 10$ $[M/L_V]_{\odot}$ even at their centers.  This fact has made dSphs the focus of intense scrutiny in the effort to understand the nature of dark matter.

This article reviews the development of empirical constraints on the amount and distribution of dark matter within the Milky Way's dSph satellites.  These results follow from the application of a rich variety of analyses applied to observations conducted by many individuals and groups.  All analyses described here are formulated within the Newtonian dynamical framework.  The reader is welcome to interpret `dark matter' in terms of the substance hypothesis or more generally as a quantification of the discrepancy between dynamical and baryonic mass.  The focus here is on the relationship between data and constraint, and one hopes that this information translates meaningfully into alternative dynamical frameworks \citep[as formulated, e.g. by][]{bergmann68,milgrom83,bekenstein04,moffat06}.

\section{Observations}
\label{sec:observations}

Bright red giant stars are detectable as point sources with $m_V\la 21$ mag out to distances of $\sim 0.5$ Mpc.  Within this range, the Milky Way's dSph galactic satellites appear as localised overdensities of individually resolved stars.  Empirical information about the number, stellar structure and internal kinematics of these systems has steadily accumulated for the past eight decades.

\subsection{Census}
\label{subsec:census}

\citet{shapley38} discovered the Sculptor dSph upon visual examination of a photographic plate exposed for three hours with the 24-inch Bruce telescope at (what was then) Harvard's Boyden observatory in South Africa.  In this original dSph discovery paper, Shapley notes that Sculptor is visible---in hindsight---as a faint patch of light on a plate taken for the purpose of site testing, in 1908, during a series of five exposures totalling nearly 24 hours with a 1-inch refracting telescope.  

Reporting shortly thereafter the similar discovery of Fornax, \citet{shapley38b} notes that the new type of `Sculptor-Fornax' cluster shares properties with both globular clusters and elliptical galaxies, then speculates that `At the distance of the Andromeda system these objects would, in fact, have long escaped discovery.  There may be several others in the Local Group of galaxies; such objects may be of frequent occurence in intergalactic space and of much significance both in the census and the genealogy of sidereal systems.'

\begin{figure}
  \plotone{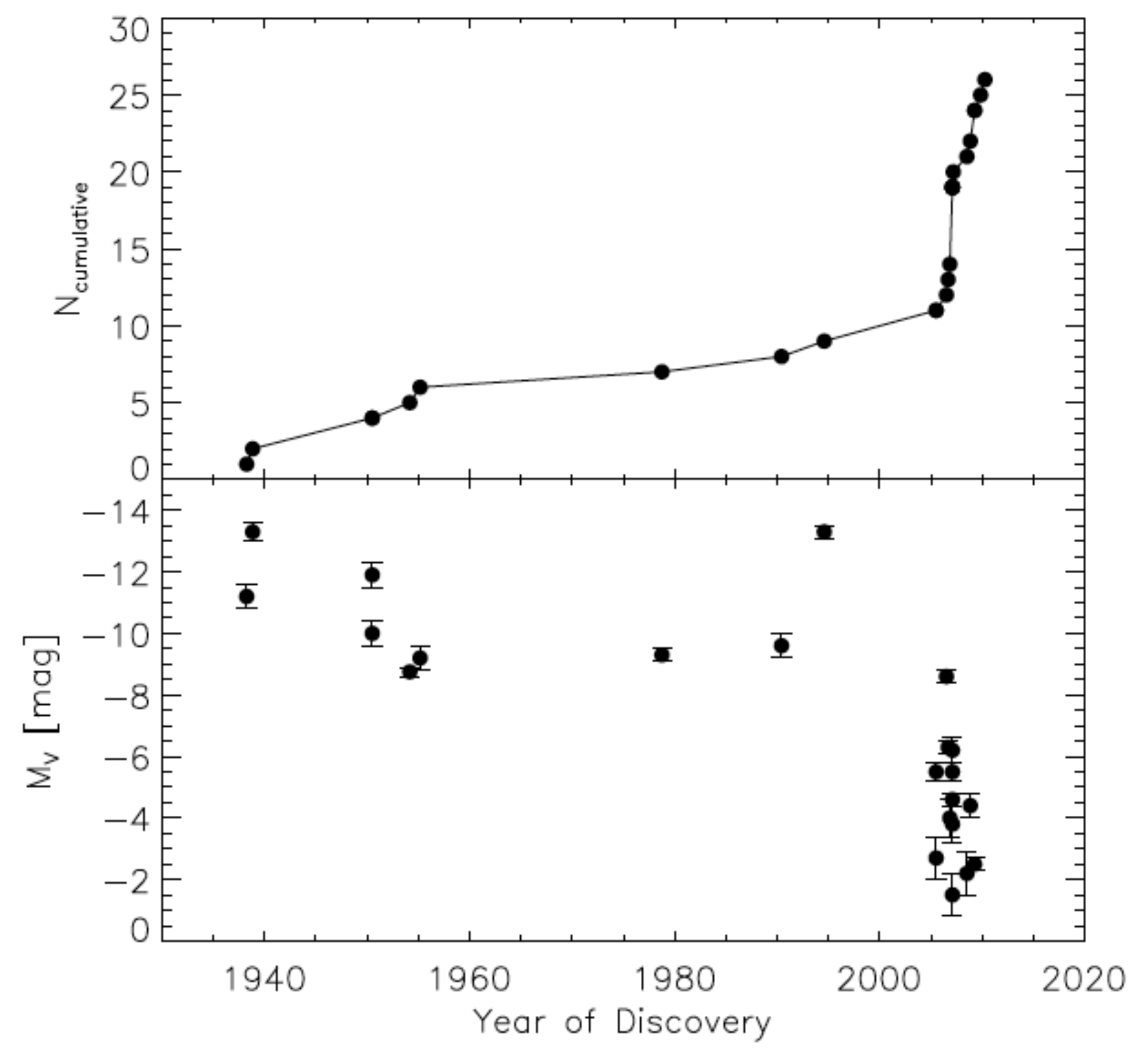}
  \caption{Cumulative number (top) and luminosity (bottom) of known dwarf spheroidal satellites of the Milky Way, versus year of publication of discovery paper \citep{shapley38,shapley38b,harrington50,wilson55,cannon78,irwin90,ibata94,willman05a,willman05b,belokurov06b,zucker06a,zucker06b,belokurov07,irwin07,walsh07,belokurov08,belokurov09,grillmair09,watkins09,belokurov10}.}
  \label{fig:discovery}
\end{figure}

Figure \ref{fig:discovery} demonstrates Shapley's prescience, plotting the cumulative number and luminosities of the Milky Way's known dSph satellites against date of discovery publication.  \citet{harrington50} and \citet{wilson55} found the next four `Sculptor-type' (as they were then called) systems---Leo I, Leo II, Draco and Ursa Minor---on photographic plates taken for the Palomar Observatory Sky Survey with the 48-inch Schmidt telescope.  \citet{cannon78} spotted Carina by eye on a plate taken with the 1.2-meter UK-Schmidt Telescope.  \citet{irwin90} used the Automated Photographic Measuring (APM) facility at the University of Cambridge to detect Sextans, again on a UK-Schmidt plate, thereby completing the census of the Milky Way's eight so-called `classical' dSphs.  

The next discovery was unique in that it came via spectroscopy rather than imaging.  During a stellar kinematic survey of the Galactic bulge, \citet{ibata94} noticed Sagittarius (Sgr) in color-velocity space as an excess of stars following a narrow velocity distribution offset from that of bulge members.  Subsequent observations showed that Sgr, located just $\sim 15$ kpc behind the Galactic center, spews stars along tidal tails that wrap spectacularly around the entire sky (e.g., \citealt{mateo96,ibata97b,majewski03,belokurov06,koposov11b}).  The tidal disruption of Sgr provides a useful tracer of the Galactic potential (e.g., \citealt{johnston05,law10,penarrubia10b}), but disqualifies Sgr from a simple equilibrium-dynamical analysis.  For this reason Sgr shall not be considered further here.

In a span of seven years, deep multi-color photometry from the Sloan Digital Sky Survey (SDSS) has now tripled the number of known Milky Way satellites \citep{willman05a,willman05b,zucker06a,zucker06b,belokurov06b,belokurov07,irwin07,walsh07,belokurov08,belokurov09,grillmair09,watkins09,belokurov10}.  The 17 satellites discovered with SDSS data have extended the floor of the observed galaxy luminosity function from $M_V\ga -8$ to $M_V\ga -2$ (bottom panel of Figure \ref{fig:discovery}), such that some galaxies are less luminous than some stars!  Unlike their brighter `classical' siblings, the `ultrafaint' satellites\footnote{As Figure \ref{fig:discovery} and the terms themselves suggest, the distinction between `classical' and `ultrafaint' dSphs involves a mixture of intrinsic luminosity with sequence of discovery.  Here this distinction (which is meaningless in the sense that members of both classes trace smooth scaling relationships involving luminosity, size, metallicity and stellar kinematics) is preserved only because observational studies of these objects---for both practical and accidental reasons---tend to be separable along the same lines.  Following common practice, dSphs known before SDSS (Carina, Draco, Fornax, Leo I, Leo II, Sculptor, Sextans, Ursa Minor) are referred to as `classical', and the rest as `ultrafaint'.} discovered with SDSS data are not apparent to the eye, even in deep images.  Rather, they are detected only by correlating spatial overdensities with overdensities in color-magnitude space (e.g., \citealt{belokurov06b,walsh09}).  In order to confirm the faintest satellites, the SDSS catalog must be supplemented by deeper, follow-up photometry as well as spectroscopy (e.g., \citealt{belokurov09,belokurov10}).  Given the rate of false positives expected for candidates remaining in SDSS data ($\ga 4/5$ in the author's experience), the expense of follow-up observations  can be prohibitive.  However, the next generation of sky surveys (Pan-STARRS, SkyMapper, DES, Gaia, LSST, etc.) will almost certainly bring a new flurry of discoveries, particularly in the relatively unexplored southern sky.  

Although they are not considered further below, it is worth noting that Shapley was correct regarding the dSph satellites of M31.  \citet{vandenbergh72} discovered the first example with the Palomar Schmidt telescope, using plates more sensitive than those used in the original Palomar Survey.  The current census includes 27 known dSph satellites of M31.  Two-thirds of this number were discovered in the past seven years \citep{zucker04,martin06,ibata07,zucker07,majewski07,irwin08,mcconnachie08,martin09,richardson11,slater11,bell11}, with SDSS data as well as photometry from the PAndAS survey conducted with the Canada-France-Hawaii Telescope \citep{mcconnachie09}.  

\subsection{Stellar Structure}
\label{subsec:structure}

Galactic dynamics is concerned with the relationship between gravitational potential and the distribution of stars in phase space.  Observers who study dark matter in dSph galaxies must gather information about the positions and velocities of dSph stars.  The largest dSphs subtend solid angles of several square degrees, making it difficult to study their stellar structure at large radius.  Complete homogenous surveys are rare and valuable. 

\subsubsection{`Classical ' dSphs}
\label{subsubsec:classical}

\citet{hodge61a,hodge61b,hodge62,hodge63,hodge64a,hodge64b} used photographic plates obtained at the Palomar (48-inch Schmidt, 100-inch and 200-inch), Lick (120-inch) and Boyden Station (24-inch Schmidt) in order to study luminous structure of the six Milky Way dSphs known at the time (Sculptor, Fornax, Leo I, Leo II, Draco and Ursa Minor).  Hodge counted stars within squares of regular grids overlaid on each plate, and provides this description of analog data reduction: `Each plate was counted at one sitting so that uniformity would be maintained.  The plates were counted to the limiting magnitude and each was counted once.  From experience ... it was decided that the reproducibility of counts on a particular plate is greater than from plate to plate, so that it is better to count many plates each once than one plate many times' \citep{hodge61a}.  

\begin{figure}
  \epsscale{1.}
  \plotone{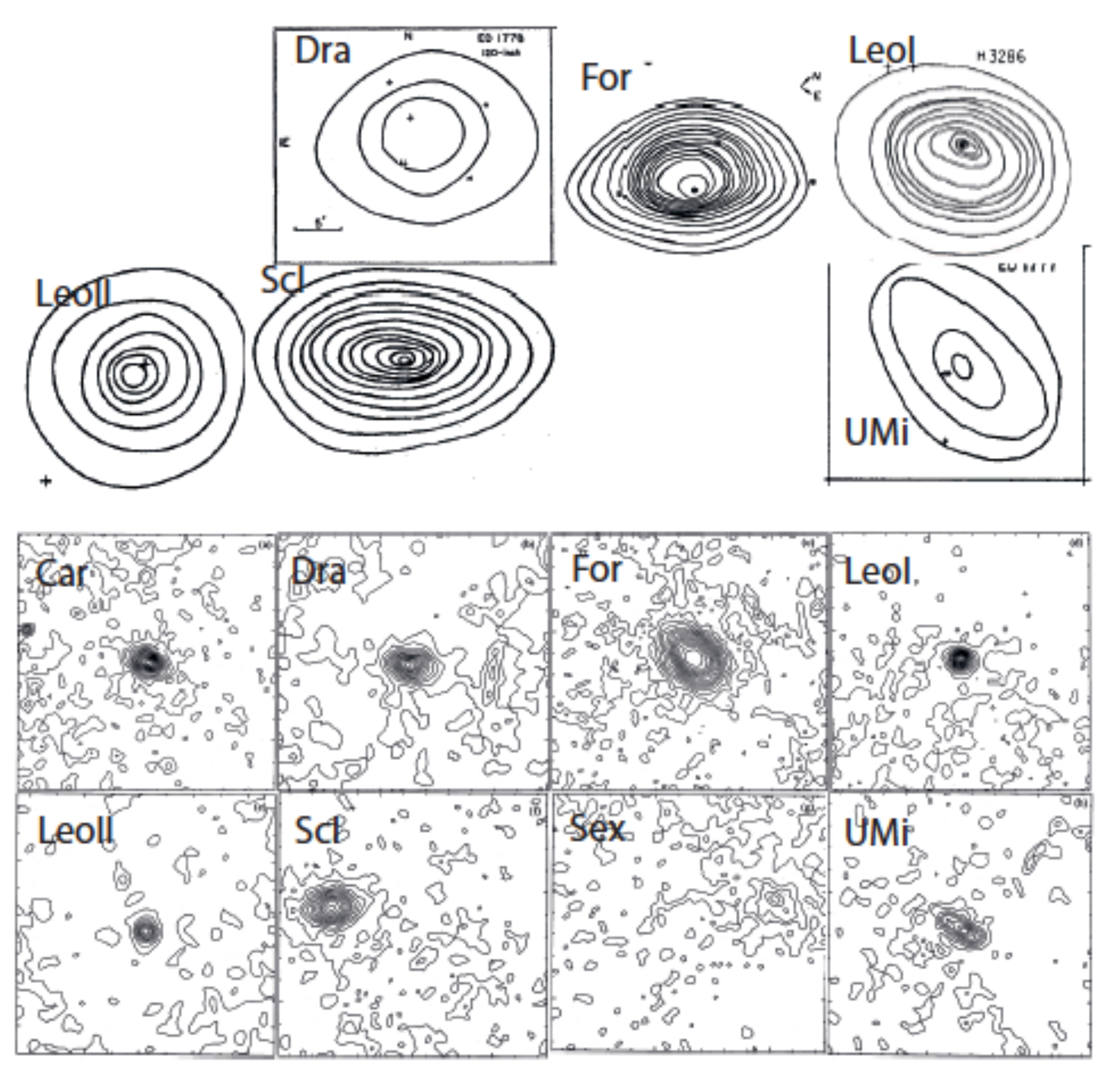}
  \caption{Stellar isodensity maps for the Milky Way's eight `classical' dSphs.  \textit{Top two rows:} from Hodge's star count studies \citep[reproduced by permission of the American Astronomical Society]{hodge61a,hodge61b,hodge62,hodge63,hodge64a,hodge64b}.  \textit{Bottom two rows:} reproduced from \textit{Structural Parameters for the Galactic Dwarf Spheroidals}, by M.\ Irwin \& D.\ Hatzidimitriou, MNRAS, 277, 1354, 1995 (by permission of John Wiley \& Sons Ltd.).}
  \label{fig:hodgeih95}
\end{figure}

Figure \ref{fig:hodgeih95} (top two rows of panels) displays isopleth maps that Hodge drew by connecting squares containing equal numbers of stars.  These maps reveal that the internal structure of dSphs is smooth.  Hodge reasoned that a well-mixed dSph requires dynamical relaxation by a process other than stellar encounters, as the low surface densities of dSphs imply internal relaxation timescales of $\ga 10^3$ Hubble times.  \citet{hodge66} and \citet{hodge69} would later suggest that encounters between stellar groups might enable exchanges of orbital energy over shorter timescales during dSph formation.  Eventually \citet{lyndenbell67} would show that the time-varying gravitational potential of a young galaxy effectively shuffles the orbital energies of its stars, generating `violent' relaxation without stellar encounters.  More recently, \citet{mayer01b} have proposed a mechanism specific to dSphs, demonstrating with N-body/hydrodynamical simulations that repeated tidal encounters with the Milky Way can effectively transform a rotating dSph progenitor into a pressure-supported spheroid in less than a Hubble time.  

Hodge discovered two other structural features common to the classical dSphs.  First, most exhibit flattened morphology, with typical ellipticities of $\epsilon\equiv 1-b/a\sim 0.3$, where $a$ and $b$ are semi-major and semi-minor axes, respectively.  Second, dSph stellar density profiles decline more steeply at large radius than do the profiles of giant elliptical galaxies.  Whereas the latter are commonly fit by formulae with relatively shallow outer profiles, e.g., $\Sigma(R)=\Sigma(0)/(1+R/a)^2$ \citep{hubble30} or $\Sigma(R)=\Sigma(0)\exp[-kR^{1/4}]$ \citep{devaucouleurs48}, Hodge found that classical dSphs all have steeper outer profiles that are better fit with the formula of \citet{king62}:
\begin{equation}
  \Sigma(R)=k\biggl [\frac{1}{\sqrt{1+(R/R_c)^2}}-\frac{1}{\sqrt{1+(R_{\mathrm{K}}/R_c)^2}}\biggr ]^2,
  \label{eq:king62}
\end{equation}
where $R_c$ is a `core' radius and $R_{\mathrm{K}}$ is a maximum, or `limiting' radius that one might expect to result from tidal truncation (Section \ref{subsubsec:extended})\footnote{In fact $R_{\mathrm{K}}$ is usually referred to as a `tidal' radius and denoted $r_t$.  The adopted nomenclature and notation avoid confusion with the tidal radius defined in Equation \ref{eq:tide}.}.

Three decades later, \citet[`IH95' hereafter]{ih95} used the APM facility at Cambridge to count stars automatically on photographic plates from the Palomar and UK Schmidt telescopes.  While confirming Hodge's findings, IH95 produced significantly deeper maps and were able to include Carina and Sextans, the two Milky Way dSphs discovered in the interim (Figure \ref{fig:hodgeih95}).  IH95 used these maps to measure the centroid, ellipticity and orientation of each dSph, and then to tabulate stellar density as a function of distance along the semi-major axis.  While the homogeneous analysis of IH95 continues to provide a valuable resource particularly for comparing dSphs in the context of scaling relations (Section \ref{sec:universal}), deeper photometric data sets now exist for most of the classical dSphs \citep[e.g.,][]{stetson98,majewski00,saviane00,odenkirchen01,palma03,walcher03,lee03,tolstoy04,coleman05a,coleman05b,battaglia06,westfall06}.  
\begin{figure}
  \epsscale{1.}
  \plotone{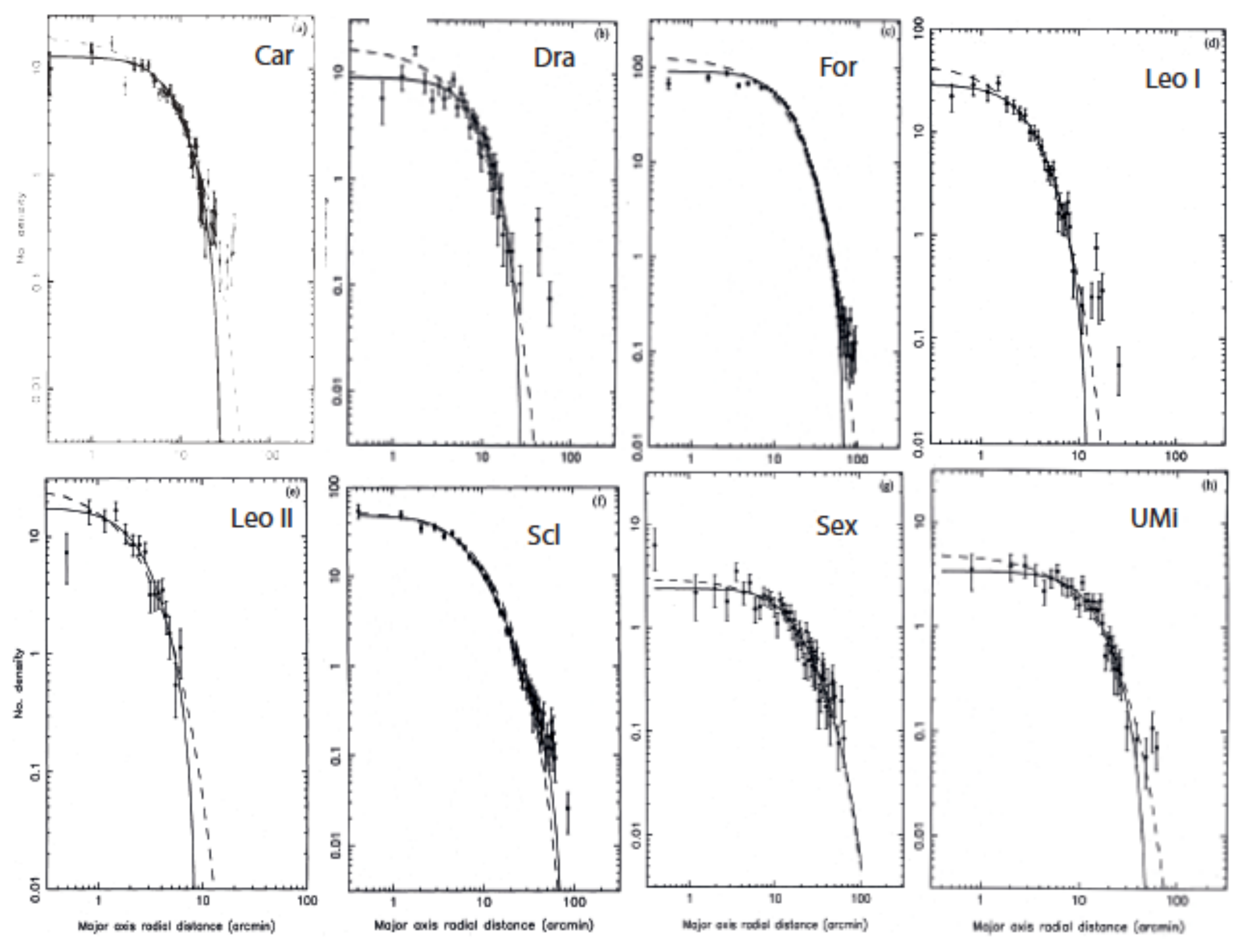}
  \caption{Stellar density profiles for the Milky Way's `classical' dSphs.  Overlaid are best-fitting \citet[solid]{king66} and exponential (dashed) surface brightness profiles.  Reproduced from \textit{Structural Parameters for the Galactic Dwarf Spheroidals}, by M.\ Irwin \& D.\ Hatzidimitriou, MNRAS, 277, 1354, 1995 (by permission of John Wiley \& Sons Ltd.). }
  \label{fig:ih95fits}
\end{figure}

In principle the structure of dSph stellar components carries information about the mechanisms that drive dSph formation and evolution.  While incompatible with shallow Hubble and de Vaucouleurs profiles, the available data often do not distinguish the King profile (Equation \ref{eq:king62}) from other commonly adopted fitting formulae---e.g., exponential and \citet{plummer11} profiles,
\begin{equation}
  \Sigma(R)=\Sigma(0)\exp[-R/R_e]
  \label{eq:exp}
\end{equation}
and 
\begin{equation}
  \Sigma(R)=\frac{\Sigma(0)}{\bigl [1+(R/R_{h})^2\bigr ]^2},
  \label{eq:plummer}
\end{equation}
respectively, where $R_{h}\approx 1.68R_e$ is the projected halflight radius (i.e., the radius of the circle enclosing half the stars as viewed in projection).  Figure \ref{fig:ih95fits} displays IH95's fits of \citet{king66} and exponential surface brightness profiles to the Milky Way's classical dSph satellites.

\subsubsection{Ultrafaint dSphs}

\begin{figure}
  \epsscale{1.}
  \plotone{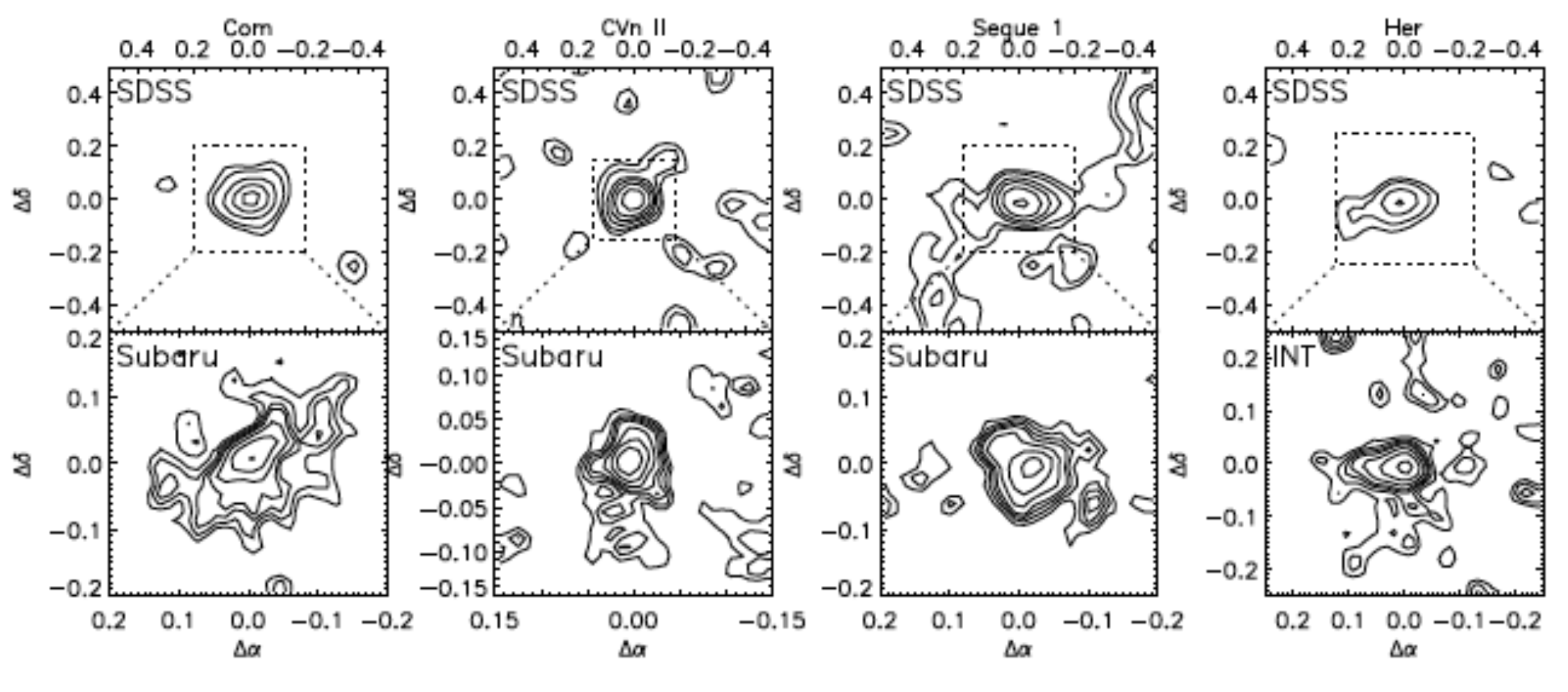}
  \caption{Stellar idodensity maps for four of the Milky Way's `ultrafaint' dSphs, from the discovery paper of \citet[reproduced by permission of the American Astronomical Society]{belokurov07}.}
  \label{fig:catsdogs}
\end{figure}

The SDSS catalog enables homogeneous studies of the structural properties of ultrafaint dSphs.  For example, Figure \ref{fig:catsdogs} displays isopleth maps constructed by \citet{belokurov07} for Coma Berenices, Canes Venatici II, Segue 1 and Hercules using  both SDSS and deeper follow-up data.  Whereas Hodge and IH95 estimated structural parameters for the classical dSphs after binning their star-count data and subtracting estimates of foreground densities, the low surface brightnesses of `ultrafaint' dSphs are conducive neither to binning nor to foreground subtraction.  Fortunately, neither procedure is necessary; indeed \citet{kleyna98} estimate structural parameters for the classical dSph Ursa Minor using a likelihood function that specifies the probability associated with each individual stellar position in terms of a parametric surface brightness profile plus constant foreground.  

\begin{figure}
  \epsscale{1}
  \plottwo{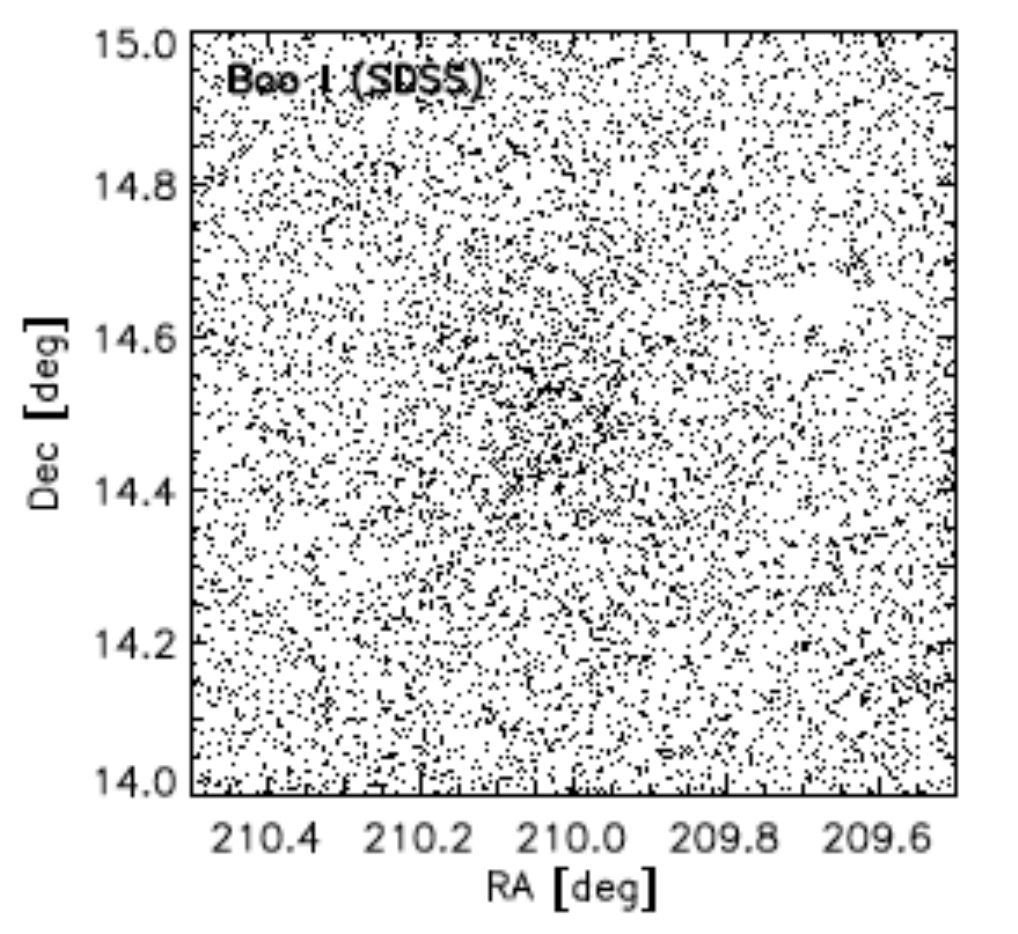}{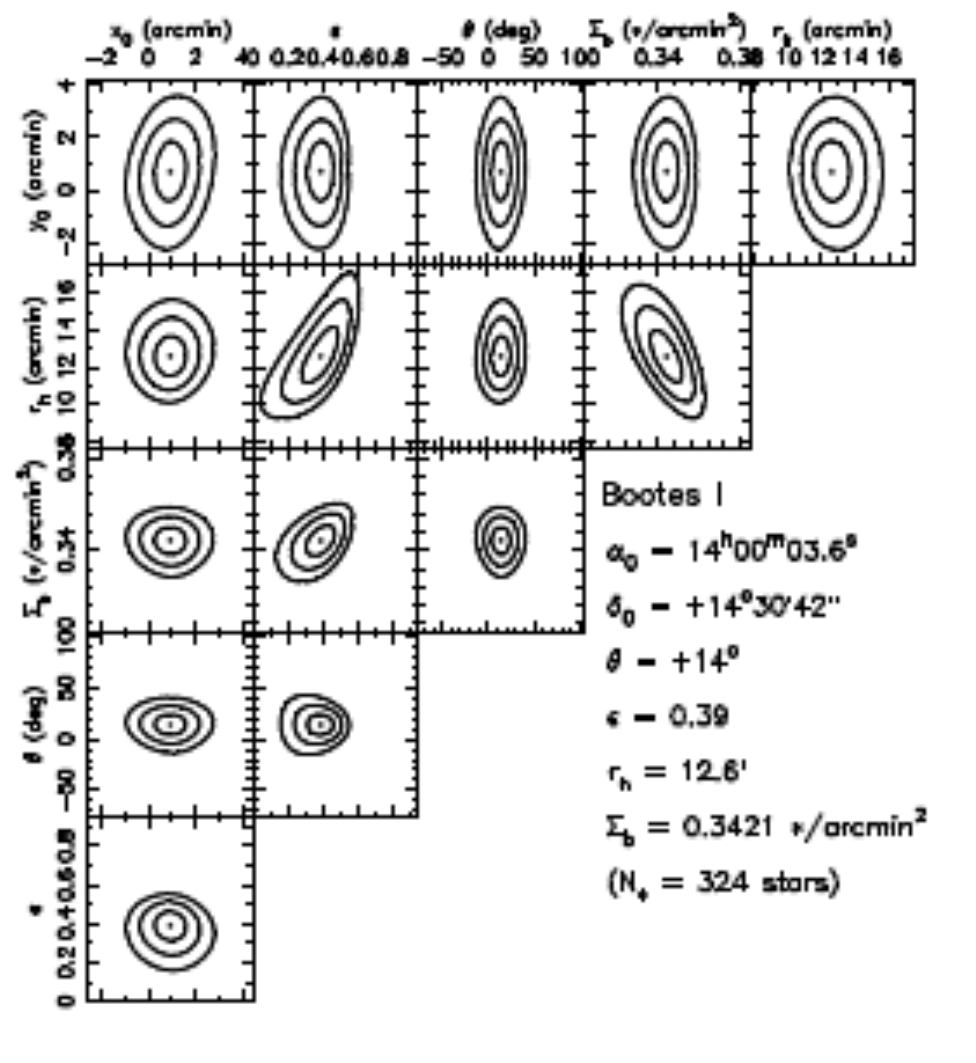}
  \caption{Measurement of structural parameters for the Bo\"otes I dSph from SDSS data \citep{martin08}.  \textit{Left:} sky positions of red giant candidates selected from the SDSS catalog.  \textit{Right:} constraints on structural parameters, from the maximum-likelihood analysis of \citet[reproduced by permission of the American Astronomical Society]{martin08}. } 
  \label{fig:boo_params}
\end{figure}

\citet{martin08} design a similar maximum-likelihood analysis that operates directly on unbinned SDSS data in order to estimate the centroid, halflight radius, luminosity, ellipticity and orientation of each ultrafaint satellite.  Figure \ref{fig:boo_params} demonstrates the efficacy of their method.  The left panel maps individual stars from the SDSS catalog that are near the line of sight to Bo\"otes I and satisfy color/magnitude criteria designed to select red giants at the distance of Bo\"otes I.  Panels on the right-hand side show maximum-likelihood estimates of each free parameter.  Marginalised error distributions include the effects of sampling errors and parameter covariances, and can be used directly in subsequent kinematic/dynamical analyses (Section \ref{sec:masses}).

\citet{martin08} show that their method recovers robust estimates even when the SDSS sample includes as few as tens of satellite members.  Subsequent tests with synthetic data by \citet{munoz11} provide reason for caution, suggesting that for objects with low surface brightness, insufficient contrast between members and foreground can generate biased estimates of structural parameters.  It is reassuring that deeper observations with CFHT \citep{munoz10} and Subaru \citep{okamoto12} yield structural parameters for subsets of ultrafaint satellites that agree well with the SDSS-derived estimates of \citet{martin08}.

\subsubsection{Extended Structure}
\label{subsubsec:extended}

The outer stellar structure of a given dSph is determined by some combination of formation processes and subsequent evolution within the external potential of the Milky Way.  In his structural analyses of the six dSphs known in the 1960s, Hodge compared the observed limiting radii, $R_{\mathrm{K}}$ (Equation \ref{eq:king62}), to simple estimates of `tidal' radii, $r_t$, beyond which stars escape into the external potential of the Milky Way \citep{vonhoerner57,king66}:
\begin{equation}
  r_t=R_{\mathrm{D}}\biggl[\frac{M_D}{(3+e)M_{\mathrm{MW}}}\biggr]^{1/3}.
  \label{eq:tide}
\end{equation}
Here, $M_{\mathrm{MW}}$ and $M_{\mathrm{D}}$ are the Milky Way and dSph (point) masses, respectively, $R_{\mathrm{D}}$ is the pericentric distance of the dSph's orbit and $e$ is orbital eccentricity.  Assuming circular orbits and considering only luminous masses, \citet{hodge66} noticed that while the two radii are similar for the nearest dSphs (Draco, Sculptor, Ursa Minor), for the three most distant dSphs (Leo I, Leo II, Fornax), $R_{\mathrm{K}}\la 0.5 r_t$.  \citet{hodge66} concluded that tidal forces play a significant role in shaping the outer structures of the nearest dSphs.

Low surface brightness in the outer regions of dSphs makes extended structural studies difficult to conduct and interpret.  For example, \citet{coleman05a} use deep wide-field photometry to conclude that Sculptor's surface brightness profile is well described as the superposition of two equilibrium stellar components with different scale radii.  Using an independent data set, \citet{westfall06} fit a single profile and identify a `break' in Sculptor's surface brightness profile, which makes an apparent transition from a King profile in the inner parts to a shallower power law in the outer parts.  \citet{westfall06} cite this transition not as evidence for a second component, but rather as a signal that Sculptor is losing stars to tidal disruption. 

\begin{figure}
  \epsscale{0.5}
  \plotone{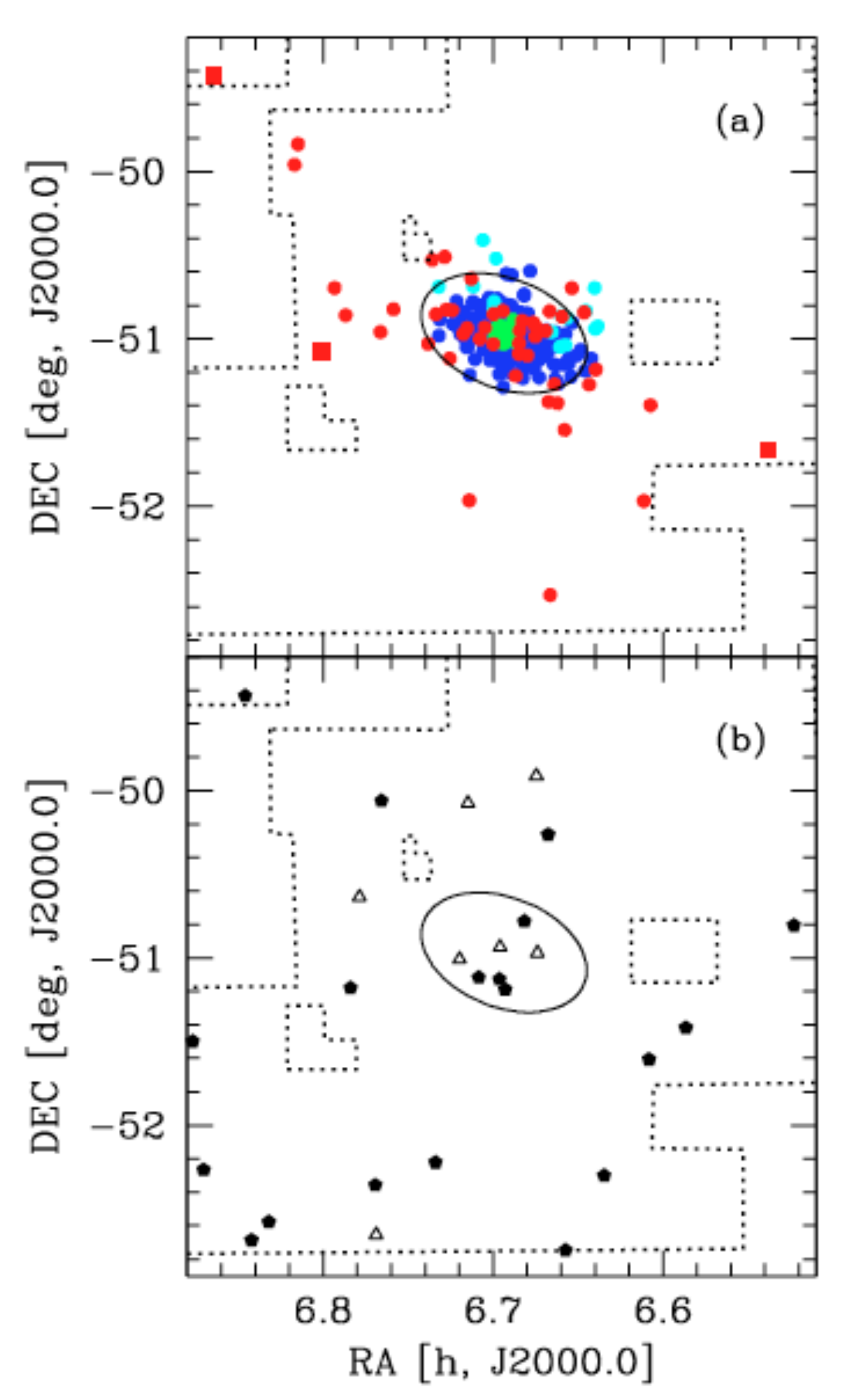}
  \caption{Extended stellar structure of the Carina dSph, from narrow-band photometry and spectroscopy of \citet[reproduced by permission of the American Astronomical Society]{munoz06}.  The top/bottom panels show sky positions of red giant candidates confirmed as Carina members/nonmembers.  Ellipses mark Carina's limiting radius, $R_{\mathrm{K}}$ (Equation \ref{eq:king62}), determined from smooth fits to star count data.  The extension of faint stellar structure at $R\ga 3R_{\mathrm{K}}$ suggests tidal interaction (Section \ref{subsec:tides}).  \citet{munoz05} report similar results for Ursa Minor.}
  \label{fig:munoz06}
\end{figure}

In many cases, the use of narrow-band filters that are sensitive to stellar surface gravity (e.g., DDO51) can help to distinguish dSph red giants from foreground dwarf stars, providing a valuable boost in contrast \citep[e.g.,][]{majewski00,majewski05,palma03,westfall06}.  Having used narrow-band photometry to select spectroscopic targets at large dSph radi, \citet{munoz05} and \citet{munoz06} present velocities that confirm the membership of stars in Ursa Minor and Carina (Figure \ref{fig:munoz06}) out to radii $\sim 3$ and $\sim 5$ times larger, respectively, than estimates of $R_{\mathrm{K}}$.  These detections imply that tides can play a significant role in shaping the outer parts of dSphs (Section \ref{subsec:tides}).

\subsubsection{Structural Peculiarities of Individual dSphs}
\label{subsubsec:peculiar}

Most kinematic analyses of dSphs proceed from the assumption that dSphs host a single, spherically symmetric stellar component in dynamic equilibrium (Section \ref{sec:masses}).  However, real dSphs are all flattened ($0.1\la \epsilon\equiv 1-b/a\la 0.7$, \citealt{ih95,martin08,sand11}), and it is not clear how severely this violation of spherical symmetry affects conclusions regarding dSph dynamics.  In fact most dSphs exhibit individual peculiarities that further violate the simplistic assumptions (Section \ref{sec:masses}) employed in kinematic analyses.  

\begin{figure}
  \epsscale{1}
  \plotone{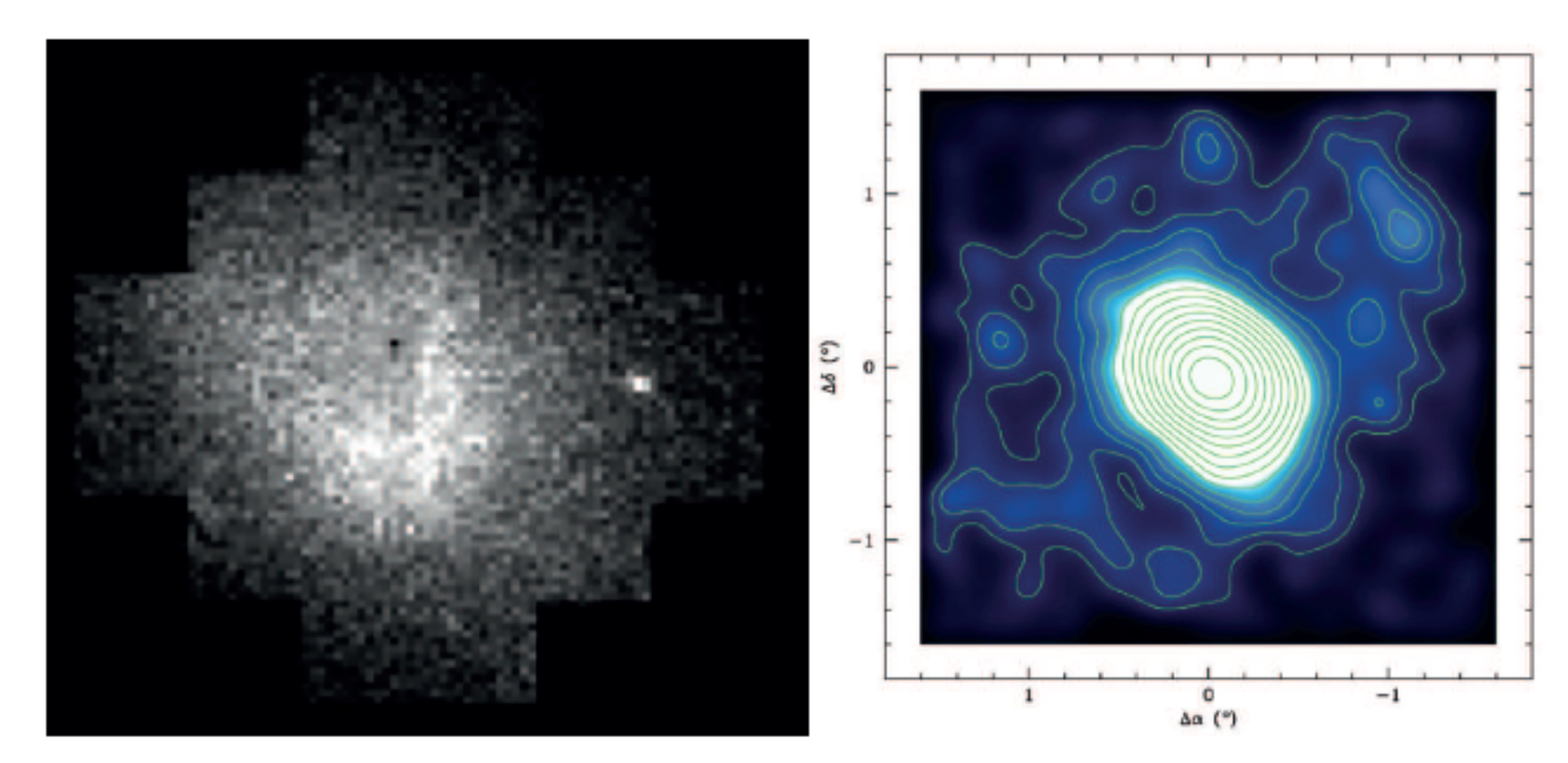}
  \caption{Structural irregularities in Fornax.  \textit{Left:} Grayscale map ($\sim 42'$ per side) of surface density of stars with $18.5\leq V\leq 23$, from the photometry of \citet[reproduced with permission from The University of Chicago Press]{stetson98}.  Notice the crescent-like feature near the center.  \textit{Right:} Smoothed stellar map from $V$, $I$ photometry of \citet[reproduced by permission of the American Astronomical Society]{coleman05b}, who detect two shell-like features (one is visible $\sim 1.3$ deg northwest of the center) aligned with lobes along the morphological minor axis.  }
  \label{fig:stetsoncoleman}
\end{figure}

For example, Sculptor, Fornax and Sextans all display evidence for chemo-dynamically independent stellar \textit{sub-}populations \citep[][respectively]{tolstoy04,battaglia06,battaglia11}.  In all three cases, a relatively metal-rich, kinematically cold population has smaller scale radius than does a metal-poor, kinematically hot population (Section \ref{subsubsec:models}, Figure \ref{fig:battaglia}).  Fornax also has irregular stellar structure in the form of a crescent-shaped feature near its center \citep[and Figure \ref{fig:stetsoncoleman}, left panel]{stetson98}, two shell-like features  \citep{coleman04} and lobes along its morphological minor axis \citep[and Figure \ref{fig:stetsoncoleman}, right panel]{coleman05b}.  These features suggest that Fornax may have undergone a recent merger, an event that might be related to the presence of a young (age $\sim 100$ Myr), centrally concentrated main sequence in Fornax \citep{battaglia06}.  Ursa Minor exhibits clumpy stellar substructure, most dramatically in the form of a secondary peak in its luminosity distribution, offset by $\sim 20'$ from the central peak \citep{olszewski85}.  The region near the secondary peak is kinematically colder than the rest of Ursa Minor \citep{kleyna03}, and so may represent a bound star cluster (Section \ref{subsubsec:indirect}).  

Such peculiarities extend to the ultrafaint satellites as well.  Spectroscopic surveys reveal that Segue 1 is superimposed on at least one stream of stellar debris \citep{geha09,niederste-ostholt09,simon11}, offset by $\sim 100$ km s$^{-1}$ from Segue 1 in velocity space.  Segue 1 also shows hints of extended stellar structure superimposed on Sgr debris, although interpretation of these extremely low-surface-brightness features in terms of tidal disruption remains controversial \citep{belokurov07,geha09,niederste-ostholt09,simon11}.  Segue 2 appears to be embedded in---and comoving with---a stream of stellar debris, perhaps from a tidally disrupted parent system \citep{belokurov09}.  Ursa Major II has flattened morphology and distorted outer morphology that suggest ongoing tidal disruption \citep{zucker06b,munoz10}.  Another strong candidate for disruption is Bo\"otes III, which has irregular, clumpy morphology and a large measured velocity dispersion of $\sigma\sim 14$ km s$^{-1}$ \citep{carlin09}.  The alignment of stellar distortions in Leo IV and Leo V hints at a low surface brightness `bridge' spanning the $\sim 20$ kpc between these systems (\citealt{belokurov08,walker09c,dejong10}; note, however, that \citealt{sand10} detect no such feature in deep photometry around Leo IV).  Stars nearest the center of the Willman 1 satellite exhibit near-zero velocity dispersion and a mean velocity that is offset from the rest of Willman 1 members by $\sim 8$ km s$^{-1}$ \citep{willman11}.  Each newly discovered galaxy seems to present a new quirk of its own.

\subsection{Stellar Velocities}
\label{subsec:velocities}

Assuming that dSphs are purely stellar systems truncated by tidal interaction with the Milky Way, such that $r_t\sim R_{\mathrm{K}}$, \citet{ostriker74} applied Equation \ref{eq:tide} to estimate the mass of the Milky Way.  Inverting the calculation by assuming an isothermal Galactic halo with $v_{\mathrm{circ}}=225$ km s$^{-1}$, \citet{faber83} estimated masses of dSphs.  For the nearest dSphs (Draco, Ursa Minor, Carina, Sculptor), Faber \& Lin estimated mass-to-light ratios of $M/L_V\ga 10 [M/L_V]_{\odot}$, suggesting dSph dark matter.  Faber \& Lin further used their estimates to predict, via the virial theorem, values of $\ga 10$ km s$^{-1}$ for the internal stellar velocity dispersions of dSphs.  

\begin{figure}
  \begin{center}$
    \begin{array}{ll}
      \includegraphics[scale=0.5]{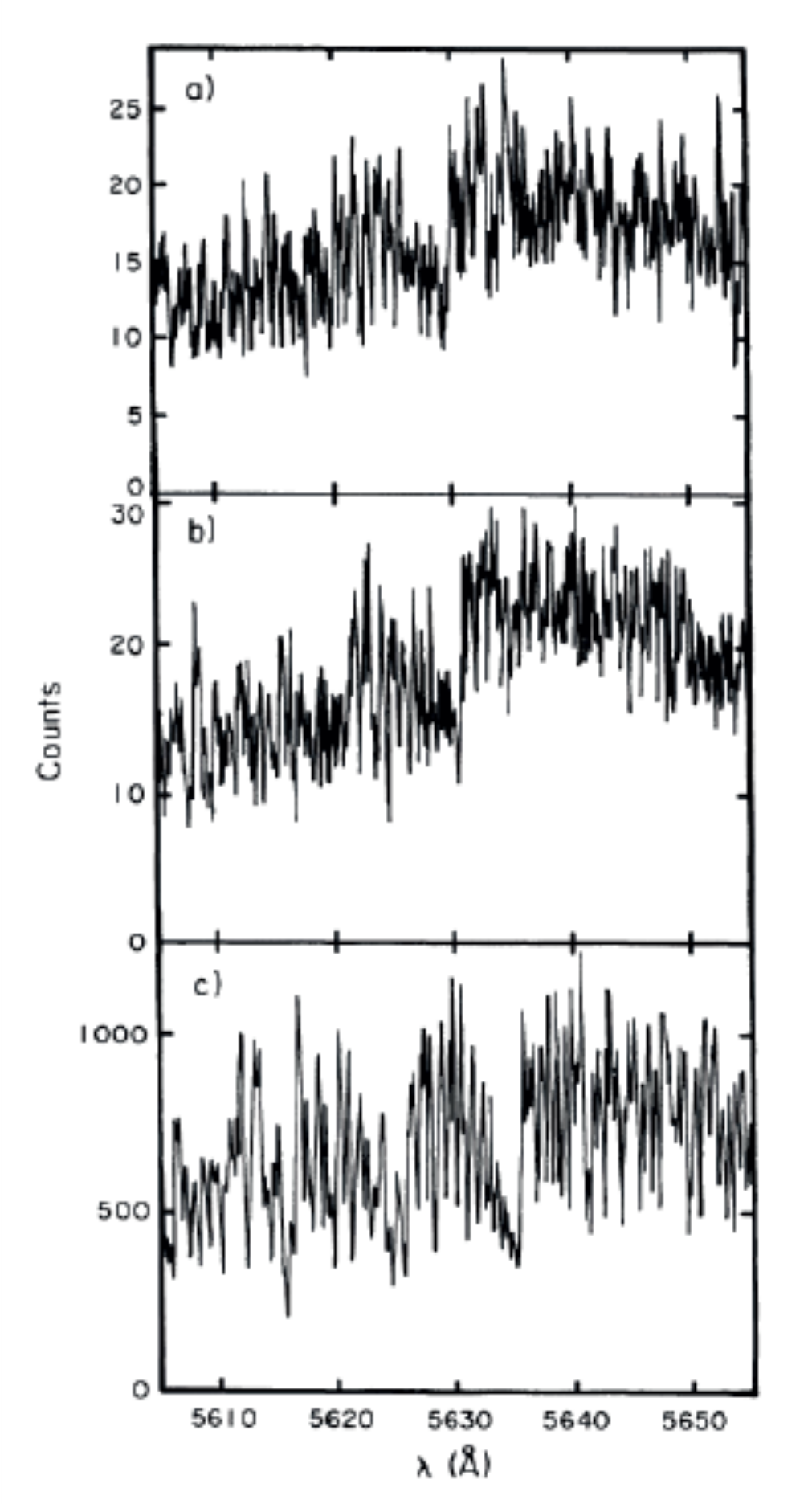}&\hspace{0.25in}\includegraphics[scale=0.5]{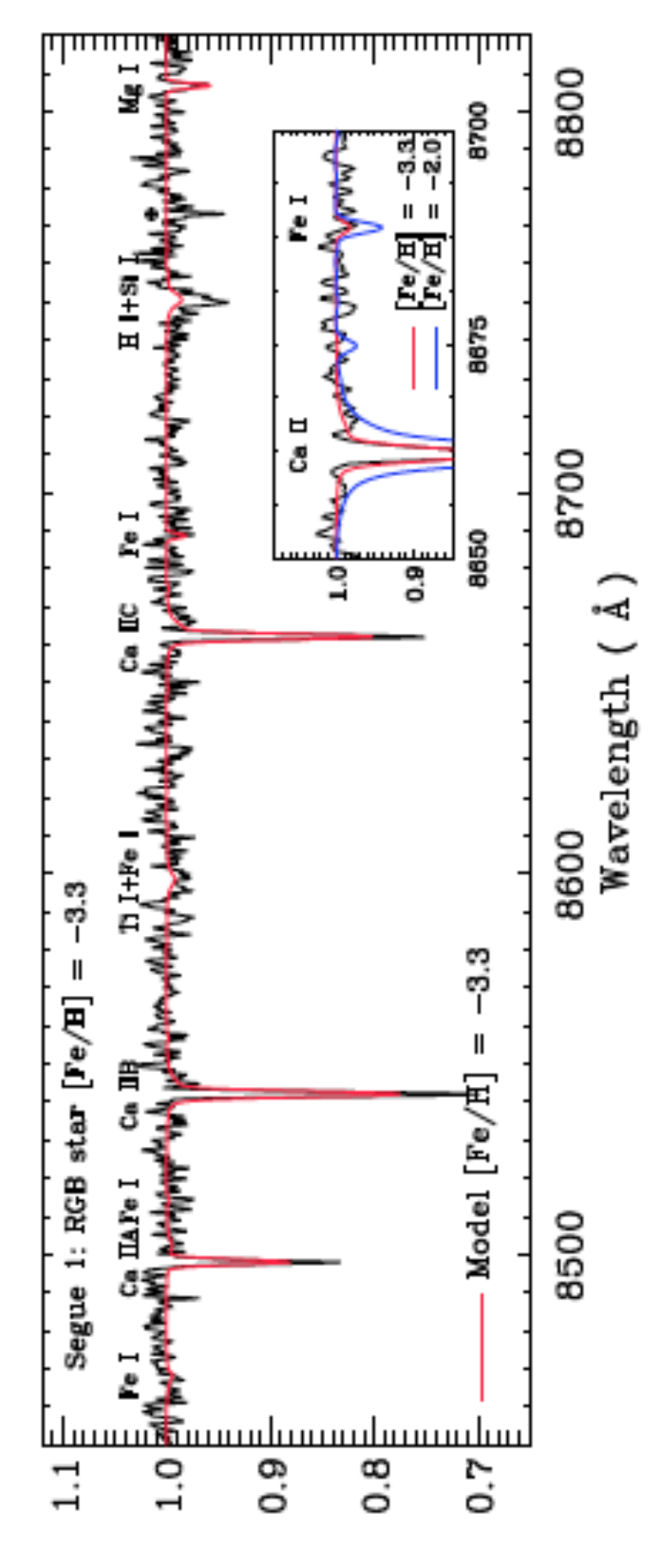}\\
    \end{array}$
  \end{center}
  \caption{Examples of spectra for individual dSph stars.  \textit{Left:} MMT echelle spectra ($R\sim 30000$) for three carbon stars in Draco, used for the first measurement of a dSph velocity dispersion \citep[reproduced by permission of the American Astronomical Society]{aaronson83}.  \textit{Right:} Keck/DEIMOS spectrum ($R\sim 6000$) for the brightest red giant in Segue 1 \citep[reproduced by permission of the American Astronomical Society]{geha09}, with absorption features labeled and best-fitting model overplotted.}
  \label{fig:aaronson}
\end{figure}

At the same time, \citet{aaronson83} provided the first actual measurement of a dSph's internal velocity dispersion.  Aaronson used the Multiple Mirror Telescope (MMT, which then consisted of six 1.8-meter mirrors working in tandem) to acquire high-resolution ($R\sim 30000$) spectra for three individual carbon stars in Draco.  Figure \ref{fig:aaronson} displays the spectra, which were sufficient for Aaronson to measure precise line-of-sight velocities of $-298.7\pm 0.9$ km s$^{-1}$ (with a follow-up measurement of $-297.6\pm 0.6$ km s$^{-1}$), $-300.2\pm 0.6$ km s$^{-1}$, and $-279.7$ km s$^{-1}$.  Aaronson calculated that such measurements require, at the $95\%$ confidence level, an intrinsic velocity dispersion of $\sigma\ga 6.5$ km s$^{-1}$.\footnote{Aaronson added to the final article proof a measurement of $-285.6\pm 1.1$ km s$^{-1}$ for a fourth, non-carbon star, further supporting a large dispersion.}  From dynamical arguments based on the virial theorem \citep[Section \ref{subsubsec:mfl}]{illingworth76,richstone86}, such a large dispersion indicates a large dynamical mass-to-light ratio $M/L_V \ga 30 [M/L_V]_{\odot}$, confirming the prediction of \citet{faber83} and indicating the presence of dark matter.  

\begin{figure}
  \epsscale{0.6}
  \plotone{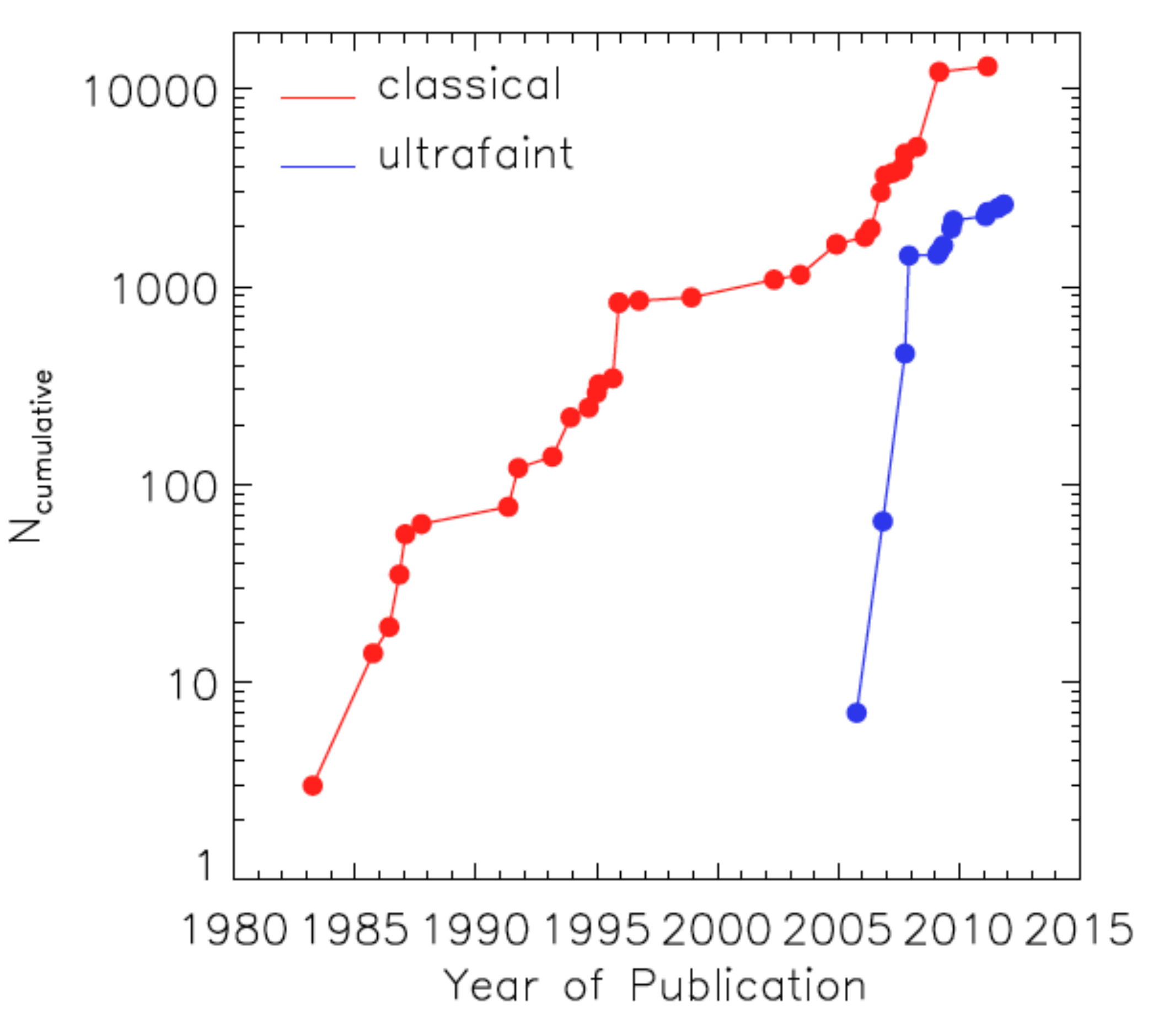}
  \caption{Growth of sample sizes available for internal kinematic studies of the Milky Way's dSph satellites.  Plotted as a function of time is the cumulative number of line-of-sight velocities measured for individual stars (including foreground contamination) targeted in dSph spectroscopic surveys.  References: \citet{aaronson83,seitzer85,suntzeff86,armandroff86,aaronson87,aaronson87b,mateo91,dacosta91,suntzeff93,mateo93,hargreaves94a,hargreaves94b,armandroff95,vogt95,queloz95,olszewski95,hargreaves96a,mateo98b,kleyna02,kleyna03,tolstoy04,kleyna05,munoz06,munoz06b,battaglia06,westfall06,walker06a,walker07,koch07,koch07b,martin07,simon07,sohn07,mateo08,koch09,geha09,walker09a,walker09c,belokurov09,carlin09,willman11,simon11,aden11,koposov11,battaglia11}.}
  \label{fig:numbers}
\end{figure}

Figure \ref{fig:numbers} plots the number of stars observed in dSph stellar velocity surveys as a function of time.  The three decades of observations separate neatly into `epochs' defined by the available instrumentation.  

\subsubsection{Small-Number Statistics}
\label{subsubsec:1980s}

The 1980s yielded the first precise velocity measurements for individual stars in Draco \citep{aaronson83}, Carina, Sculptor, and Fornax \citep{seitzer85}, Leo I and Leo II \citep{suntzeff86} and Ursa Minor \citep{aaronson87b}, and follow-up observations of Sculptor \citep{armandroff86} and Draco \citep{aaronson87}.  These samples typically included $\la 10$ stars per galaxy and indicated velocity dispersions of $\sigma \sim 6-10$ km s$^{-1}$, suggesting that dSph dynamical mass-to-light ratios reach at least double the values estimated for globular clusters.  Strong implications for the particle nature of dark matter \citep[Section \ref{subsec:particle}]{tremaine79,lin83} drew immediate attention.  Nevertheless, skepticism regarding small samples, velocity precision and the unknown contribution of binary orbital motions to the measured velocity dispersions demanded further observations and better statistics.

\subsubsection{Confirmation}
\label{subsubsec:1990s}

In the 1990s, several groups accumulated velocity samples for tens of stars per dSph.  Including Aaronson's original work, nine seasons of observations with the MMT eventually produced velocity samples that reached $\sim 20$ members in each of the Draco and Ursa Minor dSphs \citep{olszewski95,armandroff95}, yielding velocity dispersion measurements of $\sigma \sim 10$ km s$^{-1}$ and dynamical mass-to-light ratios $M/L_V\sim 75 [M/L_V]_{\odot}$ for both galaxies.  Meanwhile, medium-resolution ($R\sim 12000$) spectra from the William Herschel Telescope gave velocity samples for tens of stars in each of Draco, Ursa Minor and the newly discovered Sextans \citep{hargreaves94a,hargreaves94b,hargreaves96a}.  High-resolution spectra from ESO's 3.6-m and NTT telescopes \citep{queloz95} and Keck/HIRES \citep{mateo98b} delivered precise velocities for 23 and 33 stars in Sculptor and Leo I, respectively.  In all cases, velocity dispersions of $\sigma \ga 6$ km s$^{-1}$ indicated $M/L_V\ga 10 [M/L_V]_{\odot}$.  

Providing an early demonstration of the efficiency of multi-object fiber spectroscopy, \citet{armandroff95} compiled samples of $\sim 100$ velocities in each of Draco and Ursa Minor using the HYDRA multi-fiber spectrograph at the KPNO 4-meter telescope.  This data set included many repeat measurements, which \citet{edo96} used to estimate a binary fraction of $\sim 0.2 - 0.3$ for periods of $\sim 1$ year.  Based on Monte Carlo simulations, \citet{edo96} and \citet{hargreaves96b} concluded that binary motions contribute negligibly to the velocity dispersions measured for classical dSphs (Section \ref{subsec:binaries}).  

In the southern hemisphere, \citet{mateo91} used Las Campanas Observatory's 2.5-meter telescope and `2D Frutti' echelle photon counter to measure velocities for 44 Fornax stars, including an outer field that showed the same velocity dispersion ($\sigma\sim 10$ km s$^{-1}$) as the central stars, indicating $M/L_V\sim 10 [M/L_V]_{\odot}$.  After measuring a velocity dispersion of $\sigma\sim 7$ km s$^{-1}$ and $M/L_V\sim 40 [M/L_V]_{\odot}$ from the velocities of 17 Carina stars, \citet{mateo93} noted a scaling relation among dSphs: the dynamical mass-to-light ratios of classical dSphs are inversely proportional to luminosity, suggesting similar dynamical masses of $\sim 10^7M_{\odot}$ (Section \ref{sec:universal}).  

\subsubsection{Large Samples} 
\label{subsubsec:2000s}

The gap in Figure \ref{fig:numbers} between $1998-2002$ signifies a period not of inactivity but rather of construction.  During this time, wide-field multi-object spectrographs were built for the world's largest telescopes.  Over the past decade, surveys with these new instruments have increased stellar velocity samples from tens to thousands per dSph.  \citet{kleyna02,kleyna03,kleyna04} used the Wide-Field Fibre Optic Spectrograph (WYFFOS) at the 2.5-meter Isaac Newton Telescope to measure velocities for $\sim 100$ stars in each of Draco, Ursa Minor and Sextans, respectively.  These samples were sufficiently large to examine the velocity distribution as a function of distance from the dSph center (e.g., \citealt{wilkinson02,kleyna02,kleyna04,wilkinson04}; Section \ref{sec:masses}).  

\begin{figure}
  \epsscale{1.1}
  \plottwo{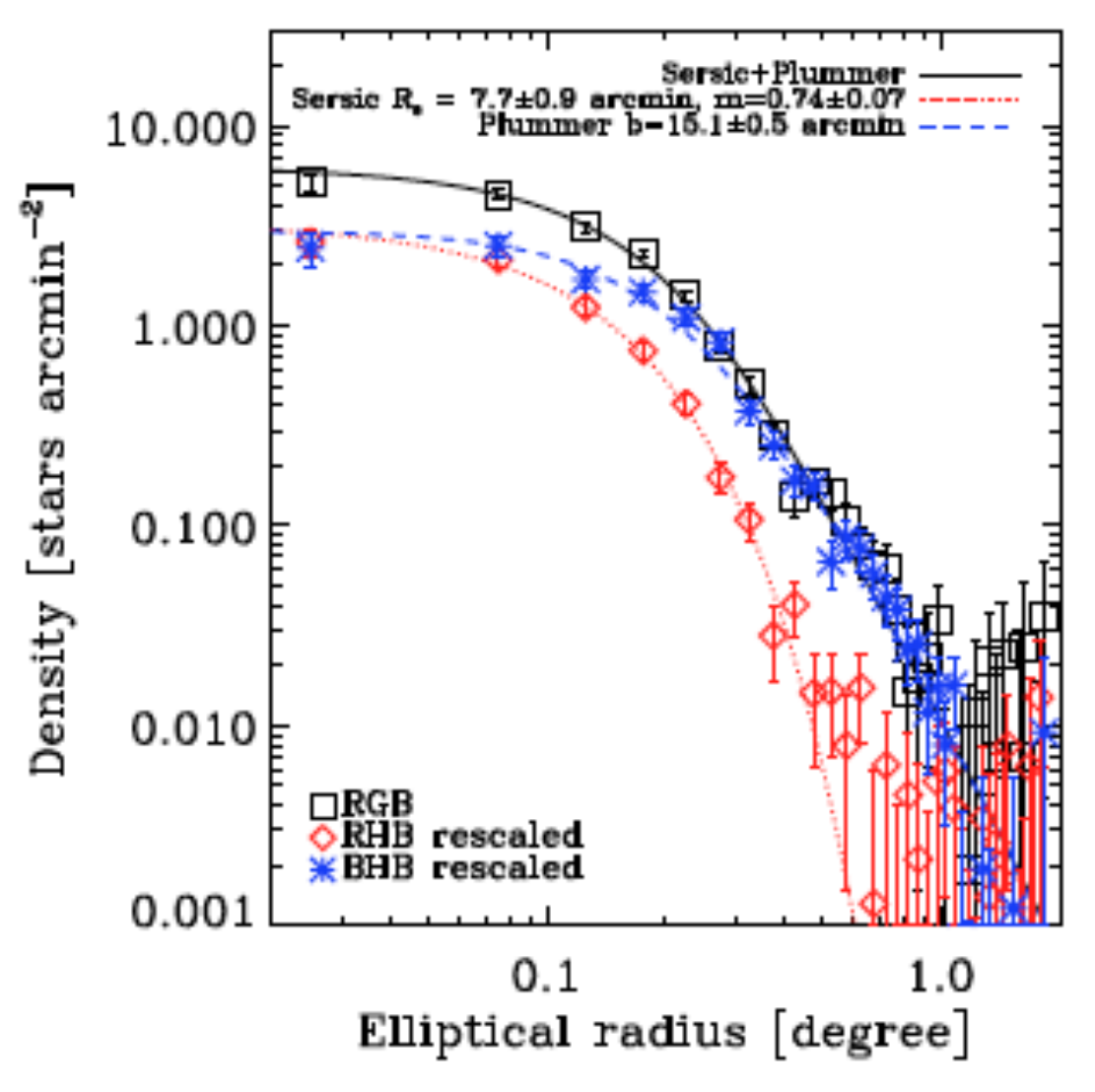}{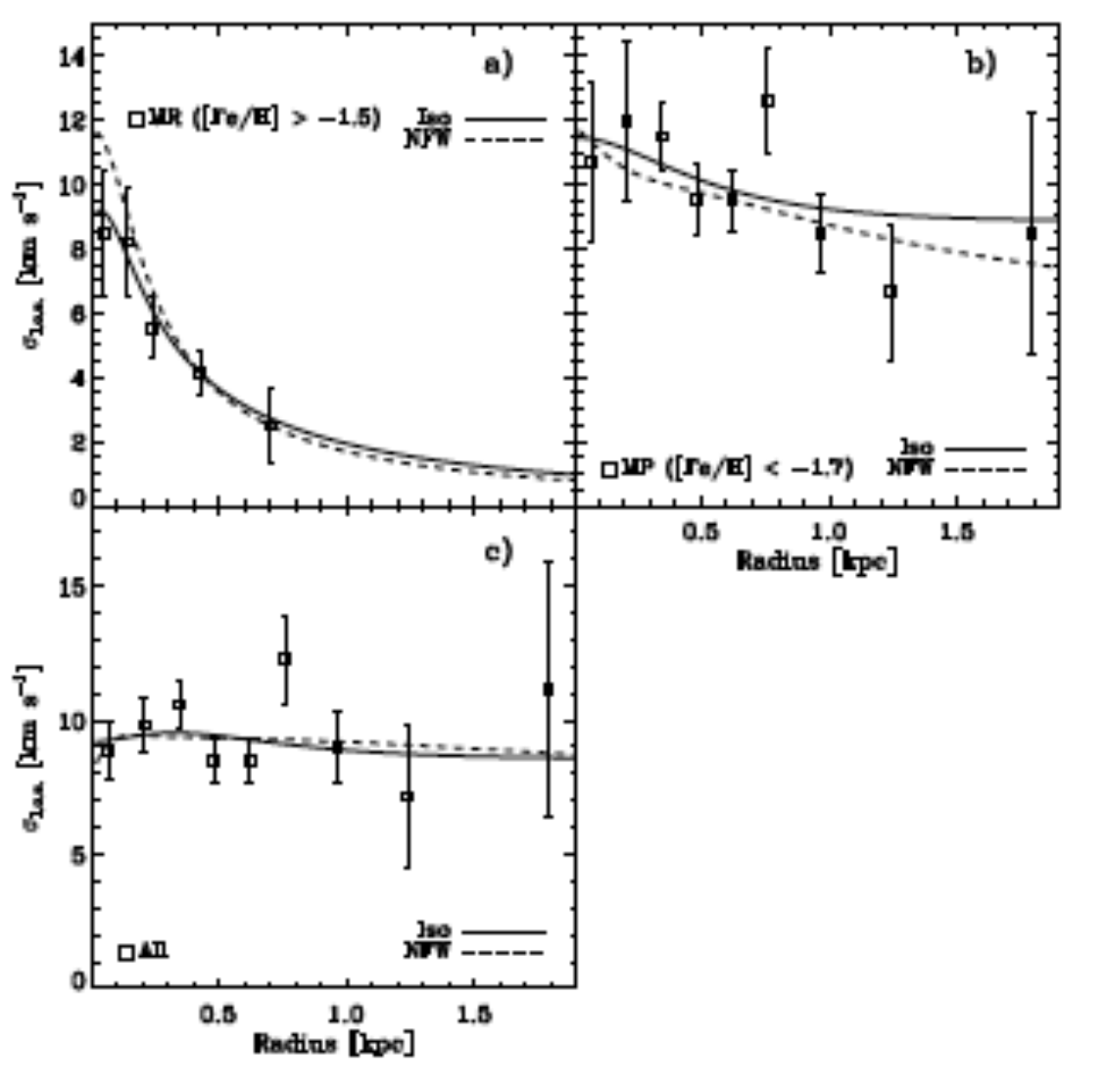}
  \caption{Sculptor's two chemodynamically independent stellar sub-populations \citep[reproduced by permission of the American Astronomical Society ; see also \citealt{tolstoy04}]{battaglia08}.  \textit{Left:} Surface brightness profiles for Sculptor's red giants (black), red-horizontal branch (red) and blue-horizontal branch stars.  \textit{Right:} Velocity dispersion profiles, calculated for subsets of relatively metal-rich (top left) and metal-poor (top right), as determined from the VLT/FLAMES spectroscopic sample of \citet{tolstoy04}.  For comparison, the lower-left panel plots the velocity dispersion profile measured from the composite population.}
  \label{fig:battaglia}
\end{figure}

Soon thereafter, the Dwarf Abundances and Radial Velocities Team (DART) used the FLAMES fiber spectrograph at the 8.2-meter Very Large Telescope (VLT; UT2)  to measure velocities and metallicities (derived from the strength of the calcium-triplet absoprtion feature at $\sim 8500 $ \AA) for $\sim 310$, $\sim 560$ and $\sim 175$ members of Sculptor, Fornax and Sextans, respectively \citep{tolstoy04,battaglia06,battaglia11}.  Theses samples yielded the discovery that all three of these dSphs contain multiple, chemodynamically independent stellar populations (Figure \ref{fig:battaglia} and Section \ref{subsubsec:models}).  Also with VLT/FLAMES, \citet{koch07b} measured a velocity dispersion of $\sigma \sim 7$ km s$^{-1}$ from $\sim 170$ members of Leo II, providing what remains the largest published sample for this galaxy.  

Meanwhile \citet{munoz06} used archival VLT/FLAMES spectra (see also \citealt{fabrizio11}) to measure velocities for $\sim 300$ Carina members and added another $\sim 45$ members from spectra obtained sequentially with the MIKE spectrograph at the Magellan/Clay 6.5-meter telescope.  The extra members observed with MIKE extend to $\sim 5$ times Carina's limiting radius as determined from photometry (Figure \ref{fig:munoz06}), indicating that Carina has lost mass to tidal interactions with the Milky Way.  

\citet{koch07b} used a pair of multi-slit spectrographs---the Gemini Multiobject Spectrograph (GMOS) at the Gemini-North 8-meter telescope and the Deep Imaging Multi-object Spectrograph (DEIMOS) at the Keck 10-meter telescope---to measure velocities for $\sim 100$ members of Leo I.  \citet{sohn07} added another $\sim 100$ members from their own observations with Keck/DEIMOS, and \citet{mateo08} contributed velocities for $\sim 300$ Leo I members using the Hectochelle multi-fiber spectrograph at the MMT.  The latter two studies explored larger radii and both found kinematic evidence for tidal streaming motions in the outskirts of Leo I.  This result is surprising given Leo I's current distance of $\sim 250$ kpc \citep{ih95}, leading \citet{mateo08} to suggest that Leo I's orbit is nearly radial.  

\begin{figure}
  \epsscale{1}
  \plotone{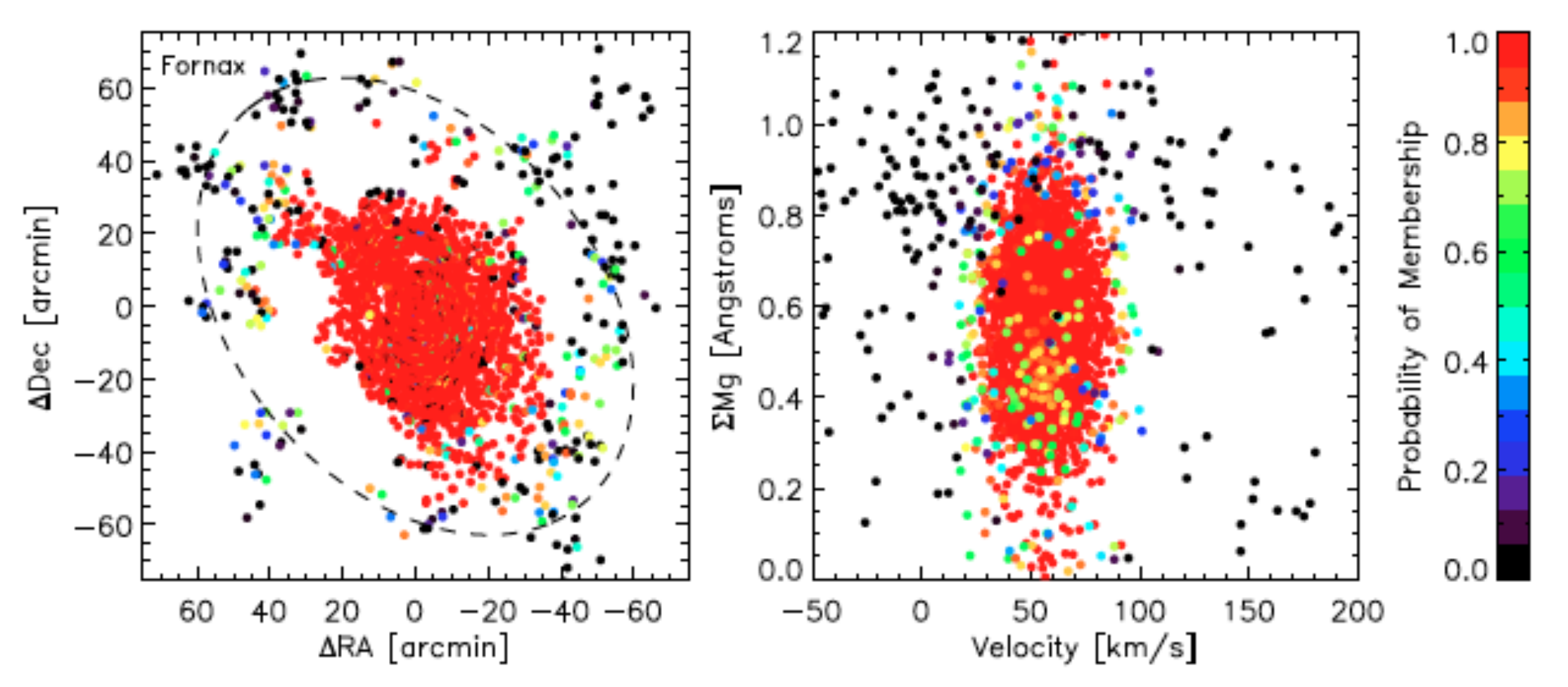}
  \caption{Magellan/MMFS spectroscopic data for Fornax \citep[updated to include the full sample of $\sim 3200$ members]{walker09a}.  \textit{Left:} Sky positions of individual stars.  \textit{Right:} velocities and spectral indices (pseudo-equivalent widths of the Mg-triplet absorption feature.  Color indicates membership probability, as estimated from position, velocity and spectral index distributions.  The ellipse indicates the limiting radius, $R_{\mathrm{K}}$ (Equation \ref{eq:king62}).}
  \label{fig:mmfsdata}
\end{figure}

\begin{figure}
  \epsscale{1}
  \plotone{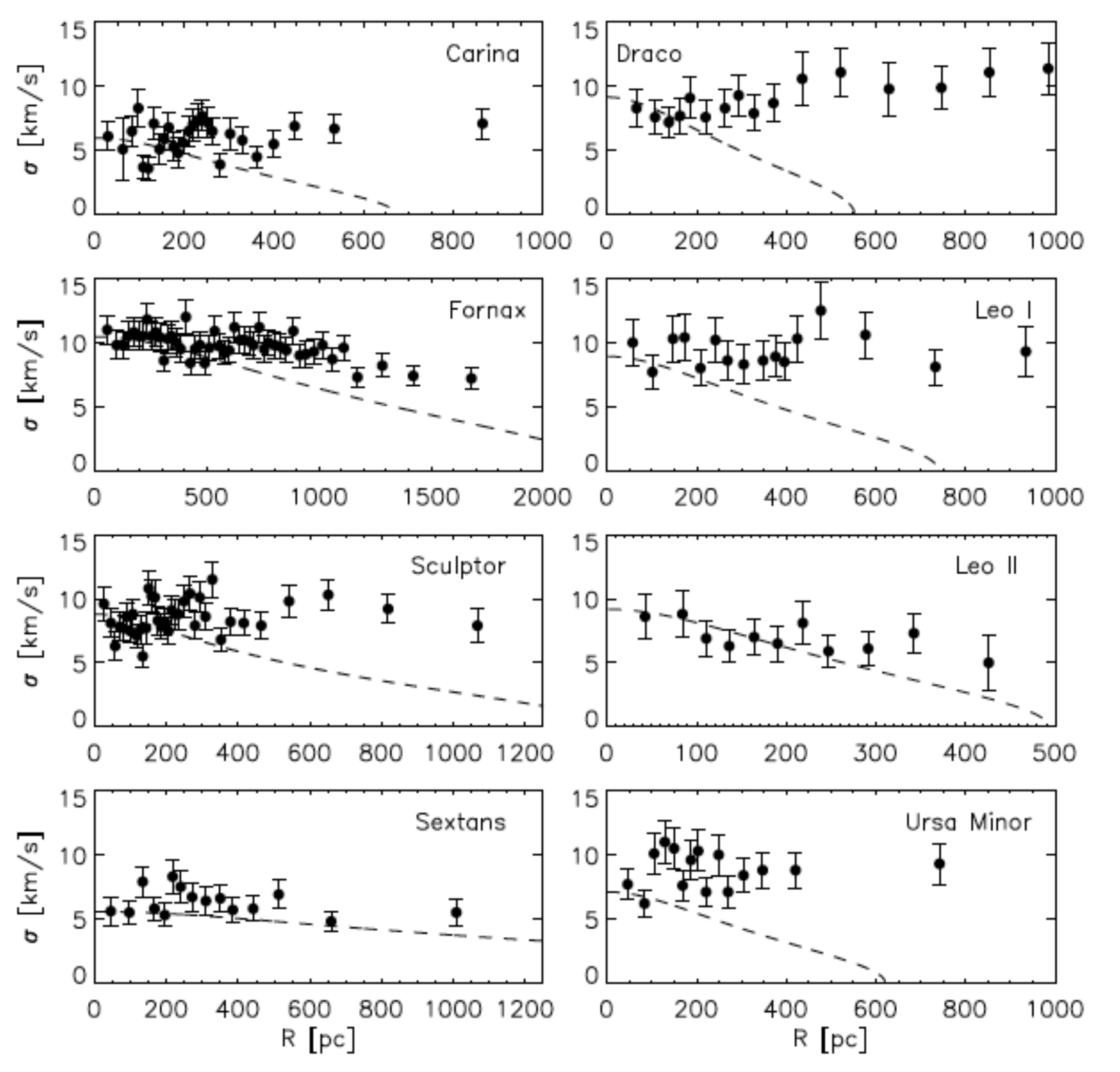}
  \caption{Velocity dispersion profiles observed for the Milky Way's eight `classical' dSphs \citep{walker07,mateo08,walker09d}.  See also \citet{kleyna02,kleyna04,wilkinson04,munoz05,munoz06,sohn07,koch07,koch07b,battaglia08,battaglia11}.  Overplotted are mass-follows-light \citet{king66} models (Section \ref{subsubsec:mfl}) normalised to reproduce the observed central dispersions.  Failure of these models to reproduce the large velocity dispersions at large radius provides the strongest available evidence that dSphs have dominant and extended dark matter halos.}
  \label{fig:mmfsmmtprofiles}
\end{figure}

Operating from $2004-2011$, the Michigan-MIKE Fiber Spectrograph (MMFS; $R\sim 20000$), built by Mario Mateo for the Magellan/Clay 6.5-meter telescope, provided what remain the largest homogeneous velocity samples for `classical' dSphs.  The public catalog of Magellan/MMFS velocities includes $\sim 775$, $\sim 2500$, $\sim 1365$ and $\sim 440$ members of Carina, Fornax, Sculptor and Sextans, respectively \citep{walker09a}.  Figure \ref{fig:mmfsdata} displays the Fornax data, including sky positions as well as the two quantities measured from each spectrum: line-of-sight velocity and a spectral index that indicates the pseudo-equivalent width of the Mg-triplet feature at $\sim 5170$ \AA.  Data from a similar survey conducted in the North with MMT/Hectochelle will soon become public.  Figure \ref{fig:mmfsmmtprofiles} displays velocity dispersion profiles calculated from the Magellan/MMFS and MMT/Hectochelle data, demonstrating that the luminous regions of classical dSphs have approximately constant velocity dispersion.

\subsubsection{(Necessarily) Small Samples}

With kinematic samples for `classical' dSphs growing exponentially, discoveries of `ultrafaint' dSphs with SDSS data generated a wave of interest in the faintest Milky Way satellites, and efforts to obtain spectroscopic follow-up began immediately.  \citet{kleyna05} contributed a first result echoing Aaronson's original study of Draco: from Keck/HIRES velocities for 5 members of Ursa Major I, \citet{kleyna05} estimate that $\sigma>6.5$ km s$^{-1}$ with $95\%$ confidence.  Given UMaI's low luminosity, simple dynamical models imply $M/L_V\ga 500 [M/L_V]_{\odot}$.  \citet{munoz06b} used the HYDRA multi-fiber spectrograph at the 3.5-meter WIYN telescope to measure velocities for 7 members of the Bo\"otes I dSph, obtaining a velocity dispersion of $\sim 6.5$ km s$^{-1}$ and $M/L_V\ga 130 [M/L_V]_{\odot}$.  

Whereas spectroscopic surveys of `classical' dSphs target bright red giant branch (RGB) stars, the least luminous `ultrafaint' satellites host few RGBs.  Samples for even tens of stars for such objects require observations of faint stars near the main sequence turnoff and are feasible only with the largest telescopes.  \citet{martin07} and \citet{simon07} used Keck/DEIMOS to observe tens of velocities in 10 of the 11 `ultrafaints' known at the time, measuring $\sigma\ga 3$ km s$^{-1}$ and concluding that these objects indeed have extremely large dynamical mass-to-light ratios, of order $M/L_V\ga 100 [M/L_V]_{\odot}$ and larger.  \citet[see example spectrum in the right panel of Figure \ref{fig:aaronson}]{geha09} and \citet{simon11} followed with a Keck/DEIMOS survey of Segue 1, measuring $\sigma \sim 4$ km s$^{-1}$ and concluding that this object is the `darkest galaxy', with $M/L_V\sim 3400 [M/L_V]_{\odot}$.  

\citet{aden09} and \citet{koposov11} used VLT/FLAMES to measure velocity dispersions for Hercules and Bo\"otes I, respectively, that while indicative of large dynamical mass-to-light ratios, are both smaller than previously measured with Keck/DEIMOS.  \citet{aden09} obtained a smaller velocity dispersion after using Str\"omgren photometric criteria to remove foreground interlopers.  \citet{koposov11} used a novel observing strategy that included $\sim 15$ individual 45-60 minute exposures taken over a month.  After measuring velocities for each exposure, \citet{koposov11} were able to resolve binary orbital motions directly and to exclude stars that showed significant velocity variability (Section \ref{subsec:binaries}).

\subsection{The Smallest Galaxies}
\label{subsec:smallest}

\begin{figure}
  \epsscale{1}
  \plotone{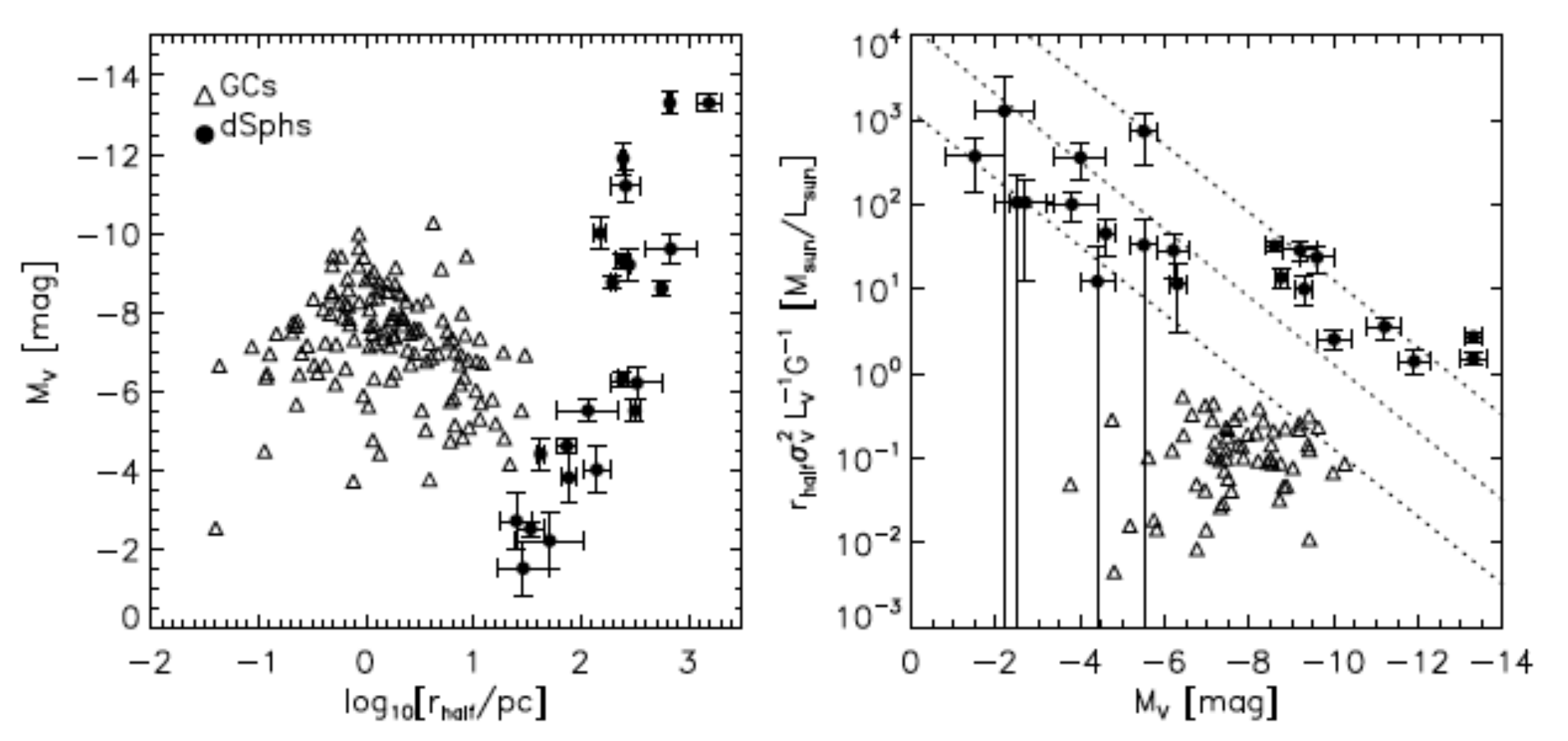}
  \caption{\textit{Left:} Luminosity versus size (updated from \citealt{belokurov07,gilmore07,martin08}), for Milky Way satellites including objects classified as globular clusters (open triangles; data from \citet[2010 edition]{harris96}) and dwarf spheroidals (filled circles with error bars; data from compilations by \citealt{ih95,mateo98,martin08,sand11}).  The apparent lack of objects toward low luminosity and large size is a selection effect that reflects the surface-brightness limit of the SDSS survey \citep{koposov08}.  \textit{Right:} Dynamical mass-to-light ratio (modulo a constant scale factor) versus luminosity.  Updated from \citet{mateo93,mateo98,simon07,geha09}.  Dotted lines correspond to constant masses of $10^5 M_{\odot}$, $10^6M_{\odot}$ and $10^7M_{\odot}$.}
  \label{fig:satellites}
\end{figure}

Figure \ref{fig:satellites} displays two scaling relations defined by the structural and kinematic observations discussed above, and lets one compare the properties of objects classified as dSphs directly with those of objects classified as globular clusters.  Parameters for globular clusters are adopted from the catalog of \citet[2010 edition]{harris96}.  

The left-hand panel of Figure \ref{fig:satellites} plots luminosity against size, characterised by the projected halflight radius \citep{belokurov07,gilmore07,martin08,sand11}.  Here one can appreciate another conclusion of Shapley's (1938b) regarding the first known dSphs: `The Sculptor and Fornax systems might be called greatly expanded giant clusters'.  Indeed, while the luminosity distributions of dSphs and globular clusters overlap substantially, most dSphs have $R_{h}\ga 100$ pc while most globular clusters have $R_{h}\la 10$ pc \citep{gilmore07}.\footnote{M31 hosts several `extended' globular clusters with halflight radii as large as several tens of pc \citep[e.g.,][]{huxor05}, but no similar population within the Milky Way has yet been discovered.}    The region between $10\la R_{h}/\mathrm{pc}\la 100$ is populated only by globular clusters with $M_V\la -4$ or by dSphs with $M_V\ga -4$.  

Current kinematic results suggest that this separation in luminosity between the smallest dSphs and the largest globular clusters is not merely an artifact of classification, but points to a fundamental structural difference.  The right-hand panel of Figure \ref{fig:satellites} plots the product $R_{h}\sigma^2/(L_VG)$ (dimensionally a mass-to-light ratio) against luminosity.  In the region of overlapping size, the less luminous objects tend to have larger velocity dispersions, amplifying the separation in luminosity such the smallest, faintest dSphs have the largest dynamical mass-to-light ratios of any known galaxies \citep{kleyna05,munoz06b,martin07,simon07,geha09,simon11}.\footnote{Some ambiguity regarding the masses of the smallest, faintest dSphs results from the convergence of three relevant quantities---the typical velocity measurement error, the measured velocity dispersions, and the potential contribution to the measured dispersions from binary orbital motions---on the same value, $\sim 3-4$ km s$^{-1}$ \citep{simon07,martin07,mcconnachie10,simon11,koposov11}.  For some faint dSphs the most compelling evidence for large amounts of dark matter comes from stellar-atmospheric chemistry rather than kinematics.  The faintest objects classified as dSphs tend to have metallicity dispersions ($\sigma_{\mathrm{[Fe/H]}}\ga 0.4$ dex, \citealt{geha09,norris10,kirby11,willman11}) that are indicative of prolonged and perhaps multiple episodes of star formation, thereby requiring gravitational potentials sufficiently deep to retain interstellar media despite pressures generated by stellar feedback.  An adequate discussion of the relationships between dSph kinematics and stellar chemistry is beyond the scope of the present work; \citet{tolstoy09} provide an excellent, recent review.}  Furthermore, the ultrafaint dSphs extend a relation under which less luminous dSphs have larger $M/L_V$ \citep[Section \ref{sec:universal}]{mateo93,mateo98}.  The discontinuity in dynamical $M/L_V$ between dSphs and globular clusters seems to mark a boundary between objects with dark matter and those without.

\section{Stellar Velocity Dispersion as a Proxy for Mass}
\label{sec:dispersion}

In rotating spiral galaxies, the circular velocity at radius $r$ relates directly to enclosed mass via $v_{\mathrm{circ}}^2=GM(r)/r$.  Ordered rotation in dSphs is dynamically negligible (see next section); instead, dSphs receive support against gravity primarily from the random motions of their stars.  Therefore the estimation of dSph masses is a fundamentally statistical enterprise.  The simplest statistic that characterises dSph stellar dynamics is the dispersion of velocities along the line of sight, $\sigma$.  For a relaxed system of characteristic size $R$, the virial theorem implies $\sigma^2\propto GM/R$.  In principle, measurements of a dSph's size and velocity dispersion provide a simple estimate of its mass (Section \ref{sec:masses}).  In practice, one must be aware of effects that can inflate measured values of the velocity dispersion above equilibrium values.  

\subsection{Rotation}
\label{subsec:rotation}

\begin{figure}
  \epsscale{0.6}
  \plotone{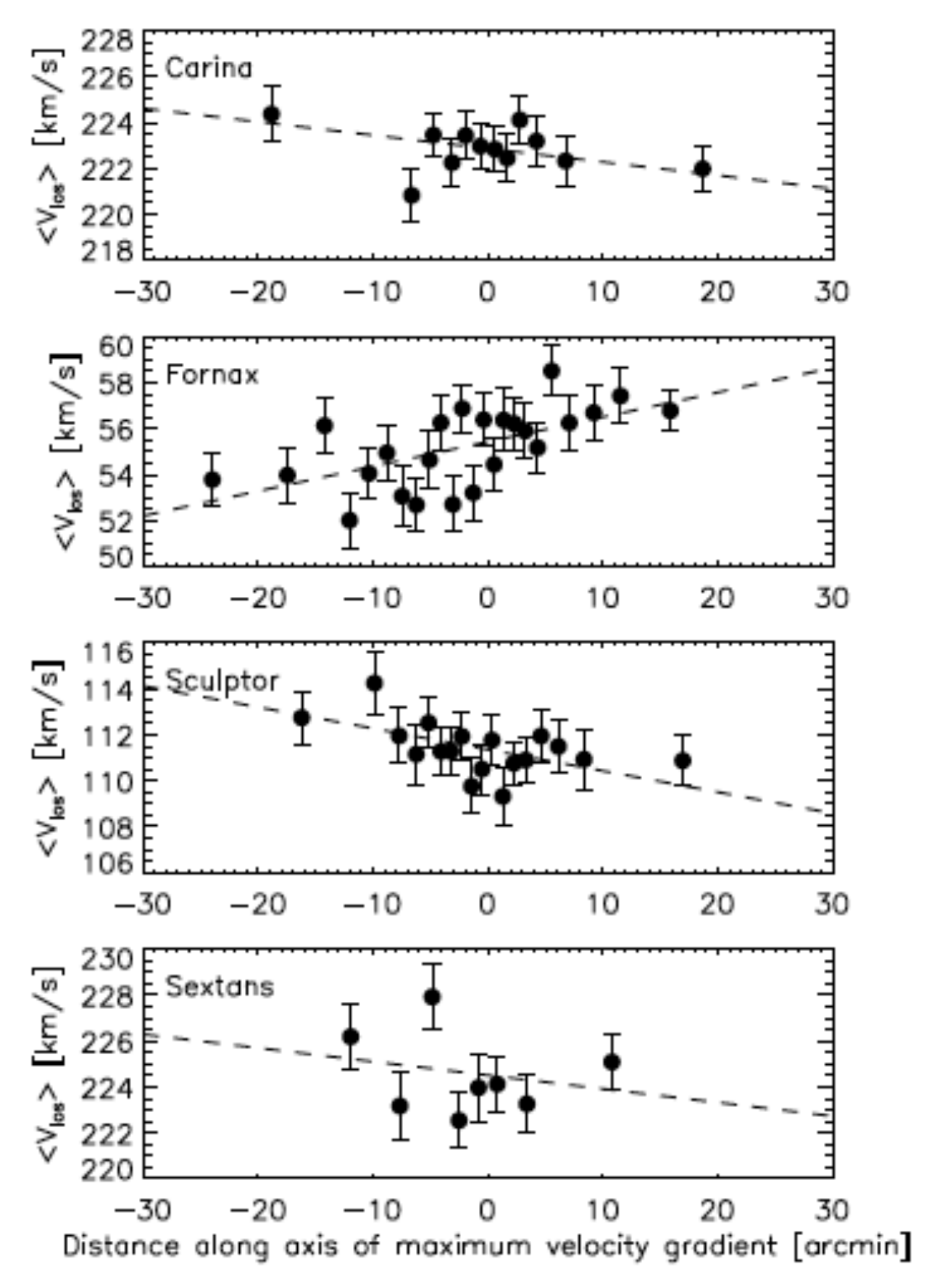}
  \caption{Line-of-sight velocity gradients in the Milky Way's `classical' dSph satellites.  Panels display mean velocity as a function of distance along the axis corresponding to the maximum velocity gradient, from the samples of \citet{walker09a}.  The observed gradients give rise to maximum amplitudes $R_{\mathrm{max}}dV/dR\la 3$ km s$^{-1}$, significantly less than the velocity dispersions ($\sigma \sim 10$ km s$^{-1}$).  Using VLT spectra from the DART survey, \citet{battaglia08} report similar results for Sculptor, measuring a gradient of $7.6_{-2.2}^{+3.0}$ km s$^{-1}$deg$^{-1}$ along the morphological major axis.}
  \label{fig:dsph_rotation}
\end{figure}

In the simplest case, solid-body rotation about an axis misaligned with the line of sight will induce a gradient, $dV/dR$, in the line-of-sight velocity distribution.  For example, Figure \ref{fig:dsph_rotation} plots mean velocity along the axes for which the observed velocity gradients are maximal in Carina, Fornax, Sculptor and Sextans.  Even at the radii of the outermost observed stars, any ordered motion due to rotation is limited to $R_{\mathrm{max}}dV/dR\la 3$ km s$^{-1}$, negligible compared with the observed velocity dispersions of $\sigma \sim 10$ km s$^{-1}$.  

For Carina and Fornax the amplitude and orientation of the observed velocity gradients are consistent with a perspective effect induced not by rotation, but rather by these dSphs' systemic orbital motions transverse to the line of sight \citep{kaplinghat08,walker08}, as measured from HST astrometry \citep{piatek02,piatek03,piatek07}.  The observed signal in Sculptor cannot be attributed to its measured proper motion \citep{schweitzer95,piatek06}, and thus Sculptor may have a residual rotational component \citep{battaglia08}, albeit one that contributes weakly ($v_{\mathrm{rot}}/\sigma\la 0.5$) to the measured velocity dispersion.

\subsection{External Tides}
\label{subsec:tides}

All dSphs considered here orbit within the gravitational potential of the Milky Way and are therefore susceptible to external influence from tidal forces.  Tides can affect the structure and kinematics of a given satellite through a variety of mechanisms, including stripping, shocking, `stirring' and various orbital resonances.  Depending on the strengths of interactions and timescales of subsequent relaxation, tides might inflate observed velocity dispersions and mass estimates based thereupon.  

In the most extreme scenarios, tides have been invoked to explain dSph velocity dispersions without dark matter \citep[e.g.,][]{kuhn89,kuhn93,kroupa97,fleck03,metz07}.  Such explanations are most plausible for dSphs that are closest to the Milky Way and exhibit elongated stellar structures, e.g., Ursa Major II \citep{zucker06b,munoz10}, Hercules \citep{belokurov07,coleman07,martin10}, and possibly Segue 1 \citep{niederste-ostholt09}.  However, generalisation of purely tidal mechanisms to explain the apparent dark matter content of the entire dSph population does not account without contrivance for the wide distribution of dSph distances ($\sim 30\la D/\mathrm{kpc}\la 250$) or for a monotonic metallicity-luminosity relation \citep{mateo98,kirby08,kirby11}.  

While tidal stripping involves the transfer of mass from the satellite to the parent outside a particular boundary\footnote{Equation \ref{eq:tide} gives the tidal radius for the idealised case of point-mass potentials.  Calculations and simulations by \citet{read06} demonstrate that stars are actually lost from various depths depending on the internal mass distributions of satellite and parent, as well as on the properties of the stellar orbits themselves (e.g., prograde versus retrograde with respect to the orbit of the dSph about the Galaxy).}, tidal `shocking' involves impulsive injection of energy into the satellite as it plunges through the disk and/or near the center of the parent system \citep[e.g.,][]{gnedin99}.  While the former process removes mass preferentially from the satellite's outer regions, the latter process tends to decrease its central density \citep{read06b}.  Early numerical simulations by \citet{pp95} and \citet{oh95} suggest that even strong tidal interactions do not significantly inflate a satellite's central velocity dispersion, which therefore can remain a reliable indicator of dynamical mass.  

Many recent N-body simulations examine specific phenomenology associated with tidal interactions.  For example, \citet{read06b} use simulations to demonstrate that the projection of tidal streaming motions along most viewing angles tends to cause velocity dispersion profiles to increase at large radius, and then argue that the lack of such upturns in the classical dSphs (with the possible exception of Draco---see Figure \ref{fig:mmfsmmtprofiles}) limits the severity of current disruption events.  \citet{penarrubia08b} use simulations to show that repeated tidal encounters cause monotonic declines in equilibrium values (as evaluated at apocenter) of the satellite's central surface brightness, velocity dispersion and scale radius.  Further simulations by \citet{penarrubia09} indicate that when a dark matter halo is present, tides do \textit{not} generate a clear truncation in the surface brightness profile of the bound remnant, as otherwise is prescribed by the self-consistent dynamical model of \citet{king66}.  Rather, as the bound remnant relaxes after a pericentric encounter, tidal debris generates an `excess' of stars at radii where the local crossing time ($R_c/\sigma\sim 10$ Myr at the core radius of a typical dSph) exceeds the time elapsed since the encounter.  

Such simulations provide a context for evaluating the tidal origin of breaks and bumps in observed surface brightness profiles of individual dSphs (Section \ref{subsubsec:extended} and Figure \ref{fig:hodgeih95}), and for gauging the severity with which tides influence the observed kinematics.  In general, one can expect tides to have stronger and more enduring influence on the outer regions of dSphs, and for the degree of influence on any particular satellite to scale with orbital parameters.  A typical dSph with $R_c\sim 100$ pc and $\sigma\sim 10$ km s$^{-1}$ requires $\sim 20$ core crossing times to travel a distance of $\sim 100$ kpc (a typical Galactocentric distance) at speed $\sim 250$ km s$^{-1}$; the assumption of dynamic equilibrium should therefore hold reasonably well out to several core radii for the majority of dSphs.  In practice one must use all available information about a satellite's orbit and outer structure to evaluate the likely contribution of tides on a case-by-case basis.  

\subsection{Binary Stars}
\label{subsec:binaries}

Unresolved binary orbital motions might contribute significantly to the observed velocity dispersions of the least massive galaxies.  \citet{olszewski95} use 112 independent velocity measurements for 42 stars in Draco and Ursa Minor to identify seven stars that exhibit velocity variability.  Elimination of these stars from their samples has negligible impact on the measured velocity dispersions of Draco and Ursa Minor.  Using the larger KPNO/HYDRA sample of \citep{armandroff95} ($373$ independent velocity measurements for $185$ stars), \citet{edo96} perform Monte Carlo simulations to estimate that the binary frequency for Draco and Ursa Minor stars with periods of $\sim 1$ year is $\sim 0.2-0.3$ per decade of period.  Even though this fraction is larger than the one found in the solar neighborhood \citep[e.g.,][]{duquennoy91}, further simulations by \citet{edo96} and \citet{hargreaves96b} demonstrate that the scatter introduced by binaries is small compared to the measured dispersions of $\sigma\sim 10$ km s$^{-1}$, and thus that binaries do not significantly inflate dynamical masses of the `classical' dSphs.  

More recently, measurements of velocity dispersions as small as $\sigma\sim 3$ km s$^{-1}$ for several ultrafaint satellites \citep{simon07,martin07,walker09c,aden09,koposov11} have renewed concerns over the possible contribution of binary motions.  While simulations by \citet{minor10} find that binaries have little effect on dispersions measured for systems with intrinsic velocity dispersions $\sigma\ga 4$ km s$^{-1}$, complementary simulations by \citet{mcconnachie10} demonstrate that for systems with intrinsic dispersions near zero (as would be expected for many of the ultrafaints if they contain no dark matter), binaries can inflate measured dispersions to values as high as $\sigma\sim 4$ km s$^{-1}$.  

It is therefore necessary to verify the extreme mass-to-light ratios of the faintest (and coldest) dSphs with repeat spectroscopic measurements that constrain the velocity variability of individual stars.  \citet{simon11} present second-epoch Keck/DEIMOS velocity measurements for several of the Segue 1 stars first measured by \citet{geha09}.  They find that one of Segue 1's six red giants shows significant velocity variability, as do two fainter stars.  Including parametric binary orbital distribution functions in a Bayesian analysis of Segue 1's velocity dispersion, \citet[see also \citealt{martinez11}]{simon11} estimate a velocity dispersion of $3.7_{-1.1}^{+1.4}$ km s$^{-1}$, implying a dynamical mass-to-light ratio of $M/L_V\sim 3400 [M/L_V]_{\odot}$ and reinforcing their previous conclusion that Segue 1 is the `darkest' galaxy known.  

\begin{figure}
  \epsscale{1}
  \plotone{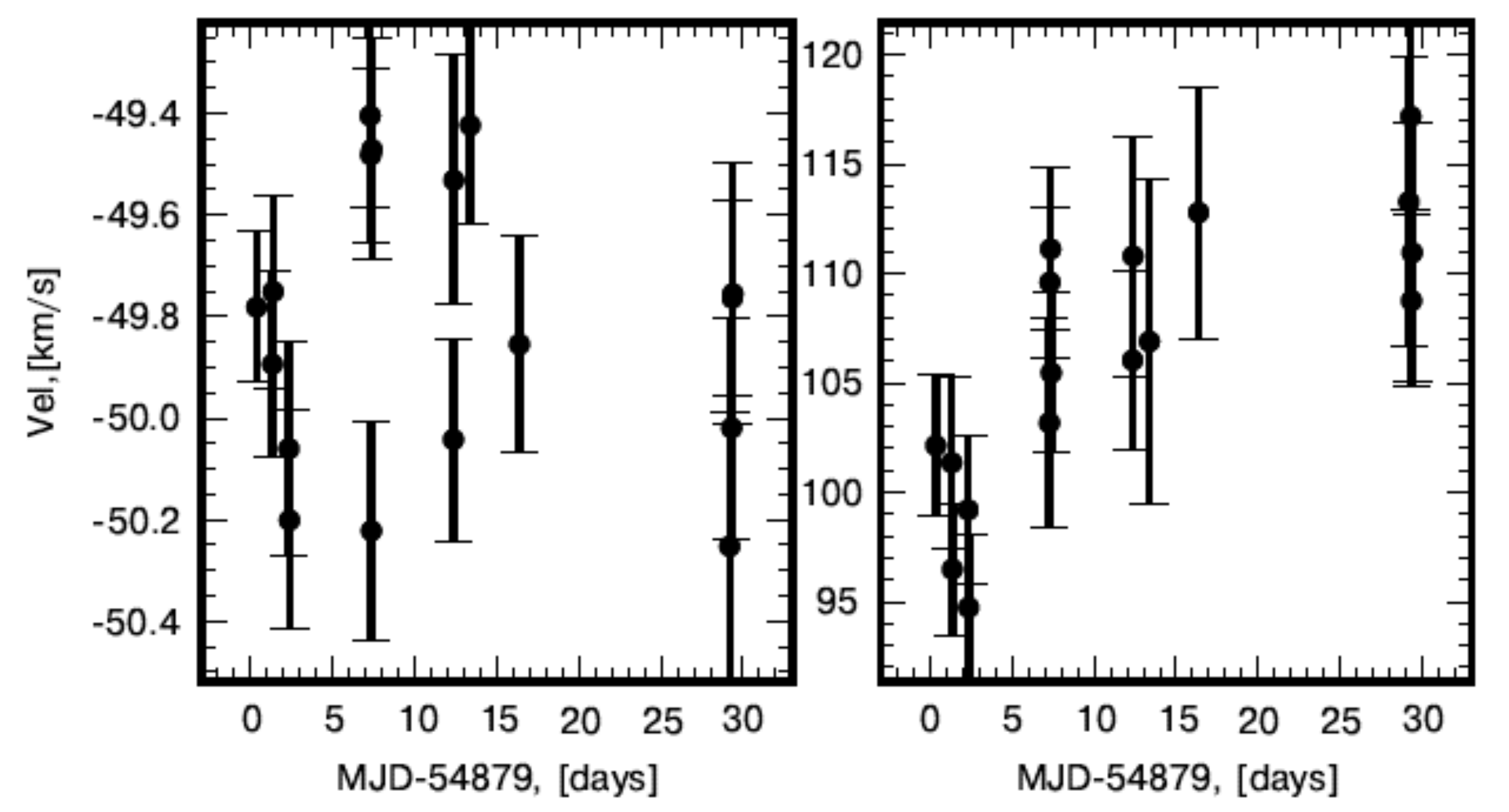}
  \caption{Direct detection of binary stars from individual velocity measurements in Bo\"otes I \citep[reproduced by permission of the American Astronomical Society]{koposov11}.  \textit{Left:} foreground star with velocities consistent with a constant value.  \textit{Right:} member star ($\langle v\rangle_{\mathrm{Boo}}\sim 105$ km s$^{-1}$) exhibiting a velocity increase of $\sim 10$ km s$^{-1}$ over one month.  See section \ref{subsec:binaries} for a discussion of the effects of binary orbital motions on measurements of dSph velocity dispersions.}
  \label{fig:boobinary}
\end{figure}

\begin{figure}
  \epsscale{1}
  \plotone{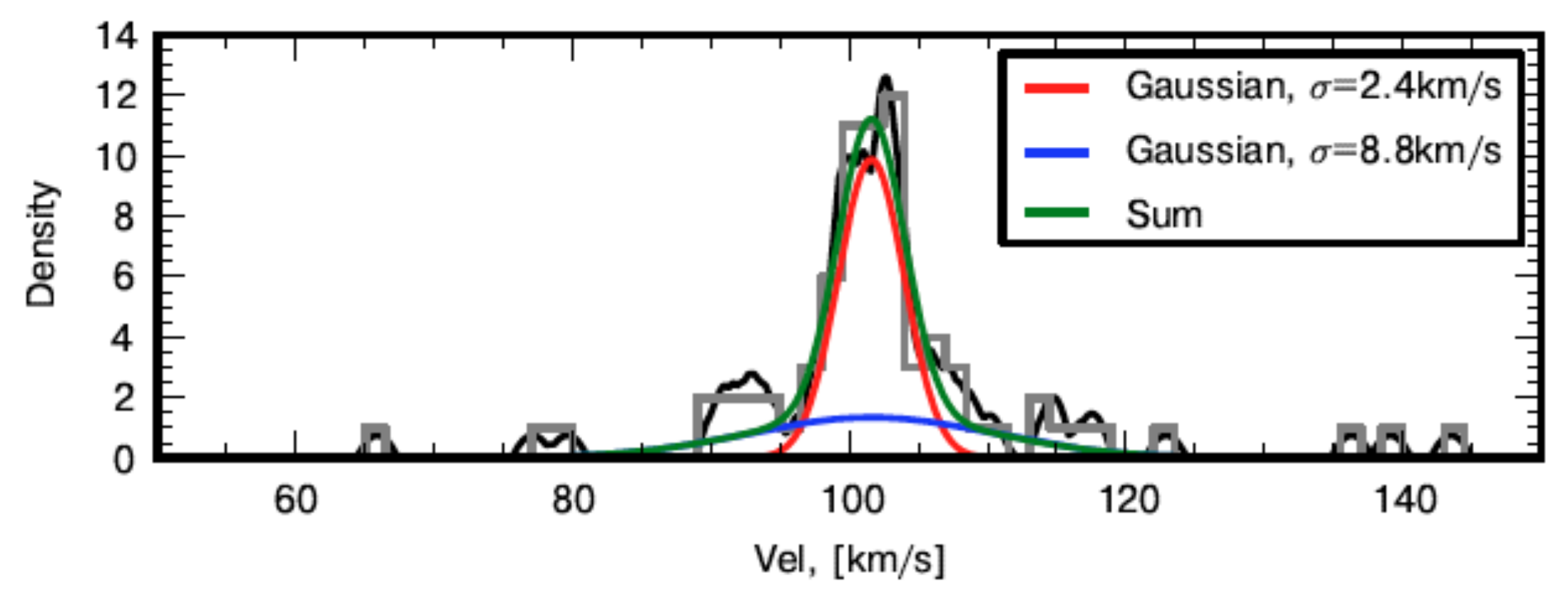}
  \caption{Velocity distribution of the Bo\"otes I dSph, from the VLT/FLAMES data of \citet[reproduced by permission of the American Astronomical Society]{koposov11}, not including $11$ stars observed to have significant velocity variability.  In the best-fitting two-Gaussian model, $70\%$ of the stars belong to a `cold' component with $\sigma\sim 2$ km s$^{-1}$ and the rest belong to a hotter component with $\sigma\sim 9$ km s$^{-1}$. }
  \label{fig:boo1}
\end{figure}

In a separate study, \citet{koposov11} use $\sim 15$ VLT/FLAMES observations obtained over one month in order to resolve binary motions directly among members of Bo\"otes I.  Figure \ref{fig:boobinary} displays independent velocity measurements for two stars as a function of time.  While velocities for the star in the left-hand panel are consistent with a constant velocity, velocities for the star in the right-hand panel change systematically by $\sim 10$ km s$^{-1}$.  After discarding probable binaries, \citet{koposov11} find that most members of Bo\"otes I belong to a cold population with dispersion $\sigma=2.4_{-0.5}^{+0.9}$ km s$^{-1}$ (Figure \ref{fig:boo1}), significantly smaller than previous single-epoch estimates of $\sigma\sim 6.5$ km s$^{-1}$ \citep{munoz06b,martin07}.  At present the contribution of binaries to the velocity dispersions measured for the coldest dSphs is poorly understood, and more multi-epoch studies are required.  

\section{dSph Masses}
\label{sec:masses}

For a collisionless stellar system in dynamic equilibrium, the gravitational potential, $\Phi$, relates to the phase-space distribution of stellar tracers\footnote{The distribution function is defined such that $f(\vec{r},\vec{v},t)d^3\vec{x}d^3\vec{v}$ specifies the number of stars inside the volume of phase space $d^3\vec{x}d^3\vec{v}$ centered on $(\vec{x},\vec{v})$ at time $t$.  }, $f(\vec{r},\vec{v},t)$, via the collisionless Boltzmann Equation (Equation 4-13c of \citealt{bt08}):
\begin{equation}
  \frac{\partial f}{\partial t}+\vec{v}\cdot \nabla f-\nabla\Phi\cdot \frac{\partial f}{\partial\vec{v}}=0.
  \label{eq:cbe}
\end{equation}

Current instrumentation resolves the internal distributions of neither distance nor proper motions for dSph stars.  The structural and kinematic observations described in section \ref{sec:observations} provide information only about the projections of phase space distributions along lines of sight, limiting knowledge about $f$ and hence also about $\Phi$.  Therefore all efforts to translate existing data sets into constraints on $\Phi$ involve simplifying assumptions.  Along with dynamic equilibrium, common assumptions include spherical symmetry and particular functional forms for the distribution function and/or the gravitational potential.  The most useful analyses identify the least restrictive assumptions that are appropriate for a given data set.  Modern structural/kinematic data contain information that is sufficient to place reasonably robust constraints not only on the amount of dSph mass, but in some cases also on its spatial distribution.

\subsection{Amount}
\label{subsec:amount}

\subsubsection{`Mass Follows Light' Models}
\label{subsubsec:mfl}

A common method for analysing dSph kinematics employs the following assumptions:

\noindent 1. dynamic equilibrium;

\noindent 2. spherical symmetry;

\noindent 3. isotropy of the velocity distribution, such that $\langle v_r^2\rangle=\langle v_{\theta}^2\rangle=\langle v_{\phi}^2\rangle$;

\noindent 4. a single stellar component;

\noindent 5. the mass density profile, $\rho(r)$, is proportional to the luminous density profile, $\nu(r)$ (i.e., $M/L$ is constant, or `mass follows light').  

Historically, assumption (5) has been adopted when the velocity dispersion profile $\sigma(R)$ is unavailable.  Examples include early analyses of classical dSphs \citep[][and references therein]{mateo98} and initial analyses of ultrafaint dSphs \citep[e.g.,][]{kleyna05,munoz06b,martin07,simon07}.  Under this assumption, the steeply falling outer surface brightness profiles of dSphs (Section \ref{subsubsec:classical} and Figure \ref{fig:hodgeih95}) motivate the use of dynamical models that allow for truncation by external tides \citep[e.g.,][]{michie63,king66}.  Consider, for example, the model of \citet[see also Chapter 4 of \citealt{bt08}]{king66}, in which the distribution function depends only on energy:
\begin{equation}
  f(\varepsilon)=k(2\pi v_s^2)^{-3/2}\bigl (\exp[\varepsilon/v_s^2]-1\bigr ).
\label{eq:king}
\end{equation}
The relative potential, $\Psi\equiv \Phi_0-\Phi$, is defined such that $\varepsilon\equiv \Psi-\frac{1}{2}v^2\geq 0$ at radii $r\leq R_{\mathrm{K}}$.  Given the assumption that mass follows light, this model is fully specified by dimensionless parameter $\Psi(0)/v_s^2$ (or equivalently, $R_{\mathrm{K}}/R_c$), the core radius $R_c$ and one of either the central velocity dispersion $\sigma_0$ or central mass density $\rho_0$, which are all related by $R_c^2=\frac{9\sigma^2_0}{4\pi G\rho_0}$.

\citet{illingworth76} shows that under assumptions (1)-(5) and a distribution function of the form specified by Equation \ref{eq:king}, the total mass is given by
\begin{equation}
  \frac{M_{\mathrm{tot}}}{M_{\odot}}=167\eta \biggl [\frac{R_c}{\mathrm{pc}}\biggr ] \biggl [\frac{\sigma^2}{\mathrm{km^2s^{-2}}}\biggr ],
  \label{eq:illingworth}
\end{equation}
where parameter $\eta$ is determined by concentration $c\equiv \log_{10}[R_{\mathrm{K}}/R_{\mathrm{c}}]$.  Following \citet{mateo98}, many authors adopt $\eta\approx 8$, which is appropriate for the low concentrations characteristic of classical dSphs but is not well-constrained for the faintest dSphs.  

\begin{figure}
  \epsscale{0.6}
  \plotone{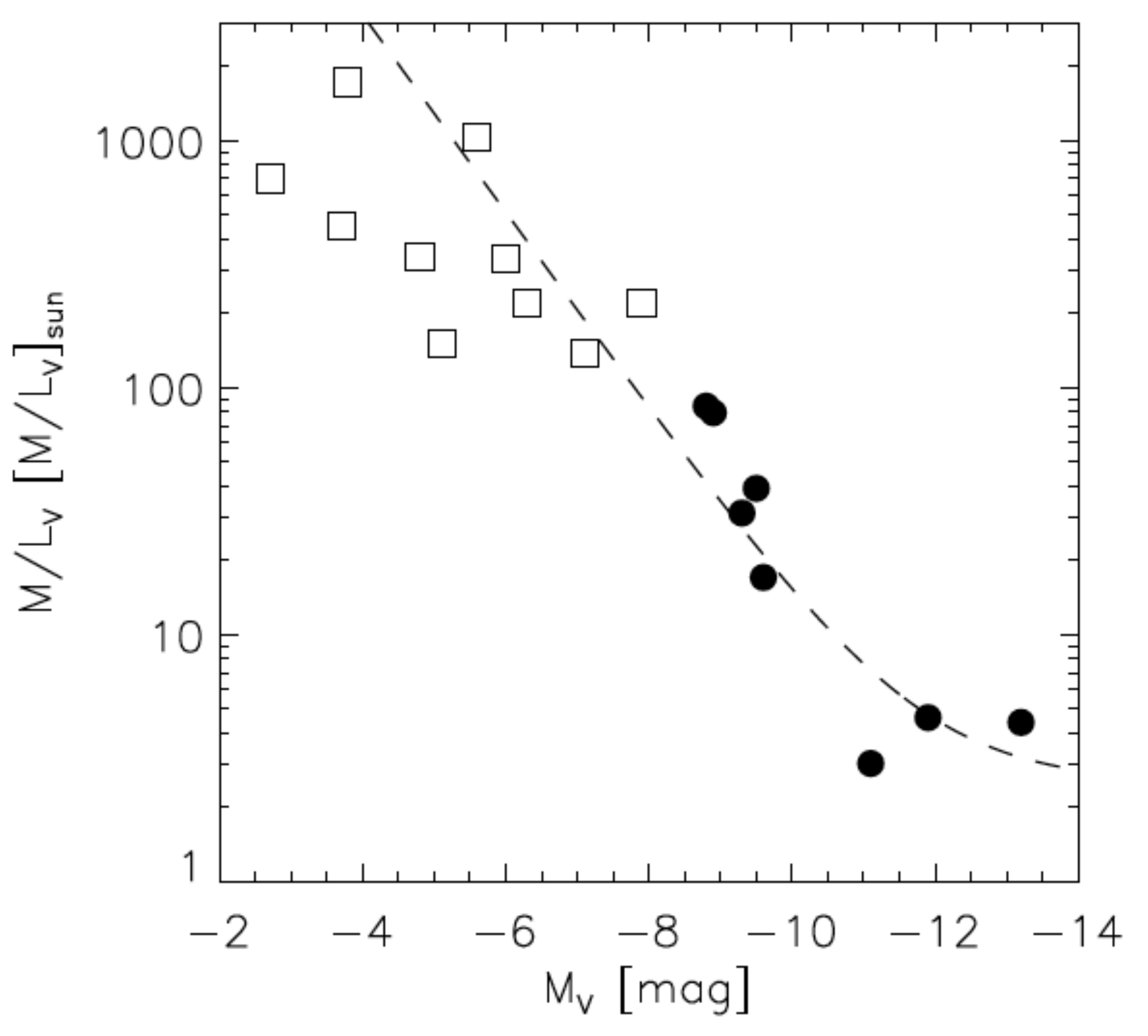}
  \caption{Dynamical mass-to-light ratio, derived from mass-follows-light models (Section \ref{subsubsec:mfl}), versus luminosity.  Data for the Milky Way's `classical' dSphs (filled circles) are from the review of \citet{mateo98}.  Data for the `ultrafaint' dSphs (open squares) are from Keck/DEIMOS observations by \citet{simon07} and \citet{martin07}.  The dotted line corresponds to $[M/L_V/[M/L_V]_{\odot}]=2.5+10^7/(L/L_{V,\odot})$ (Section \ref{sec:universal}).}
  \label{fig:mateosimongeha}
\end{figure}

More generally, \citet{richstone86} show that for `almost any' spherical, isotropic system with constant mass-to-light ratio and centrally cored luminosity profile, the dynamical mass-to-light ratio is approximately
\begin{equation}
  \frac{M}{L}\approx \frac{9\sigma_0^2}{2\pi G \Sigma(0) R_{\mathrm{hb}}},
  \label{eq:corefit}
\end{equation}
where the half-brightness radius is defined by $\Sigma(R_{\mathrm{hb}})= \frac{1}{2}\Sigma(0)$ and is often similar (within $\sim 25\%$) to $R_c$.  Figure \ref{fig:mateosimongeha} plots dynamical mass-to-light ratios calculated from Equation \ref{eq:corefit} against dSph luminosity.  Masses obtained for the Milky Way's eight classical dSphs are all $M_{\mathrm{tot}}\sim 10^7 M_{\odot}$ \citep{mateo98}.  For the ultrafaints, masses range from $10^5\la M_{\mathrm{tot}}/M_{\odot}\la 10^7$ \citep{martin07,simon07}.  Dynamical mass-to-light ratios increase monotonically with decreasing luminosity, ranging from $10\la M/L_V/[M/L_V]_{\odot}\la 1000$.  

\subsubsection{Does Mass Follow Light?}
\label{subsubsec:doesmfl}

When available, empirical velocity dispersion profiles provide simple tests of the assumptions listed above.  Dashed lines in Figure \ref{fig:mmfsmmtprofiles} display best-fitting \citet[Equation \ref{eq:king}]{king66} models constructed under assumptions (1)-(5), fit to the surface brightness profiles of \citet[Figure \ref{fig:hodgeih95}]{ih95}, and normalised to fit the central velocity dispersions.  For all of the Milky Way's classical dSphs, the central velocity dispersions imply large mass-to-light ratios $M/L_V\ga 10 [M/L_V]_{\odot}$, as plotted in Figure \ref{fig:mateosimongeha}.  However, the mass-follows-light models underpredict the observed velocity dispersions at large radii, which tend to remain approximately constant out to the outermost measured points.  

These discrepancies between empirical velocity dispersion profiles and mass-follows-light models in Figure \ref{fig:mmfsmmtprofiles} imply that at least one of assumptions (1)-(5) is invalid.  In fact \textit{all} are invalid at some level.  However, various studies indicate that it is unlikely any one of assumptions (1)-(4) alone can be the problem.  For example:
 
\noindent 1. Simulations by \citet{read06b} suggest that tidal disruption (a violation of the equilibrium assumption) is more likely to generate rising velocity dispersion profiles rather than the flat profiles observed for real dSphs.

\noindent 2. Axisymmetric models of Fornax considered by \citet[see Section \ref{subsubsec:models}]{jardel11} favor dark matter halos that, while violating the assumption of spherical symmetry, extend well beyond the luminous component and therefore also invalidate the mass-follows-light assumption.

\noindent 3. \citet{evans09} derive analytically an expression for the anisotropy profile, $\beta_a(r)\equiv 1-\langle v^2_{\theta}\rangle/\langle v^2_{r}\rangle$ in terms of surface brightness and velocity dispersion profiles.  If mass follows light, then the flat empirical velocity dispersion profiles tend to imply unphysical values $\beta_a>1$.

\noindent 4. While recent observations indicate that some dSphs contain at least two distinct stellar sub-populations (Sections \ref{subsubsec:peculiar} and \ref{subsubsec:models}), scale radii of the individual sub-populations are sufficiently well constrained \citep[e.g.,][]{battaglia06,battaglia08} that superpositions of two mass-follows-light models continue to underpredict the observed velocity dispersions at large radii.  Indeed \citet{battaglia08} find that the flat velocity dispersion profile they measure for Sculptor's more spatially extended subpopulation continues to imply an even more extended dark matter halo.

Models that allow for sufficiently extended dark matter halos (violating assumption (5) while retaining assumptions (1)-(4)) can provide good fits to surface brightness and velocity dispersion profiles simultaneously \citep[e.g.,][]{pryor90,wilkinson02}.  On these grounds, the empirical profiles shown in Figures \ref{fig:hodgeih95} and \ref{fig:mmfsmmtprofiles} provide the strongest available evidence that dSphs have dominant dark matter halos that extend beyond luminous regions.

Some scenarios for dSph formation and evolution---particularly the tidal stirring mechanism of \citet[Section \ref{subsec:tides}]{mayer01a,mayer01b} and the tidal disruption simulations of \citet{munoz08}---tend to produce configurations in which mass approximately follows light.  The results discussed above seem to rule out this configuration.  However, \citet{lokas09} finds reasonable agreement with mass-follows-light models in Carina, Fornax, Sculptor and Sextans after trimming velocity samples in order to remove member stars classified by an iterative mass estimator \citep{klimentowski07} as unbound.  This result rests in part on a circular argument, as the adopted mass estimator \citep{heisler85} is based on the virial theorem, which itself assumes mass follows light.  However, the same charge of circularity can be brought against the standard kinematic analysis, in which the inclusion of stars at large radius in the kinematic analysis implicitly assumes they are bound by a sufficiently extended dark matter halo.  Thus conclusions regarding the extended structure of dSph dark matter halos are generally sensitive to the assumptions employed when determining which stars to consider or reject in kinematic analyses.  More secure is the conclusion that dark matter dominates dSph potentials: even mass-follows-light models require central mass-to-light ratios $M/L_V\ga 10 [M/L_V]_{\odot}$ in order to fit the central velocity dispersions of dSphs \citep[and Figure \ref{fig:mmfsmmtprofiles}]{munoz08,lokas09}.  

\subsubsection{Jeans Analysis}
\label{subsec:jeans}

The methods for mass estimation described in the previous section either employ directly or are derived from specific distribution functions $f(\vec{r},\vec{v})$ that correspond to physical dynamical models restricted by particular assumptions.  Integration of Equation \ref{eq:cbe} over velocity space provides an alternative starting point in the form of the Jeans equations \citep[see][]{bt08}.  With spherical symmetry one obtains
\begin{equation}
  \frac{1}{\nu}\frac{d}{dr}(\nu \langle v_r^2\rangle)+2\frac{\beta_a\langle v_r^2\rangle}{r}=-\frac{GM(r)}{r^2},
  \label{eq:jeans}
\end{equation}
where $\nu(r)$, $\langle v_r^2\rangle (r)$, and $\beta_a(r)\equiv 1-\langle v_{\theta}^2\rangle/\langle v_r^2\rangle$ describe the 3-dimensional density, radial velocity dispersion, and orbital anisotropy, respectively, of the (stellar) tracer component.  The mass profile, $M(r)$, includes contributions from any dark matter halo.  While there is no requirement that mass follow light, there is also no guarantee that a given solution to Equation \ref{eq:jeans}---even one that fits the data---corresponds to a physical dynamical model (i.e., one for which $f(\vec{r},\vec{v})$ is non-negative).  

Equation \ref{eq:jeans} has general solution \citep{vandermarel94,mamon05}
\begin{equation}
  \nu\langle v^2_r\rangle=\frac{1}{f(r)}\displaystyle\int_{r}^{\infty}f(s)\nu(s)\frac{GM(s)}{s^2}ds,
  \label{eq:jeanssolution}
\end{equation}
where $f(r)=2f(r_1)\exp\int_{r_1}^{r}\beta_a(s)s^{-1}ds$.
Projecting along the line of sight, the mass profile relates to observable profiles, the projected stellar density, $\Sigma(R)$ (Figure \ref{fig:hodgeih95}), and velocity dispersion, $\sigma(R)$ (Figure \ref{fig:mmfsmmtprofiles}), according to \citep{bt08}
\begin{equation}
  \sigma^2(R)\Sigma(R)=2\displaystyle \int_{R}^{\infty}\biggl (1-\beta_a\frac{R^2}{r^2}\biggr ) \frac{\nu \langle v_r^2\rangle r}{\sqrt{r^2-R^2}}dr.
  \label{eq:jeansproject}
\end{equation}
Equation \ref{eq:jeansproject} forms the basis for many methods of mass estimation, including parametric \citep[e.g.,][]{strigari06,strigari08,strigari10,koch07,koch07b,battaglia08,walker07b,walker09d,martinez11} and nonparametric \citep[e.g.][]{wang05} techniques as well as algebraic inversion \citep[e.g.,][]{wilkinson04,gilmore07}.  

All methods based on Equation \ref{eq:jeansproject} are limited fundamentally by a degeneracy between the function of interest, $M(r)$, and the anisotropy profile, $\beta_a(r)$, which is poorly constrained by velocity data confined to the line of sight.\footnote{\citet{lokas05} develop a Jeans analysis that uses higher-order velocity moments (e.g., $\langle v^4\rangle$) in order to reduce degeneracy between anisotropy (assumed to be constant) and total mass (effectively normalising a cusped mass profile that is assumed to have $\gamma=1$ in the notation of Equation \ref{eq:rho}).}  Consideration of a common parametric method helps to illustrate this limitation.  For example, it is common to assume that the the gravitational potential is dominated everywhere by a dark matter halo with mass density profile 
\begin{equation}
  \rho(r)=\rho_s\biggl (\frac{r}{r_s}\biggr )^{-\gamma}\biggl [1+\biggl (\frac{r}{r_s}\biggr )^{\alpha}\biggr ]^{\frac{\gamma-\beta}{\alpha}},
  \label{eq:rho}
\end{equation}
i.e., the generalisation by \citet{zhao96} of the \citet{hernquist90} profile.  Equation \ref{eq:rho} provides a flexible halo model in the form of a split power-law, with free parameter $\alpha$ controlling the transition from index $-\gamma$ at small radii ($r\ll r_s$) to a value of $-\beta$ at large radii ($r\gg r_s$).  

From spherical symmetry, the density profile specifies the mass profile via 
\begin{equation}
  M(r)=4\pi\int_0^r s^2\rho(s)ds
  \label{eq:mass}
\end{equation}
and the surface brightness profile specifies the (deprojected) stellar density profile via
\begin{equation}
  \nu(r)=-\frac{1}{\pi}\displaystyle\int_{r}^{\infty}\frac{d\Sigma}{dR}\frac{dR}{\sqrt{R^2-r^2}}.
  \label{eq:nu}
\end{equation}
Given values for free parameters $\rho_s$, $r_s$, $\alpha$, $\beta$, $\gamma$ and an assumption about the otherwise unconstrained anisotropy profile\footnote{Typical assumptions about anisotropy range in simplicity from $\beta_a=0$ or $\beta_a=$constant to $\beta_a(r)=(\beta_{\infty}-\beta_0)r^2/(r_{\beta}^2+r^2)+\beta_0$ \citep[e.g.,][]{strigari10}, introducing as many as three new free parameters.}, Equation \ref{eq:jeansproject} specifies a product $\Sigma(R)\sigma^2(R)$ that can then be compared with observations.  

\begin{figure}
  \epsscale{0.5}
  \plotone{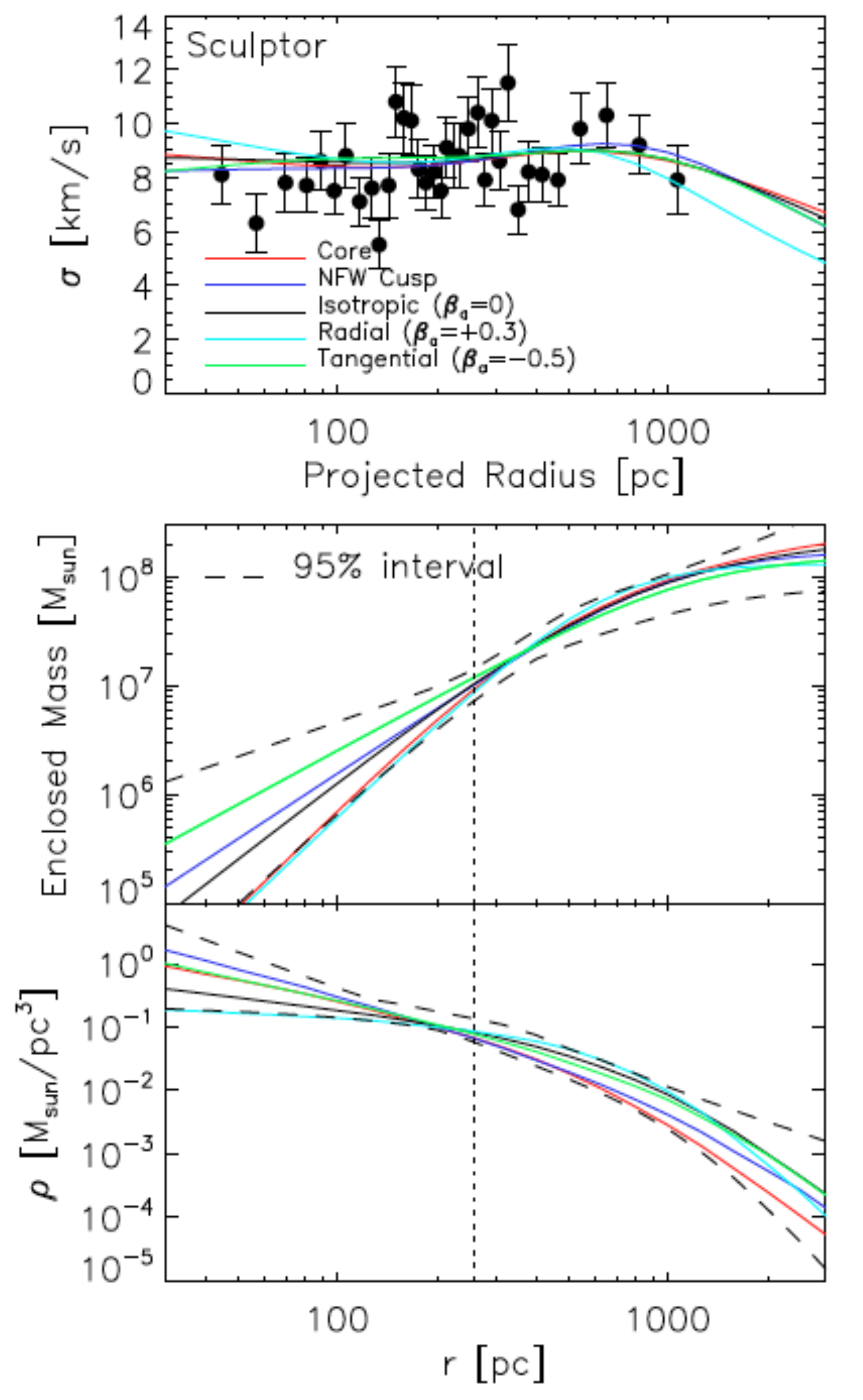}
  \caption{Degeneracy in the Jeans analysis of dSph mass profiles (Section \ref{subsec:jeans}).  In the top panel, overplotted on Sculptor's empirical velocity dispersion profile are best-fitting cored ($\gamma=0$ in the notation of Equation \ref{eq:rho}) and cusped ($\gamma=1$) density profiles, as well as models that assume isotropic ($\beta_a=0$), radially anisotropic ($\beta_a=+0.3$) and tangentially anisotropic ($\beta_a=-0.5$) velocity distributions.  All of these models provide equivalent fits to the data.  The bottom two panels plot mass and density profiles corresponding to each model.  Dotted curves indicate $95\%$ confidence intervals, as determined from a Jeans/MCMC analysis that assumes constant anisotropy and lets the unspecified parameters in Equation \ref{eq:rho} vary freely (Section \ref{subsec:jeans}).  The vertical dotted line indicates Sculptor's projected halflight radius.}
  \label{fig:scl_profiles}
\end{figure}

Using Sculptor as an example, Figure \ref{fig:scl_profiles} demonstrates the degeneracy that is inherent in the standard Jeans analysis.  Overplotted on Sculptor's empirical velocity dispersion profile are best-fitting models obtained under specific assumptions either about the inner slope of the mass-density profile ($\gamma=0$ or $\gamma=1$), or about the amount of velocity anisotropy ($\beta_a=-0.5$, $\beta_a=0$ or $\beta_a=+0.3$).  Despite corresponding to a wide range of total masses and mass distributions (bottom panels of Figure \ref{fig:scl_profiles}), all of these models can provide equivalent fits to the structural and kinematic data.  

While the Jeans analysis provides useful constraints on none of the halo parameters included in Equation \ref{eq:rho}, it does imply a model-independent constraint on a basic dynamical quantity.  All models that fit the velocity dispersion profile give approximately the same value for the enclosed mass near the halflight radius (middle panel of Figure \ref{fig:scl_profiles}), indicating that this quantity is well determined by the available data.  Indeed, formal confidence intervals derived from Markov-Chain Monte Carlo scans of the full parameter space \citep[e.g.,][]{strigari07,strigari10,walker09d,wolf10,martinez11} show a characteristic `pinch' near the halflight radius (Figure \ref{fig:scl_profiles}).  

Equivalent constraints on the enclosed mass at the halflight radius can be obtained more simply by solving Equation \ref{eq:jeans} under the assumptions that $\beta_a=0$ and $\sigma(R)=$constant.  For a Plummer surface brightness profile (Equation \ref{eq:plummer}), one obtains \citep{walker09d}
\begin{equation}
  M(R_h)=\frac{5R_h\sigma^2}{2G}.
  \label{eq:walker}
\end{equation}
\citet{wolf10} show analytically that for various surface brightness and anisotropy profiles, the tightest constraint on the mass profile (provided that the velocity dispersion profile is sufficiently flat) can be approximated by 
\begin{equation}
  M(r_3)=\frac{3r_3\sigma^2}{G},
  \label{eq:wolf}
\end{equation}
where $r_3$ is the radius at which $d\ln\nu/d\ln r=-3$.  For most commonly-adopted surface brightness profiles, this radius is close to the \textit{deprojected} halflight radius (i.e., the radius of the sphere containing half of the stars), which typically exceeds the projected halflight radius by a factor of $\sim 4/3$.  Insofar as the assumption of flat velocity dispersion profiles (and the usual assumptions of dynamic equilibrium and spherical symmetry) holds, these simple mass estimators can be applied even to the relatively sparse kinematic data available for ultrafaint satellites.  \citet[see erratum for updated values]{walker09d} and \citet{wolf10} tabulate masses obtained from Equations \ref{eq:walker} and \ref{eq:wolf}, respectively, for $\sim 2$ dozen Local Group dSphs.  

The equivalence of such seemingly crude estimates to constraints from Jeans/MCMC explorations of parameter space follows from a combination of facts: 1) Equation \ref{eq:jeans} deals only with velocity moments of the phase space distribution function and not with the distribution function itself; 2) confinement of empirical velocity distributions to the line-of-sight component yields little information about $\beta_a$; 3) therefore the flat velocity dispersion profiles of dSphs effectively reduce the available kinematic information to just two numbers, $\sigma$ and the scale radius that characterises the adopted surface brightness profile.  The information extracted from the Jeans analysis naturally amounts to a simple combination of these two numbers.

\subsection{Distribution}
\label{subsec:distribution}

Some cosmological and particle physics models make specific predictions about how dark matter is distributed within individual halos (Section \ref{sec:implications}).  Insofar as dSphs represent the structures most dominated by dark matter and least affected by the presence of baryons, their internal dynamical properties provide the most straightforward tests of such predictions.  Section \ref{subsec:jeans} demonstrates that so long as dSphs have flat velocity dispersion profiles, the standard Jeans analysis constrains only one number, characterising the amount but not the distribution of dark matter.  However, the restrictive assumptions employed in the standard Jeans analysis overlook structure that is present in the data available for many dSphs (Section \ref{subsubsec:peculiar}).  Recent analyses that are devised to exploit more fully the available empirical information have been able to place useful constraints not only on the amount, but also on the distribution of dark matter within individual dSphs.  

\subsubsection{Indirect Constraints}
\label{subsubsec:indirect}

The presence of stellar substructure within dSphs can provide extra leverage not easily exploited in the context of equilibrium dynamical models.  For example, it has long been known that Ursa Minor's stellar component is `lumpy' \citep[Section \ref{subsubsec:peculiar}]{hodge64b,olszewski85,ih95,palma03}.  While Ursa Minor has a velocity dispersion of $\sigma\sim 10$ km s$^{-1}$, \citet{kleyna03} find that a secondary peak in Ursa Minor's stellar density field, offest by $\sim 20'$ ($\sim 400$ pc) from the nominal center, exhibits a cold dispersion of $\la 1$ km s$^{-1}$.  Interpreting this feature as a loosely bound star cluster captured by Ursa Minor, \citet{kleyna03} perform N-body simulations to examine how its stability depends on the external gravitational potential (assumed to be dominated by dark matter) of Ursa Minor.  Whereas simulated clusters remain intact for a Hubble time when the host potential has a central `core' with constant density ($\gamma=0$ in the notation of Equation \ref{eq:rho}) on scales larger than the orbital radius, they disrupt on timescales of $\la 1$ Gyr when the host potential has a centrally divergent, or `cusped' density profile ($\gamma>\frac{1}{2}$).  Therefore, if the observed clump indeed corresponds to a star cluster free of its own dark matter component, then its survival provides indirect evidence that the dark matter density of Ursa Minor is constant over the central few-hundred pc.

Substructure in the form of Fornax's five globular clusters provides another example of indirect evidence for a cored dSph potential.  Four of these clusters are projected near (within a factor of $\sim 2$) Fornax's halflight radius ($R_h\sim 670$ pc; \citealt{ih95}).  \citet{hernandez98} show analytically that the rate at which the orbits of such clusters decay due to dynamical friction depends on the underlying dSph potential.  Subsequent numerical simulations by \citet{sanchez06} and \citet{goerdt06} demonstrate that in a cusped potential, dynamical friction would require only a few Gyr to bring the clusters from their present positions (assuming the projected distances from Fornax's center are not much smaller than the true distances) all the way to Fornax's center.  On the other hand, in a cored potential dynamical friction would bring the clusters only as close as the core radius, where the harmonic potential inhibits further decay.  Thus these considerations provide indirect evidence for a core of constant density over the central few-hundred pc in Fornax.  Further simulations by \citet{goerdt10} and \citet{cole11} suggest that the transfer of angular momentum from a sinking cluster to the central dark matter is capable of \textit{transforming} an originally cusped into a cored potential, a possibility that might be relevant for the interpretation of evidence for cored dark matter halos in some dSphs (Section \ref{subsec:cosmology}).

\subsubsection{Constraints from Models}
\label{subsubsec:models}

Several groups have recently developed dynamical and/or kinematic models that exploit structure that is present in dSph data but is not considered in the Jeans analyses discussed in Section \ref{subsec:jeans}.  For example, the discoveries of two chemo-dynamically independent stellar sub-populations in Sculptor \citep[Section \ref{subsubsec:peculiar}]{tolstoy04}, Fornax \citep{battaglia06} and Sextans \citep{battaglia11} enable analyses of two tracer components in the same potential.  After imposing a metallicity cutoff to separate Sculptor's two sub-populations, \citet{battaglia08} find that cored rather than cusped potentials provide better simultaneous fits in a Jeans analysis (Section \ref{subsec:jeans}) of the two sets of surface brightness and velocity dispersion profiles.  Using the same empirical profiles for Sculptor, \citet{amorisco12} confirm this result by modelling both sub-populations with anisotropic King-Michie \citep{king62,michie63,king66} distribution functions.  These dynamical models again favor cored ($\gamma=0$) rather than cusped ($\gamma=1$) potentials, with a likelihood ratio sufficient to reject the hypothesis of cusped potentials with confidence $\ga 99\%$.  

\citet{jardel11} take a different approach, constructing axisymmetric three-integral \citet{schwarzschild79} models for both cored and cusped potentials that also allow for a central black hole.  For a given potential, libraries of stellar orbits are calculated and each orbit receives a weight based on fits to the observed distribution of velocities within discretely binned radii.  Notice that while the Jeans analysis is sensitive only to the variance of the velocity distribution in a given bin, the Schwarzschild method is sensitive to the \textit{shape} of the distribution.  \citet{jardel11} find that their models constructed from cored potentials fit Fornax's velocity data significantly better than those constructed from cusped potentials.  Within the context of their adopted models, they also place an upper limit of $\la 3.2\times 10^4M_{\odot}$ on the mass of any central black hole.  \citet{breddels11} independently develop Schwarzschild models for Sculptor, concluding that the available data rule out steep cusps ($\gamma \ga 1.5$) but are consistent with slopes in the range $0\la \gamma \la 1$.\footnote{\citet{chaname08} have formulated a Schwarzschild method that operates on discrete velocity measurements, avoiding the binning procedure altogether.  Efforts to apply this and other discrete Schwarzschild methods to dSph data are underway.  As these and other methods for analysing the shapes of dSph velocity distributions continue to develop, it will be beneficial and perhaps necessary to allow for contributions from multiple stellar populations.  Failure to consider distinct populations can generate spurious conclusions regarding orbital structure.  To see why this is so, consider that the superposition of two Gaussian distributions need not be Gaussian.}  

\subsubsection{Direct Measurement}
\label{subsubsec:wp11}

\begin{figure}
  \epsscale{1}
  \plotone{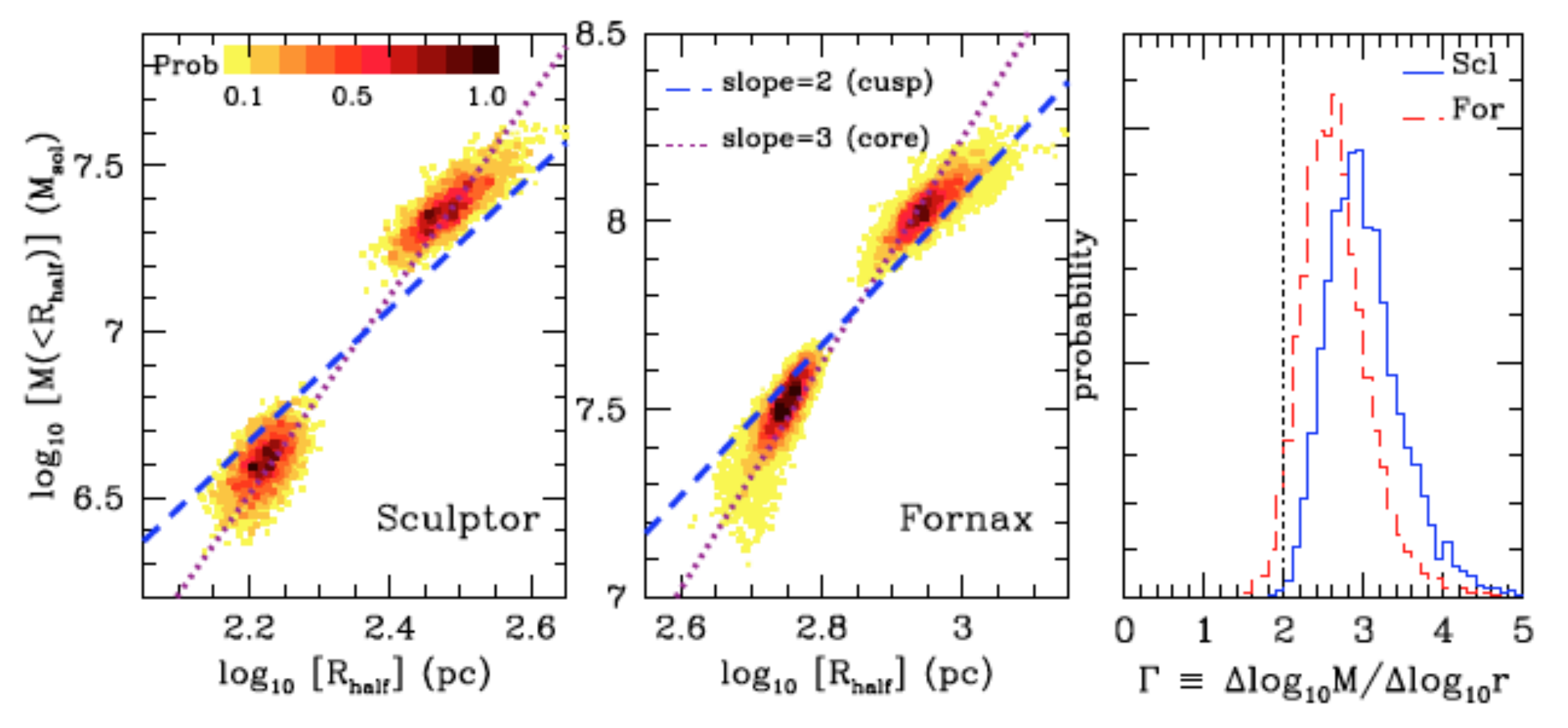}
  \caption{Empirical constraints on the slopes of mass profiles in Fornax and Sculptor, based on estimates of $M(r_h)$ for each of two chemo-dynamically independent stellar sub-populations in each galaxy (Section \ref{subsubsec:wp11}, \citealt[reproduced by permission of the American Astronomical Society]{wp11}).  Points in the left two panels indicate constraints on the two halflight radii and masses enclosed therein, with lines indicating the maximum slopes allowed for cored ($\gamma=0$ in the notation of Equation \ref{eq:rho}; $\Gamma\equiv \Delta\log M/\Delta\log r\leq 3$) and cusped ($\gamma=1$, $\Gamma\leq 2$) overplotted.  The right-hand panel indicates posterior probability distributions for the slope in each galaxy.  The data rule out the cuspy ($\gamma\ga 1$, $\Gamma\leq 2$) profiles that characterise the cold dark matter halos produced in cosmological simulations \citep[e.g.,][]{navarro97} with significance $\ga 96\%$ and $\ga 99\%$ in Fornax and Sculptor, respectively.  }
  \label{fig:wp11}
\end{figure}

\citet{wp11} introduce a method for measuring the slopes of dSph mass profiles directly from spectroscopic data and without adopting a dark matter halo model.  For a given dSph, \citet{wp11} use measurements of stellar positions, velocities and spectral indices to estimate halflight radii and velocity dispersions for as many as two chemo-dynamically independent stellar sub-populations.  Detections of two distinct sub-populations with different sizes provide mass estimates $M(R_h)\propto R_h\sigma^2$ (e.g., Equation \ref{eq:walker}) at two different radii in the same mass profile, immediately specifying a slope.  For Fornax and Sculptor, this method yields slopes of $\Gamma\equiv \Delta\log M/\Delta\log r=2.61_{-0.37}^{+0.43}$ and $\Gamma=2.95_{-0.39}^{0.51}$, respectively, on scales defined by values $\sim 0.2\la R_h/\mathrm{kpc}\la 1$ estimated for the halflight radii.  These slopes are consistent with cored ($\gamma=0$) potentials, for which $\Gamma\leq 3$ at all radii, but incompatible with cusped ($\gamma\ga 1$) potentials, for which $\Gamma\leq 2$ (Figure \ref{fig:wp11}).  

\section{A Common dSph Mass?}
\label{sec:universal}

\citet{mateo93} noticed that dSph dynamical mass-to-light ratios increase monotonically with decreasing luminosity (left panel of Figure \ref{fig:mateosimongeha}).  Based on mass-follows-light analyses (Section \ref{subsubsec:mfl}), the Milky Way's eight classical dSphs trace a relationship $\log_{10}(M/L_V/[M/L_V]_{\odot})\sim 2.5+10^7L_V/L_{V,\odot}$ \citep{mateo98}.  This relation implies that if their \textit{stellar} mass-to-light ratios are $M/L_V\sim 2.5[M/L_V]_{\odot}$, then each dSph is embedded in a dark matter halo of mass $\sim 10^7M_{\odot}$.  Allowing for extended dark matter halos, \citet{mateo93} and \citet{mateo98} interpret this value as the minimum mass that is associated empirically with dark matter.  

Kinematic studies of the newfound ultrafaint satellites alter this picture slightly.  The ultrafaint dSphs extend the luminosity floor from $M_V\sim -9$ (e.g., Draco, Ursa Minor) to $M_V\sim -2$ (e.g., Segue 1), nearly three orders of magnitude in luminosity.  Applying mass-follows-light models to their velocity data for the least luminous dSphs, \citet{martin07} and \citet{simon07} estimate dynamical masses as small as $\sim 10^5M_{\odot}$.  While these masses imply extreme mass-to-light ratios $M/L_V\ga 100 [M/L_V]_{\odot}$ that extend the monotonic increase in $M/L_V$ with decreasing luminosity, they suggest that this relationship becomes flatter toward the smallest luminosities (right panel of Figure \ref{fig:mateosimongeha}).  

\begin{figure}
  \plottwo{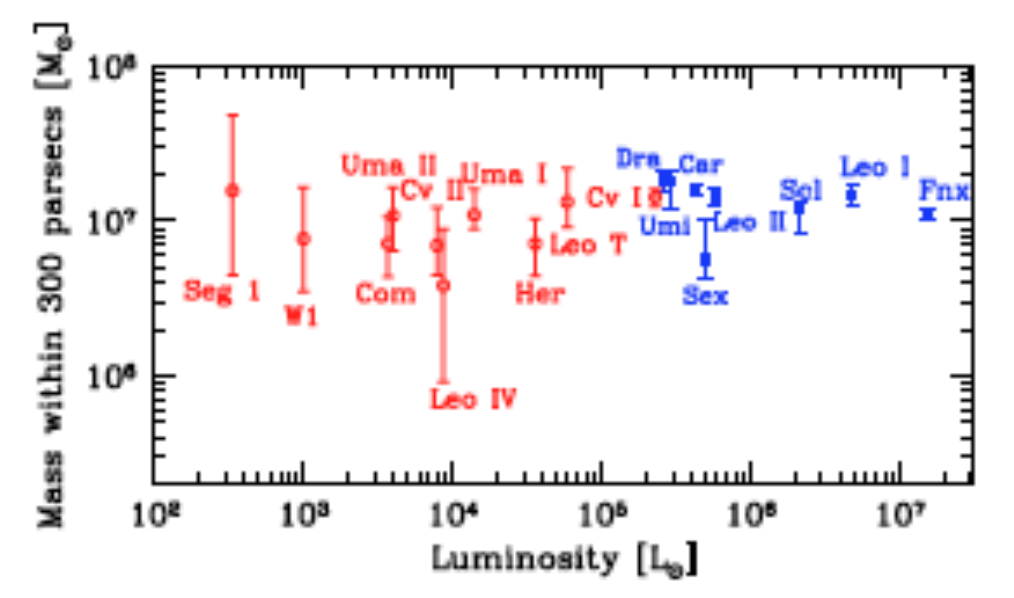}{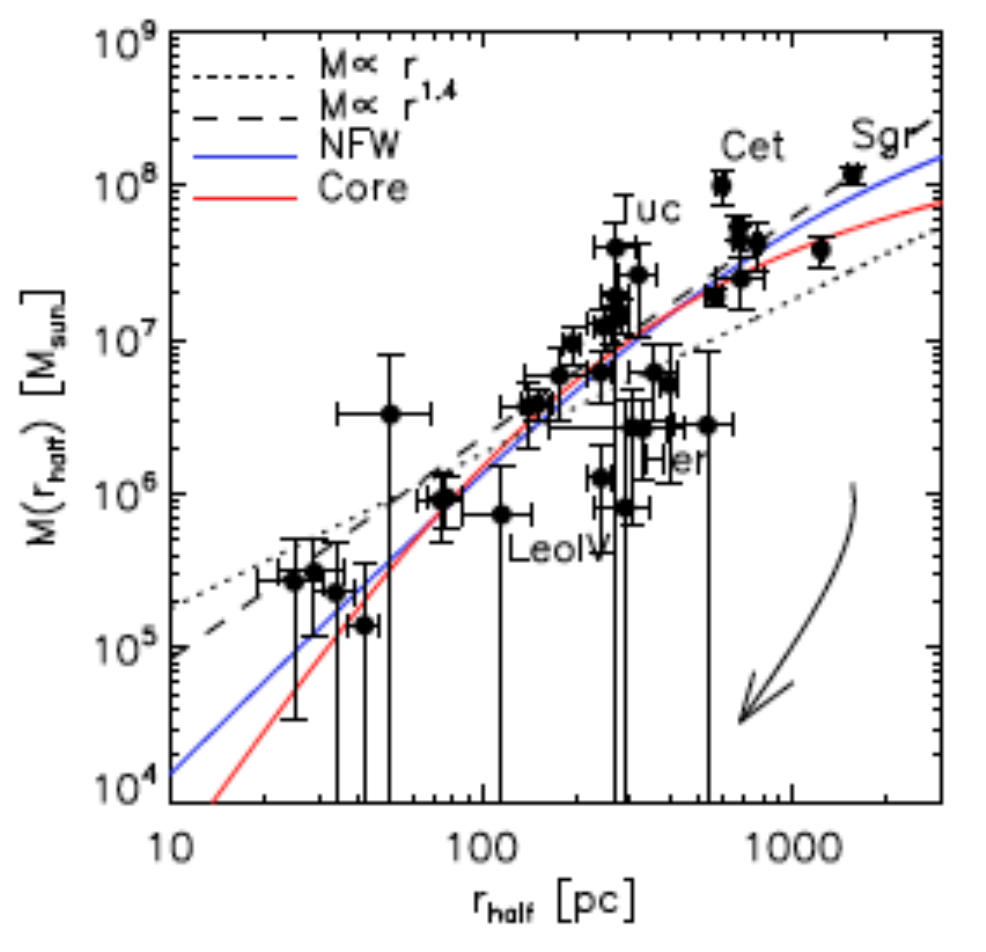}
  \caption{A common mass?  \textit{Left:} mass enclosed within a radius of $r\leq 300$ pc, estimated from a Jeans analysis (See Section \ref{subsec:jeans}), versus luminosity \citep[reproduced with permission]{strigari08}.  \textit{Right:} mass enclosed within the projected halflight radius, estimated from Equation \ref{eq:walker}, versus halflight radius \citep[reproduced by permission of the American Astronomical Society]{walker09d}.  Arrows indicate the trajectory followed by a satellites as they lose up to 99\% of their original stellar mass, from N-body simulations by \citet{penarrubia08b}.  Overplotted are various dark matter halo models \citep{walker09d}.}
  \label{fig:strigariump}
\end{figure}

\citet{strigari08} extend to ultrafaint satellites the notion of a common dSph mass by considering the amount of mass enclosed within a fixed radius of $300$ pc (implicitly assuming that all dSphs occupy dark matter halos at least this large).  Using a Jeans analysis similar to that described in Section \ref{subsec:jeans}, \citet{strigari08} estimate $M_{300}\equiv M(r\leq 300\mathrm{pc})\sim 10^7M_{\odot}$ for dSphs spanning five orders of magnitude in luminosity (Figure \ref{fig:strigariump}, left panel).  \citet{walker09d} take a different approach, using the model-independent estimates of $M(R_h)$ provided by Equation \ref{eq:walker} to evaluate the hypothesis that all dSphs are embedded in identical dark matter halos characterised by a `universal' mass profile (Figure \ref{fig:strigariump}, right panel).  While the scatter of $M(R_h)$ values about a single power law, $M(r)\propto r^{x}$, is larger (by a factor of $\sim 2$) than the scatter expected to arise from observational errors, it is similar to the scatter about a common value of $M_{300}$.  

\begin{figure}
  \plottwo{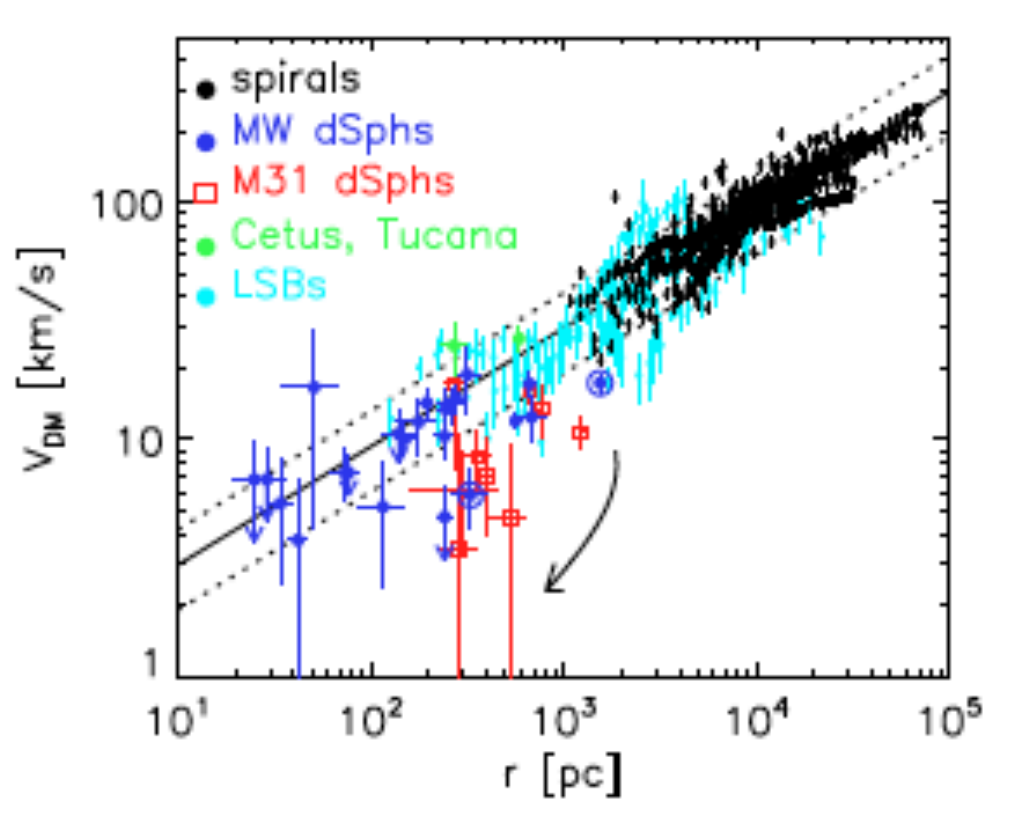}{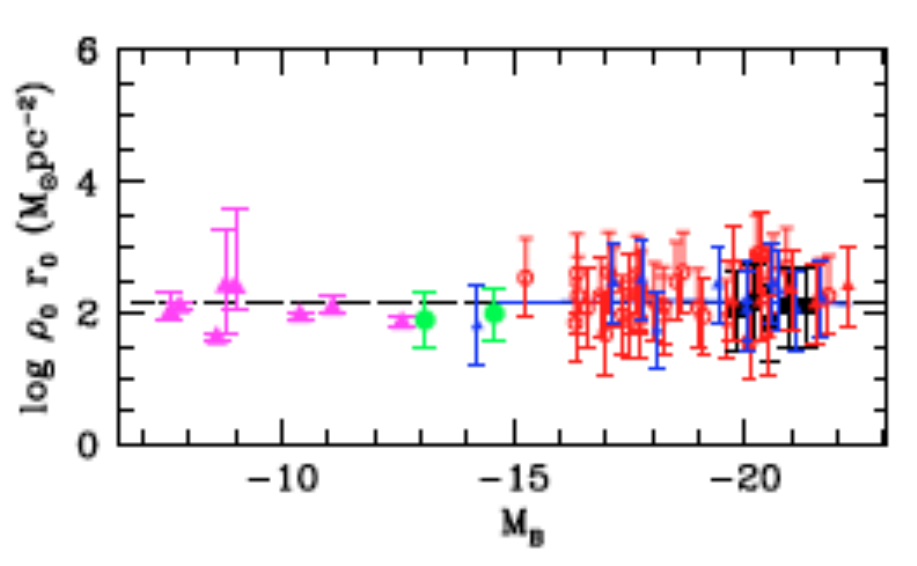}
  \caption{Similarity with spiral galaxies (Section \ref{sec:universal})?  \textit{Left:} circular velocities ($v_{\mathrm{circ}}=\sqrt{GM(R_h)/R_h}$) for the Milky Way's dSph satellites lie on the extrapolation of the mean rotation curve due to dark matter in spiral \citep{mcgaugh07} and low surface brightness \citep{kuzio08} galaxies \citep[reproduced by permission of the American Astronomical Society]{walker10}.  \textit{Right:} Dark matter surface densities inferred from fits of \citet{burkert95} halos have the same constant value inferred from spiral galaxy rotation curves (reproduced from \textit{Dwarf Spheroidal Galaxy Kinematics and Spiral Scaling Laws}, by P.\ Salucci et al., MNRAS, in press, by permission of John Wiley \& Sons Ltd.).}
  \label{fig:walkersalucci}
\end{figure}

While it has long been known that dSphs do not fit naturally onto scaling relations defined by larger elliptical galaxies \citep[e.g.,][]{kormendy85}, the roughly exponential decline of dSph surface brightness profiles (Section \ref{subsubsec:classical}) hints at a structural relationship to spirals \citep{faber83,lin83}, perhaps via an evolutionary mechanism such as tidal stirring (Sections \ref{subsubsec:classical}, \citealt{mayer01a,mayer01b}).  Converting estimates of $M(R_h)$ to circular velocities $v^2_{\mathrm{circ}}=GM(R_h)/R_h$, \citet{walker10} find that the Milky Way's dsph satellites lie on an extrapolation of the mean rotation curve estimated for dark matter in spiral galaxies (Figure \ref{fig:walkersalucci}, left panel), $\log_{10}[V_{\mathrm{circ}}/(\mathrm{km s^{-1}})]=1.5+0.5\log_{10}[r/\mathrm{kpc}]$ \citep{mcgaugh07}.  \citet{donato09} and \citet{salucci11} find similar results: assuming cored dark matter halos of the form $\rho(r)=\rho_0r_0^3(r+r_0)^{-1}(r^2+r_0^2)^{-1}$ \citep{burkert95}, fits to dSph velocity dispersion profiles extend a simple scaling relationship that is common to spiral galaxies \citep{kormendy04} and is characterised by $\log_{10}[\rho_0/(M_{\odot}\mathrm{pc^{-3}})r_0/\mathrm{pc}]\sim 2$, independently of luminosity (Figure \ref{fig:walkersalucci}, right panel).

These scaling relationships have provoked a variety of disparate interpretations.  For example, \citet{boyarsky10b} show analytically that `secondary infall' of dark matter onto isolated halos would cause surface density to vary slowly with halo mass, explaining the apparent constancy of $\rho_0r_0$ over the range spanned by spirals.  They argue that the apparent extension to dSphs agrees with results from cosmological N-body simulations \citep[e.g.,][]{springel08}, which produce satellite halos with approximately the same surface density.  On the other hand, \citet{mcgaugh10} show that dSphs deviate from another spiral scaling relation---the baryonic Tully-Fisher relation \citep[`BTF',][]{mcgaugh00}---systematically according to the amount by which their halflight radii exceed their tidal radii, provided that the latter radius is estimated using Milgrom's (1983) Modified Newtonian Dynamics (a framework that naturally implies the BTF relation).  

In any case, scatter among the Milky Way's dSphs increased when \citet{aden09} and \citet{koposov11} measured smaller velocity dispersions for Hercules and Bo\"otes I, respectively, than had previously been estimated (Section \ref{subsec:velocities}).  Furthermore, early kinematic results for the dSph satellites of M31 suggest systematically smaller masses (at a given halflight radius) than estimated for the Milky Way dSphs \citep{kalirai10,collins10,collins11}.  However, more recent, larger data sets suggest that the M31 dSphs are more similar to the MW dSphs than previously suggested \citep{tollerud11}.  Pending more detailed spectroscopic surveys, particularly of the least luminous and most distant dSphs, the interpretation of scaling relationships that connect galaxies across such wide ranges of luminosity and morphological type should proceed with caution.

\section{Implications for Cosmology and Particle Physics}
\label{sec:implications}

Observations indicate that the Milky Way's known dSph satellites have masses $10^5\la M(R_h)/M_{\odot}\la 10^7$ and that dark matter dominates their internal kinematics at all radii, $M/L_V\ga 10[M/L_V]_{\odot}$ (Section \ref{subsec:amount}).  Current observations also provide direct and/or indirect constraints on the internal distributions of dark matter in three dSphs: Fornax, Sculptor and Ursa Minor (Section \ref{subsec:distribution}).  For all three, the available evidence indicates central `cores' of constant density on scales of a few hundred pc.\footnote{In a recent preprint, \citet{wolf12} argue that the situation is more ambiguous, and specifically that the Sculptor data of \citet{walker09a} favor a dark matter halo that is centrally cusped ($\gamma=1$) rather than cored ($\gamma=0$).  Their Figure 1 demonstrates that this conclusion---and more generally the ability of their Jeans analysis to yield an apparent constraint on $\gamma$---depends critically on their assumption that the data sample a single stellar population with isotropic ($\beta_a=0$ in the notation of Equation \ref{eq:jeansproject}) velocity distribution.  This assumption is incompatible with the multi-component structure that is present in the spectroscopic data \citep[see Sections \ref{subsubsec:peculiar} and \ref{subsec:distribution}]{tolstoy04,battaglia06,wp11}.}  Taken at face value, these basic results have implications for broader areas of physics.

\subsection{Cosmology}
\label{subsec:cosmology}

Observations of structure on large scales (e.g., as inferred from redshift surveys and anisotropy of the cosmic microwave background radiation) seem to require a significant contribution to the mass budget from non-baryonic dark matter, $\Omega_{\mathrm{DM}}\sim 0.22$ \citep[e.g,][]{bennett03,spergel03}.  The `cold dark matter' (CDM) cosmological paradigm is built on the hypothesis that the dark matter consists of fundamental particles that act like a collisionless gas after decoupling from radiation at non-relativistic speeds shortly after the Big Bang.  Small cross sections and low thermal velocities allow CDM structure to form and survive at high densities in small volumes, thereby enabling the growth of structure on small scales in the early universe.  

Calculations of the matter power spectrum associated with popular `weakly interacting massive particle' (WIMP) candidates for the dark matter (e.g., neutralinos with mass $m_{\chi}\ga 10$ GeV) indicate $P(k)\propto k^{-3}$ at small scales until collisional damping and free streaming finally cause an exponential decline on sub-parsec (co-moving) scales \citep{green04,diemand05}.  The corresponding halo mass function would be approximately $dN(M)/dM\propto M^{-\alpha}$ with $\alpha\sim 1.9$, and a galaxy like the Milky Way would host roughly $\sim 10^{15}$ satellites in the form of individual, self-bound dark matter `subhalos', `sub-subhalos', ..., and `microhalos' with masses $\ga 10^{-6}M_{\odot}$ \citep{hofmann01,diemand05,springel08}.  

Observational requirements derived from the current census of Milky Way satellites seem rather modest in this context.  A viable dark matter particle needs to accommodate the formation and survival of only a few tens (or hundreds when correcting for incompleteness of sky surveys, \citealt{koposov08,tollerud08}) of dark matter halos with masses $M\ga 10^{5}M_{\odot}$ (Section \ref{sec:masses}) around the Milky Way.  These constraints allow for significantly less massive, `warmer' particle candidates (e.g., the sterile neutrino, \citealt{dodelson94}; see recent reviews and discussion by, e.g., \citealt{abazajian06,boyanovsky08,boyarsky09c,kusenko09}) whose longer free-streaming lengths might naturally truncate the matter power spectrum at scales more similar to those that characterise the smallest galaxies \citep{gilmore07,bode01,maccio10b,polisensky11,lovell12}.

Thus dark matter particle candidates and associated cosmologies can be classified in practicial terms according to whether the particles' free streaming plays a significant role in galaxy formation \citep[e.g.,][]{boehm01}.  For sufficiently massive and `cold' particles it does not, and other physical processes must be invoked to explain the suppression and/or truncation of galaxy formation in low-mass halos \citep[e.g.,][]{klypin99,koposov09,li10,maccio10,kravtsov10,font11}.  The negligible thermal velocities invoked for `standard' CDM particles also imply that N-body simulations can track the growth of structure accurately with relatively few particles, making CDM cosmological simulations the simplest, fastest and most widely practiced kind.  

Cosmological simulations demonstrate that if gravitational interactions between standard CDM particles dominate the formation and evolution of galactic structure, then galaxies ought to be embedded in dark matter halos that have central cusps characterised by $\lim_{r\rightarrow 0}\rho(r)\propto r^{-\gamma}$, with $\gamma\ga 1$ \citep[e.g.,][]{dubinski91,navarro96,navarro97,moore98,klypin01,diemand05b,springel08}.  Observations indicate that most individual galaxies with suitable measurements are not embedded in such halos.  Instead, rotation curves of spiral and low surface brightness galaxies tend to favor dark matter halos with resolved `cores' ($\gamma\sim 0$) of constant density (e.g., \citealt{moore94,flores94,deblok97,salucci00,mcgaugh01,simon05,kuzio06,kuzio08}, \citealt[and references therein]{deblok10}).  These results imply that (standard) gravitational interactions between CDM particles do not always dominate the formation and evolution of galactic structure.  

Indeed galaxies contain baryons prone to interact via forces other than gravity.  Many hydrodynamical simulations demonstrate that various poorly-understood baryon-physical mechanisms might influence the structure of galactic CDM halos \citep[e.g.,][]{blumenthal86,navarro96b,elzant01,gnedin04,tonini06,romanodiaz09,delpopolo10,governato10,pontzen11,governato12}.  Insofar as their baryons are dynamically negligible, dSphs and low surface brightness galaxies enable the most direct comparisons to structures formed in CDM-only simulations.  In this context the available evidence against cusped dark matter halos in Fornax, Sculptor and Ursa Minor (Section \ref{subsec:distribution}) becomes particularly relevant: the viability of standard CDM now requires that baryon-driven mechanisms can have \textit{reduced} the central dark matter densities in these galaxies to $\rho_0\ga 5\times 10^7 M_{\odot}$kpc$^{-3}$ while leaving behind stellar populations with low luminosities $10^5\la L_V/L_{V,\odot}\la 10^7$ and central surface brightnesses $23\la \mu_0/(\mathrm{mag/arcsec}^2)\la 25$.\footnote{\citet{boylankolchin11} identify a similar (perhaps the same) structural problem, noting that the most massive `subhalos' produced in the \textit{Aquarius} CDM simulation \citep{springel08} have central densities larger than those estimated for any of the known dSphs.}

Recent work identifies several mechanisms that might accomplish this feat on dSph scales by invoking either the dynamical coupling of the dark matter to energetic baryonic outflows (e.g., \citealt{read05,mashchenko06,mashchenko08,desouza11}) or the transfer of energy/angular momentum to dark matter from massive infalling objects (e.g., \citealt{sanchez06,goerdt06,goerdt10,cole11}).  Hydrodynamical simulations by \citet{sawala10} and \citet{parry11} indicate that the former category of solutions has difficulty reproducing other dSph observables---specifically, star formation histories as well as luminosity functions and metallicity distributions.  The latter category of solutions is difficult to evaluate observationally, as the evidence can literally be destroyed (e.g., by tidal disruption); furthermore it seems unlikely that such infall mechanisms generate cores of sufficient size.\footnote{For example, \citet{goerdt10} conclude that a sinking object of mass $M_s$ induces core formation inside a radius where the enclosed halo mass is $M(r_{\mathrm{core}})\sim M_s$.  In this scenario the sinking of Fornax's five surviving globular clusters ($M_s\sim 10^5 M_{\odot}$) cannot have formed the core inferred from estimates $M(\sim 550\mathrm{pc})\sim 5\times 10^7M_{\odot}$ and $M(\sim 900\mathrm{pc})\sim 2\times 10^8M_{\odot}$ \citep{wp11}.}  Alternatively, cosmological simulations that consider `warmer' particle candidates demonstrate that the associated suppression of small-scale power can naturally (i.e., without invoking baryon physics) produce halos with large cores; however, such scenarios seem to require fine tuning of the relative contributions from various production mechanisms in order to reproduce simultaneously the number of observed MW dSphs \citep{polisensky11,maccio12,maccio12b}.

In any case, the emerging challenge for the standard CDM paradigm is not that empirical evidence against cusped dark matter halos necessarily rules out the hypothesis that CDM particles constitute the dark matter.  The poorly understood complexities of baryon physics---along with the freedom to invoke other processes, e.g., self-scattering of CDM particles \citep{spergel00,loeb11,vogelsberger12}---leave sufficient flexibility for CDM to be rendered consistent with virtually any realistic observation of galactic structure.  In fact that is the problem.  CDM escapes falsification of perhaps its most famous prediction only by withdrawing the prediction.  While this circumstance does not imply that CDM is incorrect, it does mean that CDM currently fails to make accurate predictions regarding the stellar dynamics of galaxies, a primary piece of evidence for dark matter in the first place.  In this context a decisive outcome favorable to standard CDM seems to require the detection of either 1) gravitational interactions involving dark matter halos on sub-galactic scales (e.g., via microlensing or perturbations of loosely bound luminous structure) or 2) nongravitational interactions involving cold dark matter particles.

\subsection{Particle Physics}
\label{subsec:particle}

It has long been recognised that the small sizes and large mass densities of dSphs place strong constraints on the particle nature of dark matter.  For example, Liouville's theorem requires that the phase space densities of light, neutral lepton species do not increase after decoupling from radiation in the early Universe.  \citet{tremaine79} point out that this constraint, combined with the necessity that $\Omega_{\nu}<\Omega_{\mathrm{matter}}$, places a conservative upper limit on the neutrino mass that is summarily violated by lower limits from phase space densities inferred for galaxy halos.  Therefore neutrinos are not the dark matter in galaxies.  This exclusion is most evident on small scales, where small volumes demand heavy particles in order to satisfy phase-space requirements.  

For example, using Aaronson's (1983) initial measurement of Draco's velocity dispersion, \citet{lin83} derive a lower limit of $m_{\nu}\ga 500$ eV.  \citet{lake89} points out that this constraint is sensitive to the dubious assumption that mass follows light (Section \ref{subsubsec:mfl}).  \citet{gerhard92b} strengthen the argument by turning it around, noting that for more viable neutrino masses of $m_{\nu}\sim 30$ eV, the core radii of dSph halos would need to be unrealistically large ($\ga 10$ kpc) to accommodate model-independent lower limits of $\rho_0\ga 0.05M_{\odot}\mathrm{pc}^{-3}$ \citep{pryor90} on their central densities.  Generalising the phase-space argument of \citet{tremaine79} to relativistically decoupled warm dark matter candidates, \citet{dalcanton01} show that of all galaxies, dSphs provide the most stringent limits, $m_{\chi}\ga 700$ eV and $m_{\chi}\ga 300$ eV for thermal and degenerate fermions, respectively.  Most recently and more specifically, \citet{boyarsky09b} use phase-space arguments to conclude that $m_{\chi}\ga 1.7$ keV if the dark matter consists of sterile neutrinos produced via non-resonant mixing with active neutrinos.  

Any positive identification of a dark matter particle will require the detection of its non-gravitational interactions.  Experiments at the Large Hadron Collider might find evidence for such interactions, as might the various experiments designed to detect directly the scattering of dark matter particles in Earth's orbital path.  Alternatively, high-energy photons might be released if dark matter self-annihilates \citep{gunn78,stecker78} or decays \citep{pal82,boyarsky06,kusenko06}, providing an opportunity for indirect detection.

Their large mass-to-light ratios, low astrophysical backgrounds and close proximities make the Milky Way's dSph satellites popular targets in the search for annihilation and/or decay products \citep[e.g.,][]{evans04,strigari08b,kuhlen10}.  For annihilation, the differential $\gamma$-ray flux (units cm$^{-2}$~s$^{-1}$~sr$^{-1}$~GeV$^{-1}$) received on Earth in solid angle $\Delta\Omega$ is given by
\begin{equation}
     \frac{d\Phi_{\gamma}}{dE_{\gamma}}
        = \frac{1}{4\pi}\frac{\langle\sigma v\rangle}{2m_{\chi}^{2}}
          \cdot \frac{dN_{\gamma}}{dE_{\gamma}} \times J(\Delta\Omega),
\label{eq:flux}
\end{equation}
where $m_{\chi}$ is the particle mass, $\langle \sigma v\rangle$ is the (velocity-averaged) cross section, $dN_{\gamma}/dE_{\gamma}$ is the energy spectrum of products and 
\begin{equation}
  J(\Delta\Omega)=\int_{\Delta\Omega}\int \rho^2 (l,\Omega) \,dld\Omega.\
  \label{eq:J}
\end{equation}
This `$J$-factor' represents the astrophysical contribution to the signal and is specified by the integral of the squared dark matter density, $\rho^2(l,\Omega)$, over line of sight $l$ and solid angle $\Omega$.  The equation for the flux due to decay events is similar, except that the integral is taken over the dark matter density raised only to the first power.  Published constraints on $J$ come directly from constraints on $\rho(r)$ obtained in parametric Jeans analyses of the sort described in Section \ref{subsec:jeans} and demonstrated in Figure \ref{fig:scl_profiles} \citep[e.g.,][]{strigari07b,martinez09,charbonnier11}.  

\begin{figure}
  \plottwo{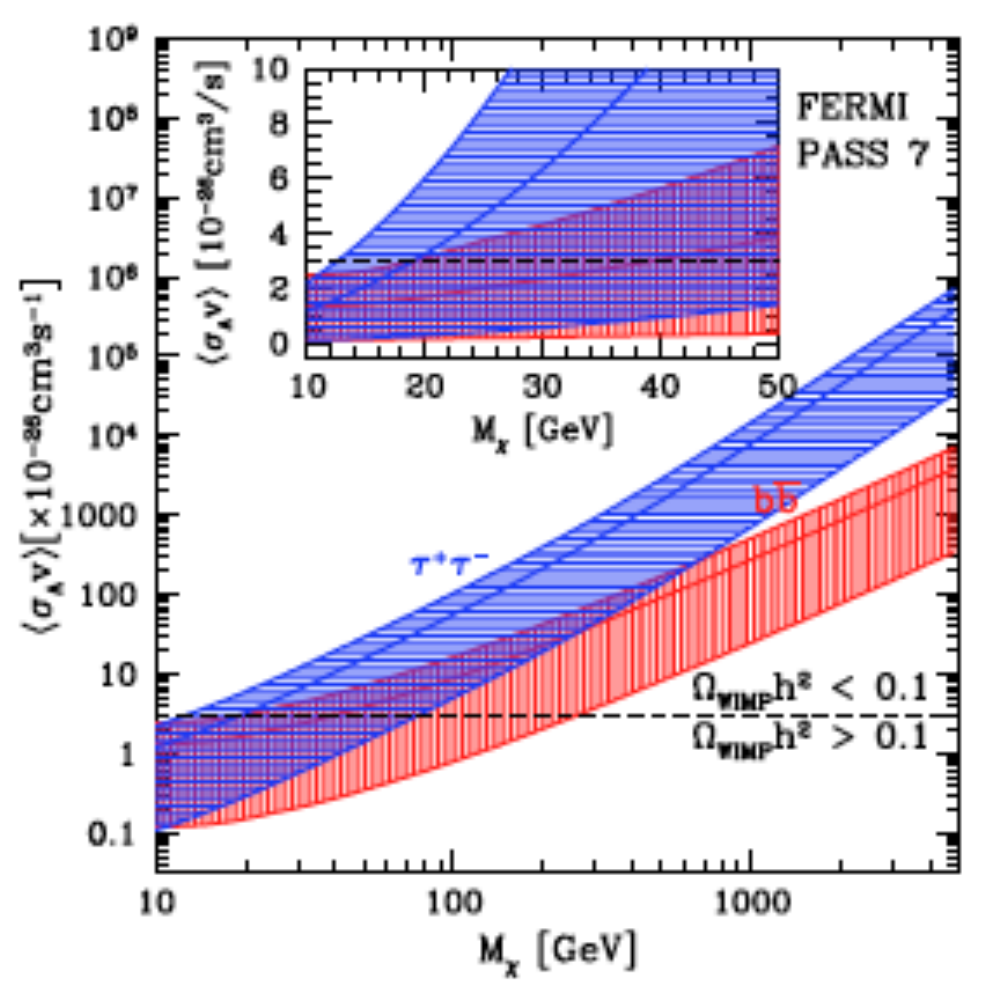}{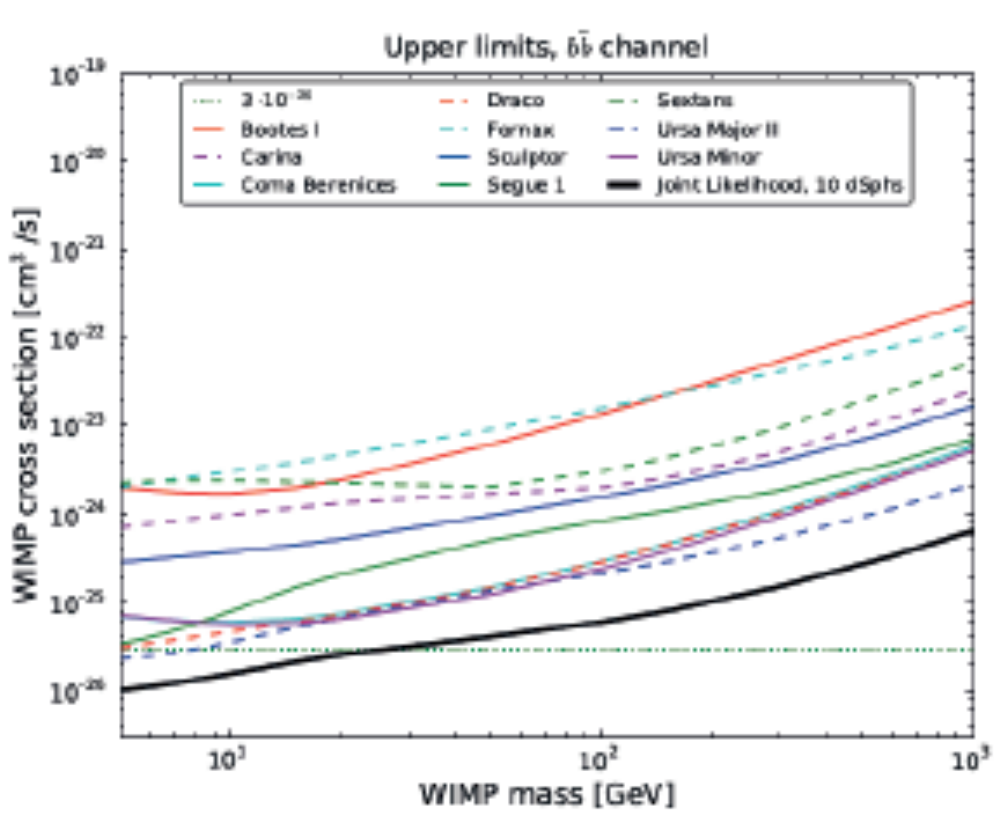}
  \caption{Exclusion of WIMP self-annihilation cross sections, based on Fermi-LAT non-detections (2-year data) of gamma-rays in the Milky Way's dSph satellites (reprinted with permission from Geringer-Sameth \& Koushiappas, Phys. Rev. Lett. 107, 241303 (2011; left) and Ackermann et al. (The Fermi Collaboration), Phys. Rev. Lett. 107, 241302 (2011; right), Copyright 2011 by the American Physical Society).  }
  \label{fig:fermi}
\end{figure}

At present, dSph surveys conducted with atmospheric Cherenkov telescopes \citep[e.g.,][]{pieri09,essig09,hess11,vivier11,aleksic11}, x-ray \citep[e.g.,][]{boyarsky07,loewenstein09,riemer09,loewenstein10,boyarsky10,loewenstein12} and gamma-ray telescopes \citep[e.g.,][]{abdo10,scott10,fermi11} yield no unambiguous detections.\footnote{\citet{loewenstein10} interpret a \textit{Chandra} detection of monochromatic ($\sim 2.5$ keV) emission from the direction of the Willman 1 satellite as a possible signal of sterile neutrino decay.  However, \citet{boyarsky10} argue that non-detections of this feature in the Galactic halo, M31 and several other dSphs rule out a dark matter origin.  Indeed, \citet{loewenstein12} report no detection of the $\sim 2.5$ keV feature in follow-up XMM-Newton observations of Willman 1; corresponding limits on the mass/mixing angle of sterile neutrinos depend on how reliably the `irregular' stellar kinematics of Willman 1 \citep[Section \ref{subsubsec:peculiar}]{willman11} trace its mass.}  From Equation \ref{eq:flux}, upper limits on photon flux translate into upper limits on the cross section $\langle \sigma v\rangle$ for a given particle mass and annihilation channel.  For example, Figure \ref{fig:fermi} plots $95\%$ upper limits on $\langle \sigma v\rangle$ derived from Fermi-LAT observations of Milky Way dSphs, based on two years of data from the planned five-year mission (\citealt{geringer11,fermi11}).  Dotted lines at $\langle \sigma v\rangle \sim 3\times 10^{-26}$ cm$^{3}$s$^{-1}$ mark the `generic' cross section expected for WIMPs with mass $m_{\chi}\sim 0.1-1$ TeV \citep[e.g.,][]{jungman96,feng10}.  WIMPs having this combination of mass and cross section would have decoupled from radiation with relic abundance $\Omega_{\mathrm{WIMP}}\sim 0.2$, the value cosmology requires of the dark matter (a coincidence sometimes referred to as the `WIMP miracle').  For particle masses below $m_{\chi}\la 100$ GeV, the combination of kinematic and high-energy data available for dSphs is now encroaching upon the cross section most readily associated with WIMPs.\footnote{\citet{charbonnier11} use published kinematic data to estimate less stringent limits of $\langle \sigma v\rangle \la 10^{-25}$ cm$^{3}$s$^{-1}$ (at $m_{\chi}\sim 10$ GeV, cf. Figure \ref{fig:fermi}) for individual dSphs.  Possible reasons for this discrepancy include different assumptions about the dark matter halo profile (\citealt{geringer11} and \citealt{fermi11} adopt $J$ values previously estimated under the assumption that dSph dark matter halos follow NFW ($\gamma=1$ in the notation of Equation \ref{eq:rho}) profiles; \citealt{charbonnier11} estimate $J$ values by marginalising over uncertain halo shape parameters), different assumptions about the energy spectrum (\citealt{geringer11} and \citealt{fermi11} explicitly consider annihilation via $b\bar{b}$ and $\tau^{+}\tau^{-}$ mechanisms; \citealt{charbonnier11} consider a conservative spectrum averaged over a variety of plausible annihilation channels) and/or different assumptions about detector sensitivity. }  Over the next decade, searches for dark matter and/or its byproducts will intensify with large-scale efforts at existing facilities and with new instrumentation that will provide unprecedented sensitivity \citep[e.g.,][]{cta10}.  

The status of dark matter as a particle will depend critically on the outcomes of direct and indirect dark matter detection experiments that are either ongoing or planned for the near future.  Also at stake is the motivation and context for studying dark matter phenomenology in dSphs.  An unambiguous, positive detection of dark matter byproducts emitted from dSphs would establish the existence of a new particle and would provide a unique means for measuring its mass and cross section; via Equation \ref{eq:flux}, constraints on such parameters would be only as good as dynamical constraints on the dark matter density profile.  In case of detections in other objects, e.g., the Galactic center \citep[see][]{hooper11} or galaxy clusters \citep[see][]{han12}, high-energy and dynamical constraints from dSphs would provide important consistency checks in the regime of minimal astrophysical background.  In the case of direct detection and characterisation of a new particle in the laboratory, dSph phenomenology would help to establish or rule out association with cosmological dark matter.  In case of failure to detect non-gravitational interactions involving dark matter particles, consideration of the extreme phenomenology exhibited by dSphs would help to inform alternative explanations for dark matter.  

\section{Some Considerations for Future Work}

Topics and/or questions for future investigations include the following:

\begin{itemize}

\item The least luminous galaxies are often disproportionately interesting.  What sorts of bound stellar/gaseous objects will the next generation of sky surveys detect?  

\item This field will continue to develop in proportion to 1) the rate at which data sets grow in size and quality and 2) the efficiency with which analytical tools can extract the information contained therein.  Promising strategies for analysis include the formulation of statistical likelihood functions that depend on discrete measurements (e.g., of individual stellar positions and velocities) rather than on binned profiles, and exploitation of multi-component structure/substructure that is present in the data.  

\item All dSphs clearly deviate from the simple models invoked to characterise their observed properties.  In order to understand dSphs as galaxies, one must consider not only the structural and kinematic data emphasised above, but also chemical abundances, star formation histories (i.e., \textit{stellar} masses), internal substructure and external environment.  

\item The outer stellar structures of dSphs carry valuable information about the gravitational competition between the Milky Way and its satellites.  Combined with deep photometric surveys that map the Galactic stellar halo, wide-field spectroscopic surveys that reach large galactocentric distances in dSphs will reveal details of the transition from satellite to host potential.  

\item The standard CDM hypothesis seems to require baryon-physical processes to have formed large constant-density cores in galactic dark matter halos.  Do the internal mass distributions of galaxies correlate with other observables directly related to baryon physics (e.g., luminosity, metallicity, star formation history)?  Constraints on core sizes for the least luminous dSphs would be extremely valuable.

\item Can observations and/or experiments provide clear evidence for non-gravitational interactions involving a dark matter particle?  Can observations provide evidence for gravitational interactions involving dark matter halos on subgalactic scales?

\item Are there cosmological and/or particle physics models for dark matter from which can be extracted accurate predictions regarding the stellar dynamics of the most dark-matter dominated galaxies?  

\end{itemize}

The author is grateful to Gerry Gilmore for the invitation to write this review and for insightful comments.  The author thanks Giuseppina Battaglia, Alexey Boyarsky, Alexander Kusenko, Oleg Ruchayskiy and Ed Olszewski for suggestions that improved the quality of this work.  MGW is supported by NASA through Hubble Fellowship grant HST-HF-51283.01-A, awarded by the Space Telescope Science Institute, which is operated by the Association of Universities for Research in Astronomy, Inc., for NASA, under contract NAS5-26555.


\begin{thebibliography}{320}
\expandafter\ifx\csname natexlab\endcsname\relax\def\natexlab#1{#1}\fi

\bibitem[{{Aaronson}(1983)}]{aaronson83}
{Aaronson}, M. 1983, \apjl, 266, L11

\bibitem[{{Aaronson} \& {Olszewski}(1987{\natexlab{a}})}]{aaronson87b}
{Aaronson}, M., \& {Olszewski}, E. 1987{\natexlab{a}}, in IAU Symposium, Vol.
  117, Dark matter in the universe, ed. {J.~Kormendy \& G.~R.~Knapp}, 153--158

\bibitem[{{Aaronson} \& {Olszewski}(1987{\natexlab{b}})}]{aaronson87}
{Aaronson}, M., \& {Olszewski}, E.~W. 1987{\natexlab{b}}, \aj, 94, 657

\bibitem[{{Abazajian} \& {Koushiappas}(2006)}]{abazajian06}
{Abazajian}, K., \& {Koushiappas}, S.~M. 2006, \prd, 74, 023527

\bibitem[{{Abdo} {et~al.}(2010){Abdo}, {Ackermann}, {Ajello}, {Atwood},
  {Baldini}, {Ballet}, {Barbiellini}, {Bastieri}, {Bechtol}, {Bellazzini},
  {Berenji}, {Bloom}, {Bonamente}, {Borgland}, {Bregeon}, {Brez}, {Brigida},
  {Bruel}, {Burnett}, {Buson}, {Caliandro}, {Cameron}, {Caraveo}, {Casandjian},
  {Cecchi}, {Chekhtman}, {Cheung}, {Chiang}, {Ciprini}, {Claus},
  {Cohen-Tanugi}, {Conrad}, {de Angelis}, {de Palma}, {Digel}, {Silva},
  {Drell}, {Drlica-Wagner}, {Dubois}, {Dumora}, {Farnier}, {Favuzzi}, {Fegan},
  {Focke}, {Fortin}, {Frailis}, {Fukazawa}, {Fusco}, {Gargano}, {Gehrels},
  {Germani}, {Giebels}, {Giglietto}, {Giordano}, {Glanzman}, {Godfrey},
  {Grenier}, {Grove}, {Guillemot}, {Guiriec}, {Gustafsson}, {Harding}, {Hays},
  {Horan}, {Hughes}, {Jackson}, {Jeltema}, {J{\'o}hannesson}, {Johnson},
  {Johnson}, {Johnson}, {Kamae}, {Katagiri}, {Kataoka}, {Kerr},
  {Kn{\"o}dlseder}, {Kuss}, {Lande}, {Latronico}, {Lemoine-Goumard}, {Longo},
  {Loparco}, {Lott}, {Lovellette}, {Lubrano}, {Madejski}, {Makeev},
  {Mazziotta}, {McEnery}, {Meurer}, {Michelson}, {Mitthumsiri}, {Mizuno},
  {Moiseev}, {Monte}, {Monzani}, {Moretti}, {Morselli}, {Moskalenko}, {Murgia},
  {Nolan}, {Norris}, {Nuss}, {Ohsugi}, {Omodei}, {Orlando}, {Ormes}, {Paneque},
  {Panetta}, {Parent}, {Pelassa}, {Pepe}, {Pesce-Rollins}, {Piron}, {Porter},
  {Profumo}, {Rain{\`o}}, {Rando}, {Razzano}, {Reimer}, {Reimer}, {Reposeur},
  {Ritz}, {Rodriguez}, {Roth}, {Sadrozinski}, {Sander}, {Saz Parkinson},
  {Scargle}, {Schalk}, {Sellerholm}, {Sgr{\`o}}, {Siskind}, {Smith}, {Smith},
  {Spandre}, {Spinelli}, {Strickman}, {Suson}, {Takahashi}, {Takahashi},
  {Tanaka}, {Thayer}, {Thayer}, {Thompson}, {Tibaldo}, {Torres}, {Tramacere},
  {Uchiyama}, {Usher}, {Vasileiou}, {Vilchez}, {Vitale}, {Waite}, {Wang},
  {Winer}, {Wood}, {Ylinen}, {Ziegler}, {Bullock}, {Kaplinghat}, {Martinez}, \&
  {Fermi-LAT Collaboration}}]{abdo10}
{Abdo}, A.~A., {Ackermann}, M., {Ajello}, M., {Atwood}, W.~B., {Baldini}, L.,
  {Ballet}, J., {Barbiellini}, G., {Bastieri}, D., {Bechtol}, K., {Bellazzini},
  R., {Berenji}, B., {Bloom}, E.~D., {Bonamente}, E., {Borgland}, A.~W.,
  {Bregeon}, J., {Brez}, A., {Brigida}, M., {Bruel}, P., {Burnett}, T.~H.,
  {Buson}, S., {Caliandro}, G.~A., {Cameron}, R.~A., {Caraveo}, P.~A.,
  {Casandjian}, J.~M., {Cecchi}, C., {Chekhtman}, A., {Cheung}, C.~C.,
  {Chiang}, J., {Ciprini}, S., {Claus}, R., {Cohen-Tanugi}, J., {Conrad}, J.,
  {de Angelis}, A., {de Palma}, F., {Digel}, S.~W., {Silva}, E.~d.~C.~e.,
  {Drell}, P.~S., {Drlica-Wagner}, A., {Dubois}, R., {Dumora}, D., {Farnier},
  C., {Favuzzi}, C., {Fegan}, S.~J., {Focke}, W.~B., {Fortin}, P., {Frailis},
  M., {Fukazawa}, Y., {Fusco}, P., {Gargano}, F., {Gehrels}, N., {Germani}, S.,
  {Giebels}, B., {Giglietto}, N., {Giordano}, F., {Glanzman}, T., {Godfrey},
  G., {Grenier}, I.~A., {Grove}, J.~E., {Guillemot}, L., {Guiriec}, S.,
  {Gustafsson}, M., {Harding}, A.~K., {Hays}, E., {Horan}, D., {Hughes}, R.~E.,
  {Jackson}, M.~S., {Jeltema}, T.~E., {J{\'o}hannesson}, G., {Johnson}, A.~S.,
  {Johnson}, R.~P., {Johnson}, W.~N., {Kamae}, T., {Katagiri}, H., {Kataoka},
  J., {Kerr}, M., {Kn{\"o}dlseder}, J., {Kuss}, M., {Lande}, J., {Latronico},
  L., {Lemoine-Goumard}, M., {Longo}, F., {Loparco}, F., {Lott}, B.,
  {Lovellette}, M.~N., {Lubrano}, P., {Madejski}, G.~M., {Makeev}, A.,
  {Mazziotta}, M.~N., {McEnery}, J.~E., {Meurer}, C., {Michelson}, P.~F.,
  {Mitthumsiri}, W., {Mizuno}, T., {Moiseev}, A.~A., {Monte}, C., {Monzani},
  M.~E., {Moretti}, E., {Morselli}, A., {Moskalenko}, I.~V., {Murgia}, S.,
  {Nolan}, P.~L., {Norris}, J.~P., {Nuss}, E., {Ohsugi}, T., {Omodei}, N.,
  {Orlando}, E., {Ormes}, J.~F., {Paneque}, D., {Panetta}, J.~H., {Parent}, D.,
  {Pelassa}, V., {Pepe}, M., {Pesce-Rollins}, M., {Piron}, F., {Porter}, T.~A.,
  {Profumo}, S., {Rain{\`o}}, S., {Rando}, R., {Razzano}, M., {Reimer}, A.,
  {Reimer}, O., {Reposeur}, T., {Ritz}, S., {Rodriguez}, A.~Y., {Roth}, M.,
  {Sadrozinski}, H.~F.-W., {Sander}, A., {Saz Parkinson}, P.~M., {Scargle},
  J.~D., {Schalk}, T.~L., {Sellerholm}, A., {Sgr{\`o}}, C., {Siskind}, E.~J.,
  {Smith}, D.~A., {Smith}, P.~D., {Spandre}, G., {Spinelli}, P., {Strickman},
  M.~S., {Suson}, D.~J., {Takahashi}, H., {Takahashi}, T., {Tanaka}, T.,
  {Thayer}, J.~B., {Thayer}, J.~G., {Thompson}, D.~J., {Tibaldo}, L., {Torres},
  D.~F., {Tramacere}, A., {Uchiyama}, Y., {Usher}, T.~L., {Vasileiou}, V.,
  {Vilchez}, N., {Vitale}, V., {Waite}, A.~P., {Wang}, P., {Winer}, B.~L.,
  {Wood}, K.~S., {Ylinen}, T., {Ziegler}, M., {Bullock}, J.~S., {Kaplinghat},
  M., {Martinez}, G.~D., \& {Fermi-LAT Collaboration}. 2010, \apj, 712, 147

\bibitem[{{Ackermann} {et~al.}(2011){Ackermann}, {Ajello}, {Albert}, {Atwood},
  {Baldini}, {Ballet}, {Barbiellini}, {Bastieri}, {Bechtol}, {Bellazzini},
  {Berenji}, {Blandford}, {Bloom}, {Bonamente}, {Borgland}, {Bregeon},
  {Brigida}, {Bruel}, {Buehler}, {Burnett}, {Buson}, {Caliandro}, {Cameron},
  {Ca{\~n}adas}, {Caraveo}, {Casandjian}, {Cecchi}, {Charles}, {Chekhtman},
  {Chiang}, {Ciprini}, {Claus}, {Cohen-Tanugi}, {Conrad}, {Cutini}, {de
  Angelis}, {de Palma}, {Dermer}, {Digel}, {Do Couto E Silva}, {Drell},
  {Drlica-Wagner}, {Falletti}, {Favuzzi}, {Fegan}, {Ferrara}, {Fukazawa},
  {Funk}, {Fusco}, {Gargano}, {Gasparrini}, {Gehrels}, {Germani}, {Giglietto},
  {Giordano}, {Giroletti}, {Glanzman}, {Godfrey}, {Grenier}, {Guiriec},
  {Gustafsson}, {Hadasch}, {Hayashida}, {Hays}, {Hughes}, {Jeltema},
  {J{\'o}hannesson}, {Johnson}, {Johnson}, {Kamae}, {Katagiri}, {Kataoka},
  {Kn{\"o}dlseder}, {Kuss}, {Lande}, {Latronico}, {Lionetto}, {Llena Garde},
  {Longo}, {Loparco}, {Lott}, {Lovellette}, {Lubrano}, {Madejski}, {Mazziotta},
  {McEnery}, {Mehault}, {Michelson}, {Mitthumsiri}, {Mizuno}, {Monte},
  {Monzani}, {Morselli}, {Moskalenko}, {Murgia}, {Naumann-Godo}, {Norris},
  {Nuss}, {Ohsugi}, {Okumura}, {Omodei}, {Orlando}, {Ormes}, {Ozaki},
  {Paneque}, {Parent}, {Pesce-Rollins}, {Pierbattista}, {Piron}, {Pivato},
  {Porter}, {Profumo}, {Rain{\`o}}, {Razzano}, {Reimer}, {Reimer}, {Ritz},
  {Roth}, {Sadrozinski}, {Sbarra}, {Scargle}, {Schalk}, {Sgr{\`o}}, {Siskind},
  {Spandre}, {Spinelli}, {Strigari}, {Suson}, {Tajima}, {Takahashi}, {Tanaka},
  {Thayer}, {Thayer}, {Thompson}, {Tibaldo}, {Tinivella}, {Torres}, {Troja},
  {Uchiyama}, {Vandenbroucke}, {Vasileiou}, {Vianello}, {Vitale}, {Waite},
  {Wang}, {Winer}, {Wood}, {Wood}, {Yang}, {Zimmer}, {Kaplinghat}, \&
  {Martinez}}]{fermi11}
{Ackermann}, M., {Ajello}, M., {Albert}, A., {Atwood}, W.~B., {Baldini}, L.,
  {Ballet}, J., {Barbiellini}, G., {Bastieri}, D., {Bechtol}, K., {Bellazzini},
  R., {Berenji}, B., {Blandford}, R.~D., {Bloom}, E.~D., {Bonamente}, E.,
  {Borgland}, A.~W., {Bregeon}, J., {Brigida}, M., {Bruel}, P., {Buehler}, R.,
  {Burnett}, T.~H., {Buson}, S., {Caliandro}, G.~A., {Cameron}, R.~A.,
  {Ca{\~n}adas}, B., {Caraveo}, P.~A., {Casandjian}, J.~M., {Cecchi}, C.,
  {Charles}, E., {Chekhtman}, A., {Chiang}, J., {Ciprini}, S., {Claus}, R.,
  {Cohen-Tanugi}, J., {Conrad}, J., {Cutini}, S., {de Angelis}, A., {de Palma},
  F., {Dermer}, C.~D., {Digel}, S.~W., {Do Couto E Silva}, E., {Drell}, P.~S.,
  {Drlica-Wagner}, A., {Falletti}, L., {Favuzzi}, C., {Fegan}, S.~J.,
  {Ferrara}, E.~C., {Fukazawa}, Y., {Funk}, S., {Fusco}, P., {Gargano}, F.,
  {Gasparrini}, D., {Gehrels}, N., {Germani}, S., {Giglietto}, N., {Giordano},
  F., {Giroletti}, M., {Glanzman}, T., {Godfrey}, G., {Grenier}, I.~A.,
  {Guiriec}, S., {Gustafsson}, M., {Hadasch}, D., {Hayashida}, M., {Hays}, E.,
  {Hughes}, R.~E., {Jeltema}, T.~E., {J{\'o}hannesson}, G., {Johnson}, R.~P.,
  {Johnson}, A.~S., {Kamae}, T., {Katagiri}, H., {Kataoka}, J.,
  {Kn{\"o}dlseder}, J., {Kuss}, M., {Lande}, J., {Latronico}, L., {Lionetto},
  A.~M., {Llena Garde}, M., {Longo}, F., {Loparco}, F., {Lott}, B.,
  {Lovellette}, M.~N., {Lubrano}, P., {Madejski}, G.~M., {Mazziotta}, M.~N.,
  {McEnery}, J.~E., {Mehault}, J., {Michelson}, P.~F., {Mitthumsiri}, W.,
  {Mizuno}, T., {Monte}, C., {Monzani}, M.~E., {Morselli}, A., {Moskalenko},
  I.~V., {Murgia}, S., {Naumann-Godo}, M., {Norris}, J.~P., {Nuss}, E.,
  {Ohsugi}, T., {Okumura}, A., {Omodei}, N., {Orlando}, E., {Ormes}, J.~F.,
  {Ozaki}, M., {Paneque}, D., {Parent}, D., {Pesce-Rollins}, M.,
  {Pierbattista}, M., {Piron}, F., {Pivato}, G., {Porter}, T.~A., {Profumo},
  S., {Rain{\`o}}, S., {Razzano}, M., {Reimer}, A., {Reimer}, O., {Ritz}, S.,
  {Roth}, M., {Sadrozinski}, H.~F.-W., {Sbarra}, C., {Scargle}, J.~D.,
  {Schalk}, T.~L., {Sgr{\`o}}, C., {Siskind}, E.~J., {Spandre}, G., {Spinelli},
  P., {Strigari}, L., {Suson}, D.~J., {Tajima}, H., {Takahashi}, H., {Tanaka},
  T., {Thayer}, J.~G., {Thayer}, J.~B., {Thompson}, D.~J., {Tibaldo}, L.,
  {Tinivella}, M., {Torres}, D.~F., {Troja}, E., {Uchiyama}, Y.,
  {Vandenbroucke}, J., {Vasileiou}, V., {Vianello}, G., {Vitale}, V., {Waite},
  A.~P., {Wang}, P., {Winer}, B.~L., {Wood}, K.~S., {Wood}, M., {Yang}, Z.,
  {Zimmer}, S., {Kaplinghat}, M., \& {Martinez}, G.~D. 2011, Physical Review
  Letters, 107, 241302

\bibitem[{{Ad{\'e}n} {et~al.}(2011){Ad{\'e}n}, {Eriksson}, {Feltzing},
  {Grebel}, {Koch}, \& {Wilkinson}}]{aden11}
{Ad{\'e}n}, D., {Eriksson}, K., {Feltzing}, S., {Grebel}, E.~K., {Koch}, A., \&
  {Wilkinson}, M.~I. 2011, \aap, 525, A153

\bibitem[{{Ad{\'e}n} {et~al.}(2009){Ad{\'e}n}, {Wilkinson}, {Read}, {Feltzing},
  {Koch}, {Gilmore}, {Grebel}, \& {Lundstr{\"o}m}}]{aden09}
{Ad{\'e}n}, D., {Wilkinson}, M.~I., {Read}, J.~I., {Feltzing}, S., {Koch}, A.,
  {Gilmore}, G.~F., {Grebel}, E.~K., \& {Lundstr{\"o}m}, I. 2009, \apjl, 706,
  L150

\bibitem[{{Aleksi{\'c}} {et~al.}(2011){Aleksi{\'c}}, {Alvarez}, {Antonelli},
  {Antoranz}, {Asensio}, {Backes}, {Barrio}, {Bastieri}, {Becerra
  Gonz{\'a}lez}, {Bednarek}, {Berdyugin}, {Berger}, {Bernardini}, {Biland},
  {Blanch}, {Bock}, {Boller}, {Bonnoli}, {Borla Tridon}, {Braun}, {Bretz},
  {Ca{\~n}ellas}, {Carmona}, {Carosi}, {Colin}, {Colombo}, {Contreras},
  {Cortina}, {Cossio}, {Covino}, {Dazzi}, {De Angelis}, {De Cea del Pozo}, {De
  Lotto}, {Delgado Mendez}, {Diago Ortega}, {Doert}, {Dom{\'{\i}}nguez},
  {Dominis Prester}, {Dorner}, {Doro}, {Elsaesser}, {Ferenc}, {Fonseca},
  {Font}, {Fruck}, {Garc{\'{\i}}a L{\'o}pez}, {Garczarczyk}, {Garrido},
  {Giavitto}, {Godinovi{\'c}}, {Hadasch}, {H{\"a}fner}, {Herrero},
  {Hildebrand}, {H{\"o}hne-M{\"o}nch}, {Hose}, {Hrupec}, {Huber}, {Jogler},
  {Klepser}, {Kr{\"a}henb{\"u}hl}, {Krause}, {La Barbera}, {Lelas}, {Leonardo},
  {Lindfors}, {Lombardi}, {L{\'o}pez}, {Lorenz}, {Makariev}, {Maneva},
  {Mankuzhiyil}, {Mannheim}, {Maraschi}, {Mariotti}, {Mart{\'{\i}}nez},
  {Mazin}, {Meucci}, {Miranda}, {Mirzoyan}, {Miyamoto}, {Mold{\'o}n},
  {Moralejo}, {Munar-Androver}, {Nieto}, {Nilsson}, {Orito}, {Oya}, {Paiano},
  {Paneque}, {Paoletti}, {Pardo}, {Paredes}, {Partini}, {Pasanen}, {Pauss},
  {Perez-Torres}, {Persic}, {Peruzzo}, {Pilia}, {Pochon}, {Prada}, {Prada
  Moroni}, {Prandini}, {Puljak}, {Reichardt}, {Reinthal}, {Rhode}, {Rib{\'o}},
  {Rico}, {R{\"u}gamer}, {Saggion}, {Saito}, {Saito}, {Salvati}, {Satalecka},
  {Scalzotto}, {Scapin}, {Schultz}, {Schweizer}, {Shayduk}, {Shore},
  {Sillanp{\"a}{\"a}}, {Sitarek}, {Sobczynska}, {Spanier}, {Spiro}, {Stamerra},
  {Steinke}, {Storz}, {Strah}, {Suri{\'c}}, {Takalo}, {Takami}, {Tavecchio},
  {Temnikov}, {Terzi{\'c}}, {Tescaro}, {Teshima}, {Thom}, {Tibolla}, {Torres},
  {Treves}, {Vankov}, {Vogler}, {Wagner}, {Weitzel}, {Zabalza}, {Zandanel},
  {Zanin}, {Fornasa}, {Essig}, {Sehgal}, \& {Strigari}}]{aleksic11}
{Aleksi{\'c}}, J., {Alvarez}, E.~A., {Antonelli}, L.~A., {Antoranz}, P.,
  {Asensio}, M., {Backes}, M., {Barrio}, J.~A., {Bastieri}, D., {Becerra
  Gonz{\'a}lez}, J., {Bednarek}, W., {Berdyugin}, A., {Berger}, K.,
  {Bernardini}, E., {Biland}, A., {Blanch}, O., {Bock}, R.~K., {Boller}, A.,
  {Bonnoli}, G., {Borla Tridon}, D., {Braun}, I., {Bretz}, T., {Ca{\~n}ellas},
  A., {Carmona}, E., {Carosi}, A., {Colin}, P., {Colombo}, E., {Contreras},
  J.~L., {Cortina}, J., {Cossio}, L., {Covino}, S., {Dazzi}, F., {De Angelis},
  A., {De Cea del Pozo}, E., {De Lotto}, B., {Delgado Mendez}, C., {Diago
  Ortega}, A., {Doert}, M., {Dom{\'{\i}}nguez}, A., {Dominis Prester}, D.,
  {Dorner}, D., {Doro}, M., {Elsaesser}, D., {Ferenc}, D., {Fonseca}, M.~V.,
  {Font}, L., {Fruck}, C., {Garc{\'{\i}}a L{\'o}pez}, R.~J., {Garczarczyk}, M.,
  {Garrido}, D., {Giavitto}, G., {Godinovi{\'c}}, N., {Hadasch}, D.,
  {H{\"a}fner}, D., {Herrero}, A., {Hildebrand}, D., {H{\"o}hne-M{\"o}nch}, D.,
  {Hose}, J., {Hrupec}, D., {Huber}, B., {Jogler}, T., {Klepser}, S.,
  {Kr{\"a}henb{\"u}hl}, T., {Krause}, J., {La Barbera}, A., {Lelas}, D.,
  {Leonardo}, E., {Lindfors}, E., {Lombardi}, S., {L{\'o}pez}, M., {Lorenz},
  E., {Makariev}, M., {Maneva}, G., {Mankuzhiyil}, N., {Mannheim}, K.,
  {Maraschi}, L., {Mariotti}, M., {Mart{\'{\i}}nez}, M., {Mazin}, D., {Meucci},
  M., {Miranda}, J.~M., {Mirzoyan}, R., {Miyamoto}, H., {Mold{\'o}n}, J.,
  {Moralejo}, A., {Munar-Androver}, P., {Nieto}, D., {Nilsson}, K., {Orito},
  R., {Oya}, I., {Paiano}, S., {Paneque}, D., {Paoletti}, R., {Pardo}, S.,
  {Paredes}, J.~M., {Partini}, S., {Pasanen}, M., {Pauss}, F., {Perez-Torres},
  M.~A., {Persic}, M., {Peruzzo}, L., {Pilia}, M., {Pochon}, J., {Prada}, F.,
  {Prada Moroni}, P.~G., {Prandini}, E., {Puljak}, I., {Reichardt}, I.,
  {Reinthal}, R., {Rhode}, W., {Rib{\'o}}, M., {Rico}, J., {R{\"u}gamer}, S.,
  {Saggion}, A., {Saito}, K., {Saito}, T.~Y., {Salvati}, M., {Satalecka}, K.,
  {Scalzotto}, V., {Scapin}, V., {Schultz}, C., {Schweizer}, T., {Shayduk}, M.,
  {Shore}, S.~N., {Sillanp{\"a}{\"a}}, A., {Sitarek}, J., {Sobczynska}, D.,
  {Spanier}, F., {Spiro}, S., {Stamerra}, A., {Steinke}, B., {Storz}, J.,
  {Strah}, N., {Suri{\'c}}, T., {Takalo}, L., {Takami}, H., {Tavecchio}, F.,
  {Temnikov}, P., {Terzi{\'c}}, T., {Tescaro}, D., {Teshima}, M., {Thom}, M.,
  {Tibolla}, O., {Torres}, D.~F., {Treves}, A., {Vankov}, H., {Vogler}, P.,
  {Wagner}, R.~M., {Weitzel}, Q., {Zabalza}, V., {Zandanel}, F., {Zanin}, R.,
  {Fornasa}, M., {Essig}, R., {Sehgal}, N., \& {Strigari}, L.~E. 2011, JCAP, 6,
  35

\bibitem[{{Amorisco} \& {Evans}(2012)}]{amorisco12}
{Amorisco}, N.~C., \& {Evans}, N.~W. 2012, \mnras, 419, 184

\bibitem[{{Armandroff} \& {Da Costa}(1986)}]{armandroff86}
{Armandroff}, T.~E., \& {Da Costa}, G.~S. 1986, \aj, 92, 777

\bibitem[{{Armandroff} {et~al.}(1995){Armandroff}, {Olszewski}, \&
  {Pryor}}]{armandroff95}
{Armandroff}, T.~E., {Olszewski}, E.~W., \& {Pryor}, C. 1995, \aj, 110, 2131

\bibitem[{{Battaglia} {et~al.}(2008){Battaglia}, {Helmi}, {Tolstoy}, {Irwin},
  {Hill}, \& {Jablonka}}]{battaglia08}
{Battaglia}, G., {Helmi}, A., {Tolstoy}, E., {Irwin}, M., {Hill}, V., \&
  {Jablonka}, P. 2008, \apjl, 681, L13

\bibitem[{{Battaglia} {et~al.}(2011){Battaglia}, {Tolstoy}, {Helmi}, {Irwin},
  {Parisi}, {Hill}, \& {Jablonka}}]{battaglia11}
{Battaglia}, G., {Tolstoy}, E., {Helmi}, A., {Irwin}, M., {Parisi}, P., {Hill},
  V., \& {Jablonka}, P. 2011, \mnras, 411, 1013

\bibitem[{{Battaglia et al.}(2006)}]{battaglia06}
{Battaglia et al.} 2006, \aap, 459, 423

\bibitem[{{Bekenstein}(2004)}]{bekenstein04}
{Bekenstein}, J.~D. 2004, \prd, 70, 083509

\bibitem[{{Bell} {et~al.}(2011){Bell}, {Slater}, \& {Martin}}]{bell11}
{Bell}, E.~F., {Slater}, C.~T., \& {Martin}, N.~F. 2011, \apjl, 742, L15

\bibitem[{{Belokurov} {et~al.}(2010){Belokurov}, {Walker}, {Evans}, {Gilmore},
  {Irwin}, {Just}, {Koposov}, {Mateo}, {Olszewski}, {Watkins}, \&
  {Wyrzykowski}}]{belokurov10}
{Belokurov}, V., {Walker}, M.~G., {Evans}, N.~W., {Gilmore}, G., {Irwin},
  M.~J., {Just}, D., {Koposov}, S., {Mateo}, M., {Olszewski}, E., {Watkins},
  L., \& {Wyrzykowski}, L. 2010, \apjl, 712, L103

\bibitem[{{Belokurov} {et~al.}(2009){Belokurov}, {Walker}, {Evans}, {Gilmore},
  {Irwin}, {Mateo}, {Mayer}, {Olszewski}, {Bechtold}, \&
  {Pickering}}]{belokurov09}
{Belokurov}, V., {Walker}, M.~G., {Evans}, N.~W., {Gilmore}, G., {Irwin},
  M.~J., {Mateo}, M., {Mayer}, L., {Olszewski}, E., {Bechtold}, J., \&
  {Pickering}, T. 2009, \mnras, 397, 1748

\bibitem[{{Belokurov} {et~al.}(2006){Belokurov}, {Zucker}, {Evans},
  {Wilkinson}, {Irwin}, {Hodgkin}, {Bramich}, {Irwin}, {Gilmore}, {Willman},
  {Vidrih}, {Newberg}, {Wyse}, {Fellhauer}, {Hewett}, {Cole}, {Bell}, {Beers},
  {Rockosi}, {Yanny}, {Grebel}, {Schneider}, {Lupton}, {Barentine},
  {Brewington}, {Brinkmann}, {Harvanek}, {Kleinman}, {Krzesinski}, {Long},
  {Nitta}, {Smith}, \& {Snedden}}]{belokurov06b}
{Belokurov}, V., {Zucker}, D.~B., {Evans}, N.~W., {Wilkinson}, M.~I., {Irwin},
  M.~J., {Hodgkin}, S., {Bramich}, D.~M., {Irwin}, J.~M., {Gilmore}, G.,
  {Willman}, B., {Vidrih}, S., {Newberg}, H.~J., {Wyse}, R.~F.~G., {Fellhauer},
  M., {Hewett}, P.~C., {Cole}, N., {Bell}, E.~F., {Beers}, T.~C., {Rockosi},
  C.~M., {Yanny}, B., {Grebel}, E.~K., {Schneider}, D.~P., {Lupton}, R.,
  {Barentine}, J.~C., {Brewington}, H., {Brinkmann}, J., {Harvanek}, M.,
  {Kleinman}, S.~J., {Krzesinski}, J., {Long}, D., {Nitta}, A., {Smith}, J.~A.,
  \& {Snedden}, S.~A. 2006, \apjl, 647, L111

\bibitem[{{Belokurov et al.}(2006)}]{belokurov06}
{Belokurov et al.} 2006, \apjl, 642, L137

\bibitem[{{Belokurov et al.}(2007)}]{belokurov07}
---. 2007, \apj, 654, 897

\bibitem[{{Belokurov et al.}(2008)}]{belokurov08}
---. 2008, \apjl, 686, L83

\bibitem[{{Bennett} {et~al.}(2003){Bennett}, {Halpern}, {Hinshaw}, {Jarosik},
  {Kogut}, {Limon}, {Meyer}, {Page}, {Spergel}, {Tucker}, {Wollack}, {Wright},
  {Barnes}, {Greason}, {Hill}, {Komatsu}, {Nolta}, {Odegard}, {Peiris},
  {Verde}, \& {Weiland}}]{bennett03}
{Bennett}, C.~L., {Halpern}, M., {Hinshaw}, G., {Jarosik}, N., {Kogut}, A.,
  {Limon}, M., {Meyer}, S.~S., {Page}, L., {Spergel}, D.~N., {Tucker}, G.~S.,
  {Wollack}, E., {Wright}, E.~L., {Barnes}, C., {Greason}, M.~R., {Hill},
  R.~S., {Komatsu}, E., {Nolta}, M.~R., {Odegard}, N., {Peiris}, H.~V.,
  {Verde}, L., \& {Weiland}, J.~L. 2003, \apjs, 148, 1

\bibitem[{{Bergmann}(1968)}]{bergmann68}
{Bergmann}, P.~G. 1968, International Journal of Theoretical Physics, 1, 25

\bibitem[{{Binney} \& {Tremaine}(2008)}]{bt08}
{Binney}, J., \& {Tremaine}, S. 2008, {Galactic Dynamics: Second Edition}
  (Princeton University Press)

\bibitem[{{Blumenthal} {et~al.}(1986){Blumenthal}, {Faber}, {Flores}, \&
  {Primack}}]{blumenthal86}
{Blumenthal}, G.~R., {Faber}, S.~M., {Flores}, R., \& {Primack}, J.~R. 1986,
  \apj, 301, 27

\bibitem[{{Bode} {et~al.}(2001){Bode}, {Ostriker}, \& {Turok}}]{bode01}
{Bode}, P., {Ostriker}, J.~P., \& {Turok}, N. 2001, \apj, 556, 93

\bibitem[{{B{\oe}hm} {et~al.}(2001){B{\oe}hm}, {Fayet}, \&
  {Schaeffer}}]{boehm01}
{B{\oe}hm}, C., {Fayet}, P., \& {Schaeffer}, R. 2001, Physics Letters B, 518, 8

\bibitem[{{Boyanovsky}(2008)}]{boyanovsky08}
{Boyanovsky}, D. 2008, \prd, 78, 103505

\bibitem[{{Boyarsky} {et~al.}(2006){Boyarsky}, {Neronov}, {Ruchayskiy},
  {Shaposhnikov}, \& {Tkachev}}]{boyarsky06}
{Boyarsky}, A., {Neronov}, A., {Ruchayskiy}, O., {Shaposhnikov}, M., \&
  {Tkachev}, I. 2006, Physical Review Letters, 97, 261302

\bibitem[{{Boyarsky} {et~al.}(2010{\natexlab{a}}){Boyarsky}, {Neronov},
  {Ruchayskiy}, \& {Tkachev}}]{boyarsky10b}
{Boyarsky}, A., {Neronov}, A., {Ruchayskiy}, O., \& {Tkachev}, I.
  2010{\natexlab{a}}, Physical Review Letters, 104, 191301

\bibitem[{{Boyarsky} {et~al.}(2007){Boyarsky}, {Nevalainen}, \&
  {Ruchayskiy}}]{boyarsky07}
{Boyarsky}, A., {Nevalainen}, J., \& {Ruchayskiy}, O. 2007, \aap, 471, 51

\bibitem[{{Boyarsky} {et~al.}(2009{\natexlab{a}}){Boyarsky}, {Ruchayskiy}, \&
  {Iakubovskyi}}]{boyarsky09b}
{Boyarsky}, A., {Ruchayskiy}, O., \& {Iakubovskyi}, D. 2009{\natexlab{a}},
  JCAP, 3, 5

\bibitem[{{Boyarsky} {et~al.}(2010{\natexlab{b}}){Boyarsky}, {Ruchayskiy},
  {Iakubovskyi}, {Walker}, {Riemer-S{\o}rensen}, \& {Hansen}}]{boyarsky10}
{Boyarsky}, A., {Ruchayskiy}, O., {Iakubovskyi}, D., {Walker}, M.~G.,
  {Riemer-S{\o}rensen}, S., \& {Hansen}, S.~H. 2010{\natexlab{b}}, \mnras, 407,
  1188

\bibitem[{{Boyarsky} {et~al.}(2009{\natexlab{b}}){Boyarsky}, {Ruchayskiy}, \&
  {Shaposhnikov}}]{boyarsky09c}
{Boyarsky}, A., {Ruchayskiy}, O., \& {Shaposhnikov}, M. 2009{\natexlab{b}},
  Annual Review of Nuclear and Particle Science, 59, 191

\bibitem[{{Boylan-Kolchin} {et~al.}(2011){Boylan-Kolchin}, {Bullock}, \&
  {Kaplinghat}}]{boylankolchin11}
{Boylan-Kolchin}, M., {Bullock}, J.~S., \& {Kaplinghat}, M. 2011, \mnras, 415,
  L40

\bibitem[{{Breddels} {et~al.}(2012){Breddels}, {Helmi}, {van den Bosch}, {van
  de Ven}, \& {Battaglia}}]{breddels11}
{Breddels}, M.~A., {Helmi}, A., {van den Bosch}, R.~C.~E., {van de Ven}, G., \&
  {Battaglia}, G. 2012, Assembling the Puzzle of the Milky Way, Le
  Grand-Bornand, France, Edited by C.~Reyl{\'e}; A.~Robin; M.~Schultheis; EPJ
  Web of Conferences, Volume 19, id.03009, 19, 3009

\bibitem[{{Burkert}(1995)}]{burkert95}
{Burkert}, A. 1995, \apjl, 447, L25

\bibitem[{{Cannon} {et~al.}(1977){Cannon}, {Hawarden}, \& {Tritton}}]{cannon78}
{Cannon}, R.~D., {Hawarden}, T.~G., \& {Tritton}, S.~B. 1977, \mnras, 180, 81P

\bibitem[{{Carlin} {et~al.}(2009){Carlin}, {Grillmair}, {Mu{\~n}oz}, {Nidever},
  \& {Majewski}}]{carlin09}
{Carlin}, J.~L., {Grillmair}, C.~J., {Mu{\~n}oz}, R.~R., {Nidever}, D.~L., \&
  {Majewski}, S.~R. 2009, \apjl, 702, L9

\bibitem[{{Chanam{\'e}} {et~al.}(2008){Chanam{\'e}}, {Kleyna}, \& {van der
  Marel}}]{chaname08}
{Chanam{\'e}}, J., {Kleyna}, J., \& {van der Marel}, R. 2008, \apj, 682, 841

\bibitem[{{Charbonnier} {et~al.}(2011){Charbonnier}, {Combet}, {Daniel},
  {Funk}, {Hinton}, {Maurin}, {Power}, {Read}, {Sarkar}, {Walker}, \&
  {Wilkinson}}]{charbonnier11}
{Charbonnier}, A., {Combet}, C., {Daniel}, M., {Funk}, S., {Hinton}, J.~A.,
  {Maurin}, D., {Power}, C., {Read}, J.~I., {Sarkar}, S., {Walker}, M.~G., \&
  {Wilkinson}, M.~I. 2011, \mnras, 418, 1526

\bibitem[{{Cole} {et~al.}(2011){Cole}, {Dehnen}, \& {Wilkinson}}]{cole11}
{Cole}, D.~R., {Dehnen}, W., \& {Wilkinson}, M.~I. 2011, \mnras, 416, 1118

\bibitem[{{Coleman} {et~al.}(2004){Coleman}, {Da Costa}, {Bland-Hawthorn},
  {Mart{\'{\i}}nez-Delgado}, {Freeman}, \& {Malin}}]{coleman04}
{Coleman}, M., {Da Costa}, G.~S., {Bland-Hawthorn}, J.,
  {Mart{\'{\i}}nez-Delgado}, D., {Freeman}, K.~C., \& {Malin}, D. 2004, \aj,
  127, 832

\bibitem[{{Coleman} {et~al.}(2005{\natexlab{a}}){Coleman}, {Da Costa}, \&
  {Bland-Hawthorn}}]{coleman05a}
{Coleman}, M.~G., {Da Costa}, G.~S., \& {Bland-Hawthorn}, J.
  2005{\natexlab{a}}, \aj, 130, 1065

\bibitem[{{Coleman} {et~al.}(2005{\natexlab{b}}){Coleman}, {Da Costa},
  {Bland-Hawthorn}, \& {Freeman}}]{coleman05b}
{Coleman}, M.~G., {Da Costa}, G.~S., {Bland-Hawthorn}, J., \& {Freeman}, K.~C.
  2005{\natexlab{b}}, \aj, 129, 1443

\bibitem[{{Coleman et al.}(2007)}]{coleman07}
{Coleman et al.} 2007, \apjl, 668, L43

\bibitem[{{Collins} {et~al.}(2010){Collins}, {Chapman}, {Irwin}, {Martin},
  {Ibata}, {Zucker}, {Blain}, {Ferguson}, {Lewis}, {McConnachie}, \&
  {Pe{\~n}arrubia}}]{collins10}
{Collins}, M.~L.~M., {Chapman}, S.~C., {Irwin}, M.~J., {Martin}, N.~F.,
  {Ibata}, R.~A., {Zucker}, D.~B., {Blain}, A., {Ferguson}, A.~M.~N., {Lewis},
  G.~F., {McConnachie}, A.~W., \& {Pe{\~n}arrubia}, J. 2010, \mnras, 407, 2411

\bibitem[{{Collins} {et~al.}(2011){Collins}, {Chapman}, {Rich}, {Irwin},
  {Pe{\~n}arrubia}, {Ibata}, {Arimoto}, {Brooks}, {Ferguson}, {Lewis},
  {McConnachie}, \& {Venn}}]{collins11}
{Collins}, M.~L.~M., {Chapman}, S.~C., {Rich}, R.~M., {Irwin}, M.~J.,
  {Pe{\~n}arrubia}, J., {Ibata}, R.~A., {Arimoto}, N., {Brooks}, A.~M.,
  {Ferguson}, A.~M.~N., {Lewis}, G.~F., {McConnachie}, A.~W., \& {Venn}, K.
  2011, \mnras, 417, 1170

\bibitem[{{CTA Consortium}(2010)}]{cta10}
{CTA Consortium}. 2010, ArXiv:1008.3703

\bibitem[{{Da Costa} {et~al.}(1991){Da Costa}, {Hatzidimitriou}, {Irwin}, \&
  {McMahon}}]{dacosta91}
{Da Costa}, G.~S., {Hatzidimitriou}, D., {Irwin}, M.~J., \& {McMahon}, R.~G.
  1991, \mnras, 249, 473

\bibitem[{{Dalcanton} \& {Hogan}(2001)}]{dalcanton01}
{Dalcanton}, J.~J., \& {Hogan}, C.~J. 2001, \apj, 561, 35

\bibitem[{{de Blok}(2010)}]{deblok10}
{de Blok}, W.~J.~G. 2010, Advances in Astronomy, 2010

\bibitem[{{de Blok} \& {McGaugh}(1997)}]{deblok97}
{de Blok}, W.~J.~G., \& {McGaugh}, S.~S. 1997, \mnras, 290, 533

\bibitem[{{de Jong} {et~al.}(2010){de Jong}, {Martin}, {Rix}, {Smith}, {Jin},
  \& {Macci{\`o}}}]{dejong10}
{de Jong}, J.~T.~A., {Martin}, N.~F., {Rix}, H.-W., {Smith}, K.~W., {Jin}, S.,
  \& {Macci{\`o}}, A.~V. 2010, \apj, 710, 1664

\bibitem[{{de Souza} {et~al.}(2011){de Souza}, {Rodrigues}, {Ishida}, \&
  {Opher}}]{desouza11}
{de Souza}, R.~S., {Rodrigues}, L.~F.~S., {Ishida}, E.~E.~O., \& {Opher}, R.
  2011, \mnras, 415, 2969

\bibitem[{{de Vaucouleurs}(1948)}]{devaucouleurs48}
{de Vaucouleurs}, G. 1948, Annales d'Astrophysique, 11, 247

\bibitem[{{Del Popolo}(2010)}]{delpopolo10}
{Del Popolo}, A. 2010, \mnras, 408, 1808

\bibitem[{{Diemand} {et~al.}(2005{\natexlab{a}}){Diemand}, {Moore}, \&
  {Stadel}}]{diemand05}
{Diemand}, J., {Moore}, B., \& {Stadel}, J. 2005{\natexlab{a}}, \nat, 433, 389

\bibitem[{{Diemand} {et~al.}(2005{\natexlab{b}}){Diemand}, {Zemp}, {Moore},
  {Stadel}, \& {Carollo}}]{diemand05b}
{Diemand}, J., {Zemp}, M., {Moore}, B., {Stadel}, J., \& {Carollo}, C.~M.
  2005{\natexlab{b}}, \mnras, 364, 665

\bibitem[{{Dodelson} \& {Widrow}(1994)}]{dodelson94}
{Dodelson}, S., \& {Widrow}, L.~M. 1994, Physical Review Letters, 72, 17

\bibitem[{{Donato} {et~al.}(2009){Donato}, {Gentile}, {Salucci}, {Frigerio
  Martins}, {Wilkinson}, {Gilmore}, {Grebel}, {Koch}, \& {Wyse}}]{donato09}
{Donato}, F., {Gentile}, G., {Salucci}, P., {Frigerio Martins}, C.,
  {Wilkinson}, M.~I., {Gilmore}, G., {Grebel}, E.~K., {Koch}, A., \& {Wyse}, R.
  2009, \mnras, 397, 1169

\bibitem[{{Dubinski} \& {Carlberg}(1991)}]{dubinski91}
{Dubinski}, J., \& {Carlberg}, R.~G. 1991, \apj, 378, 496

\bibitem[{{Duquennoy} \& {Mayor}(1991)}]{duquennoy91}
{Duquennoy}, A., \& {Mayor}, M. 1991, \aap, 248, 485

\bibitem[{{El-Zant} {et~al.}(2001){El-Zant}, {Shlosman}, \&
  {Hoffman}}]{elzant01}
{El-Zant}, A., {Shlosman}, I., \& {Hoffman}, Y. 2001, \apj, 560, 636

\bibitem[{{Essig} {et~al.}(2009){Essig}, {Sehgal}, \& {Strigari}}]{essig09}
{Essig}, R., {Sehgal}, N., \& {Strigari}, L.~E. 2009, \prd, 80, 023506

\bibitem[{{Evans} {et~al.}(2004){Evans}, {Ferrer}, \& {Sarkar}}]{evans04}
{Evans}, N.~W., {Ferrer}, F., \& {Sarkar}, S. 2004, \prd, 69, 123501

\bibitem[{{Evans, An \& Walker}(2009)}]{evans09}
{Evans, An \& Walker}. 2009, \mnras, 393, L50

\bibitem[{{Faber} \& {Lin}(1983)}]{faber83}
{Faber}, S.~M., \& {Lin}, D.~N.~C. 1983, \apjl, 266, L17

\bibitem[{{Fabrizio} {et~al.}(2011){Fabrizio}, {Nonino}, {Bono}, {Ferraro},
  {Fran{\c c}ois}, {Iannicola}, {Monelli}, {Th{\'e}venin}, {Stetson}, {Walker},
  {Buonanno}, {Caputo}, {Corsi}, {Dall'Ora}, {Gilmozzi}, {James}, {Merle},
  {Pulone}, \& {Romaniello}}]{fabrizio11}
{Fabrizio}, M., {Nonino}, M., {Bono}, G., {Ferraro}, I., {Fran{\c c}ois}, P.,
  {Iannicola}, G., {Monelli}, M., {Th{\'e}venin}, F., {Stetson}, P.~B.,
  {Walker}, A.~R., {Buonanno}, R., {Caputo}, F., {Corsi}, C.~E., {Dall'Ora},
  M., {Gilmozzi}, R., {James}, C.~R., {Merle}, T., {Pulone}, L., \&
  {Romaniello}, M. 2011, \pasp, 123, 384

\bibitem[{{Feng}(2010)}]{feng10}
{Feng}, J.~L. 2010, \araa, 48, 495

\bibitem[{{Fleck} \& {Kuhn}(2003)}]{fleck03}
{Fleck}, J.-J., \& {Kuhn}, J.~R. 2003, \apj, 592, 147

\bibitem[{{Flores} \& {Primack}(1994)}]{flores94}
{Flores}, R.~A., \& {Primack}, J.~R. 1994, \apjl, 427, L1

\bibitem[{{Font} {et~al.}(2011){Font}, {Benson}, {Bower}, {Frenk}, {Cooper},
  {De Lucia}, {Helly}, {Helmi}, {Li}, {McCarthy}, {Navarro}, {Springel},
  {Starkenburg}, {Wang}, \& {White}}]{font11}
{Font}, A.~S., {Benson}, A.~J., {Bower}, R.~G., {Frenk}, C.~S., {Cooper}, A.,
  {De Lucia}, G., {Helly}, J.~C., {Helmi}, A., {Li}, Y.-S., {McCarthy}, I.~G.,
  {Navarro}, J.~F., {Springel}, V., {Starkenburg}, E., {Wang}, J., \& {White},
  S.~D.~M. 2011, \mnras, 417, 1260

\bibitem[{{Geha} {et~al.}(2009){Geha}, {Willman}, {Simon}, {Strigari}, {Kirby},
  {Law}, \& {Strader}}]{geha09}
{Geha}, M., {Willman}, B., {Simon}, J.~D., {Strigari}, L.~E., {Kirby}, E.~N.,
  {Law}, D.~R., \& {Strader}, J. 2009, \apj, 692, 1464

\bibitem[{{Gerhard} \& {Spergel}(1992)}]{gerhard92b}
{Gerhard}, O.~E., \& {Spergel}, D.~N. 1992, \apjl, 389, L9

\bibitem[{{Geringer-Sameth} \& {Koushiappas}(2011)}]{geringer11}
{Geringer-Sameth}, A., \& {Koushiappas}, S.~M. 2011, Physical Review Letters,
  107, 241303

\bibitem[{{Gilmore} {et~al.}(2007){Gilmore}, {Wilkinson}, {Wyse}, {Kleyna},
  {Koch}, {Evans}, \& {Grebel}}]{gilmore07}
{Gilmore}, G., {Wilkinson}, M.~I., {Wyse}, R.~F.~G., {Kleyna}, J.~T., {Koch},
  A., {Evans}, N.~W., \& {Grebel}, E.~K. 2007, \apj, 663, 948

\bibitem[{{Gnedin} {et~al.}(1999){Gnedin}, {Hernquist}, \&
  {Ostriker}}]{gnedin99}
{Gnedin}, O.~Y., {Hernquist}, L., \& {Ostriker}, J.~P. 1999, \apj, 514, 109

\bibitem[{{Gnedin} {et~al.}(2004){Gnedin}, {Kravtsov}, {Klypin}, \&
  {Nagai}}]{gnedin04}
{Gnedin}, O.~Y., {Kravtsov}, A.~V., {Klypin}, A.~A., \& {Nagai}, D. 2004, \apj,
  616, 16

\bibitem[{{Goerdt} {et~al.}(2010){Goerdt}, {Moore}, {Read}, \&
  {Stadel}}]{goerdt10}
{Goerdt}, T., {Moore}, B., {Read}, J.~I., \& {Stadel}, J. 2010, \apj, 725, 1707

\bibitem[{{Goerdt} {et~al.}(2006){Goerdt}, {Moore}, {Read}, {Stadel}, \&
  {Zemp}}]{goerdt06}
{Goerdt}, T., {Moore}, B., {Read}, J.~I., {Stadel}, J., \& {Zemp}, M. 2006,
  \mnras, 368, 1073

\bibitem[{{Governato} {et~al.}(2010){Governato}, {Brook}, {Mayer}, {Brooks},
  {Rhee}, {Wadsley}, {Jonsson}, {Willman}, {Stinson}, {Quinn}, \&
  {Madau}}]{governato10}
{Governato}, F., {Brook}, C., {Mayer}, L., {Brooks}, A., {Rhee}, G., {Wadsley},
  J., {Jonsson}, P., {Willman}, B., {Stinson}, G., {Quinn}, T., \& {Madau}, P.
  2010, \nat, 463, 203

\bibitem[{{Governato} {et~al.}(2012){Governato}, {Zolotov}, {Pontzen},
  {Christensen}, {Oh}, {Brooks}, {Quinn}, {Shen}, \& {Wadsley}}]{governato12}
{Governato}, F., {Zolotov}, A., {Pontzen}, A., {Christensen}, C., {Oh}, S.~H.,
  {Brooks}, A.~M., {Quinn}, T., {Shen}, S., \& {Wadsley}, J. 2012, \mnras, 2697

\bibitem[{{Green} {et~al.}(2004){Green}, {Hofmann}, \& {Schwarz}}]{green04}
{Green}, A.~M., {Hofmann}, S., \& {Schwarz}, D.~J. 2004, \mnras, 353, L23

\bibitem[{{Grillmair}(2009)}]{grillmair09}
{Grillmair}, C.~J. 2009, \apj, 693, 1118

\bibitem[{{Gunn} {et~al.}(1978){Gunn}, {Lee}, {Lerche}, {Schramm}, \&
  {Steigman}}]{gunn78}
{Gunn}, J.~E., {Lee}, B.~W., {Lerche}, I., {Schramm}, D.~N., \& {Steigman}, G.
  1978, \apj, 223, 1015

\bibitem[{{H.~E.~S.~S.~Collaboration}
  {et~al.}(2011){H.~E.~S.~S.~Collaboration}, {Abramowski}, {Acero},
  {Aharonian}, {Akhperjanian}, {Anton}, {Barnacka}, {Barres de Almeida},
  {Bazer-Bachi}, {Becherini}, {Becker}, {Behera}, {Bernl{\"o}hr}, {Bochow},
  {Boisson}, {Bolmont}, {Bordas}, {Borrel}, {Brucker}, {Brun}, {Brun}, {Bulik},
  {B{\"u}sching}, {Carrigan}, {Casanova}, {Cerruti}, {Chadwick}, {Charbonnier},
  {Chaves}, {Cheesebrough}, {Chounet}, {Clapson}, {Coignet}, {Conrad},
  {Dalton}, {Daniel}, {Davids}, {Degrange}, {Deil}, {Dickinson},
  {Djannati-Ata{\"i}}, {Domainko}, {Drury}, {Dubois}, {Dubus}, {Dyks}, {Dyrda},
  {Egberts}, {Eger}, {Espigat}, {Fallon}, {Farnier}, {Fegan}, {Feinstein},
  {Fernandes}, {Fiasson}, {Fontaine}, {F{\"o}rster}, {F{\"u}{\ss}ling},
  {Gallant}, {Gast}, {G{\'e}rard}, {Gerbig}, {Giebels}, {Glicenstein},
  {Gl{\"u}ck}, {Goret}, {G{\"o}ring}, {Hague}, {Hampf}, {Hauser}, {Heinz},
  {Heinzelmann}, {Henri}, {Hermann}, {Hinton}, {Hoffmann}, {Hofmann},
  {Hofverberg}, {Horns}, {Jacholkowska}, {de Jager}, {Jahn}, {Jamrozy}, {Jung},
  {Kastendieck}, {Katarzy{\'n}ski}, {Katz}, {Kaufmann}, {Keogh}, {Kerschhaggl},
  {Khangulyan}, {Kh{\'e}lifi}, {Klochkov}, {Klu{\'z}niak}, {Kneiske}, {Komin},
  {Kosack}, {Kossakowski}, {Laffon}, {Lamanna}, {Lennarz}, {Lohse}, {Lopatin},
  {Lu}, {Marandon}, {Marcowith}, {Masbou}, {Maurin}, {Maxted}, {McComb},
  {Medina}, {M{\'e}hault}, {Moderski}, {Moulin}, {Naumann}, {Naumann-Godo}, {de
  Naurois}, {Nedbal}, {Nekrassov}, {Nguyen}, {Nicholas}, {Niemiec}, {Nolan},
  {Ohm}, {Olive}, {de O{\~n}a Wilhelmi}, {Opitz}, {Ostrowski}, {Panter}, {Paz
  Arribas}, {Pedaletti}, {Pelletier}, {Petrucci}, {Pita}, {P{\"u}hlhofer},
  {Punch}, {Quirrenbach}, {Raue}, {Rayner}, {Reimer}, {Reimer}, {Renaud}, {de
  Los Reyes}, {Rieger}, {Ripken}, {Rob}, {Rosier-Lees}, {Rowell}, {Rudak},
  {Rulten}, {Ruppel}, {Ryde}, {Sahakian}, {Santangelo}, {Schlickeiser},
  {Sch{\"o}ck}, {Sch{\"o}nwald}, {Schwanke}, {Schwarzburg}, {Schwemmer},
  {Shalchi}, {Sikora}, {Skilton}, {Sol}, {Spengler}, {Stawarz}, {Steenkamp},
  {Stegmann}, {Stinzing}, {Sushch}, {Szostek}, {Tavernet}, {Terrier},
  {Tibolla}, {Tluczykont}, {Valerius}, {van Eldik}, {Vasileiadis}, {Venter},
  {Vialle}, {Viana}, {Vincent}, {Vivier}, {V{\"o}lk}, {Volpe}, {Vorobiov},
  {Vorster}, {Wagner}, {Ward}, {Wierzcholska}, {Zajczyk}, {Zdziarski}, {Zech},
  {Zechlin}, \& {H.E.S.S.~Collaboration}}]{hess11}
{H.~E.~S.~S.~Collaboration}, {Abramowski}, A., {Acero}, F., {Aharonian}, F.,
  {Akhperjanian}, A.~G., {Anton}, G., {Barnacka}, A., {Barres de Almeida}, U.,
  {Bazer-Bachi}, A.~R., {Becherini}, Y., {Becker}, J., {Behera}, B.,
  {Bernl{\"o}hr}, K., {Bochow}, A., {Boisson}, C., {Bolmont}, J., {Bordas}, P.,
  {Borrel}, V., {Brucker}, J., {Brun}, F., {Brun}, P., {Bulik}, T.,
  {B{\"u}sching}, I., {Carrigan}, S., {Casanova}, S., {Cerruti}, M.,
  {Chadwick}, P.~M., {Charbonnier}, A., {Chaves}, R.~C.~G., {Cheesebrough}, A.,
  {Chounet}, L.-M., {Clapson}, A.~C., {Coignet}, G., {Conrad}, J., {Dalton},
  M., {Daniel}, M.~K., {Davids}, I.~D., {Degrange}, B., {Deil}, C.,
  {Dickinson}, H.~J., {Djannati-Ata{\"i}}, A., {Domainko}, W., {Drury},
  L.~O.~C., {Dubois}, F., {Dubus}, G., {Dyks}, J., {Dyrda}, M., {Egberts}, K.,
  {Eger}, P., {Espigat}, P., {Fallon}, L., {Farnier}, C., {Fegan}, S.,
  {Feinstein}, F., {Fernandes}, M.~V., {Fiasson}, A., {Fontaine}, G.,
  {F{\"o}rster}, A., {F{\"u}{\ss}ling}, M., {Gallant}, Y.~A., {Gast}, H.,
  {G{\'e}rard}, L., {Gerbig}, D., {Giebels}, B., {Glicenstein}, J.~F.,
  {Gl{\"u}ck}, B., {Goret}, P., {G{\"o}ring}, D., {Hague}, J.~D., {Hampf}, D.,
  {Hauser}, M., {Heinz}, S., {Heinzelmann}, G., {Henri}, G., {Hermann}, G.,
  {Hinton}, J.~A., {Hoffmann}, A., {Hofmann}, W., {Hofverberg}, P., {Horns},
  D., {Jacholkowska}, A., {de Jager}, O.~C., {Jahn}, C., {Jamrozy}, M., {Jung},
  I., {Kastendieck}, M.~A., {Katarzy{\'n}ski}, K., {Katz}, U., {Kaufmann}, S.,
  {Keogh}, D., {Kerschhaggl}, M., {Khangulyan}, D., {Kh{\'e}lifi}, B.,
  {Klochkov}, D., {Klu{\'z}niak}, W., {Kneiske}, T., {Komin}, N., {Kosack}, K.,
  {Kossakowski}, R., {Laffon}, H., {Lamanna}, G., {Lennarz}, D., {Lohse}, T.,
  {Lopatin}, A., {Lu}, C.-C., {Marandon}, V., {Marcowith}, A., {Masbou}, J.,
  {Maurin}, D., {Maxted}, N., {McComb}, T.~J.~L., {Medina}, M.~C.,
  {M{\'e}hault}, J., {Moderski}, R., {Moulin}, E., {Naumann}, C.~L.,
  {Naumann-Godo}, M., {de Naurois}, M., {Nedbal}, D., {Nekrassov}, D.,
  {Nguyen}, N., {Nicholas}, B., {Niemiec}, J., {Nolan}, S.~J., {Ohm}, S.,
  {Olive}, J.-F., {de O{\~n}a Wilhelmi}, E., {Opitz}, B., {Ostrowski}, M.,
  {Panter}, M., {Paz Arribas}, M., {Pedaletti}, G., {Pelletier}, G.,
  {Petrucci}, P.-O., {Pita}, S., {P{\"u}hlhofer}, G., {Punch}, M.,
  {Quirrenbach}, A., {Raue}, M., {Rayner}, S.~M., {Reimer}, A., {Reimer}, O.,
  {Renaud}, M., {de Los Reyes}, R., {Rieger}, F., {Ripken}, J., {Rob}, L.,
  {Rosier-Lees}, S., {Rowell}, G., {Rudak}, B., {Rulten}, C.~B., {Ruppel}, J.,
  {Ryde}, F., {Sahakian}, V., {Santangelo}, A., {Schlickeiser}, R.,
  {Sch{\"o}ck}, F.~M., {Sch{\"o}nwald}, A., {Schwanke}, U., {Schwarzburg}, S.,
  {Schwemmer}, S., {Shalchi}, A., {Sikora}, M., {Skilton}, J.~L., {Sol}, H.,
  {Spengler}, G., {Stawarz}, {\L}., {Steenkamp}, R., {Stegmann}, C.,
  {Stinzing}, F., {Sushch}, I., {Szostek}, A., {Tavernet}, J.-P., {Terrier},
  R., {Tibolla}, O., {Tluczykont}, M., {Valerius}, K., {van Eldik}, C.,
  {Vasileiadis}, G., {Venter}, C., {Vialle}, J.~P., {Viana}, A., {Vincent}, P.,
  {Vivier}, M., {V{\"o}lk}, H.~J., {Volpe}, F., {Vorobiov}, S., {Vorster}, M.,
  {Wagner}, S.~J., {Ward}, M., {Wierzcholska}, A., {Zajczyk}, A., {Zdziarski},
  A.~A., {Zech}, A., {Zechlin}, H.-S., \& {H.E.S.S.~Collaboration}. 2011,
  Astroparticle Physics, 34, 608

\bibitem[{{Han} {et~al.}(2012){Han}, {Frenk}, {Eke}, {Gao}, \& {White}}]{han12}
{Han}, J., {Frenk}, C.~S., {Eke}, V.~R., {Gao}, L., \& {White}, S.~D.~M. 2012,
  ArXiv:1201.1003

\bibitem[{{Hargreaves} {et~al.}(1996{\natexlab{a}}){Hargreaves}, {Gilmore}, \&
  {Annan}}]{hargreaves96b}
{Hargreaves}, J.~C., {Gilmore}, G., \& {Annan}, J.~D. 1996{\natexlab{a}},
  \mnras, 279, 108

\bibitem[{{Hargreaves} {et~al.}(1994{\natexlab{a}}){Hargreaves}, {Gilmore},
  {Irwin}, \& {Carter}}]{hargreaves94b}
{Hargreaves}, J.~C., {Gilmore}, G., {Irwin}, M.~J., \& {Carter}, D.
  1994{\natexlab{a}}, \mnras, 269, 957

\bibitem[{{Hargreaves} {et~al.}(1994{\natexlab{b}}){Hargreaves}, {Gilmore},
  {Irwin}, \& {Carter}}]{hargreaves94a}
---. 1994{\natexlab{b}}, \mnras, 271, 693

\bibitem[{{Hargreaves} {et~al.}(1996{\natexlab{b}}){Hargreaves}, {Gilmore},
  {Irwin}, \& {Carter}}]{hargreaves96a}
---. 1996{\natexlab{b}}, \mnras, 282, 305

\bibitem[{{Harrington} \& {Wilson}(1950)}]{harrington50}
{Harrington}, R.~G., \& {Wilson}, A.~G. 1950, \pasp, 62, 118

\bibitem[{{Harris}(1996)}]{harris96}
{Harris}, W.~E. 1996, \aj, 112, 1487

\bibitem[{{Heisler} {et~al.}(1985){Heisler}, {Tremaine}, \&
  {Bahcall}}]{heisler85}
{Heisler}, J., {Tremaine}, S., \& {Bahcall}, J.~N. 1985, \apj, 298, 8

\bibitem[{{Hernandez} \& {Gilmore}(1998)}]{hernandez98}
{Hernandez}, X., \& {Gilmore}, G. 1998, \mnras, 297, 517

\bibitem[{{Hernquist}(1990)}]{hernquist90}
{Hernquist}, L. 1990, \apj, 356, 359

\bibitem[{{Hodge}(1961{\natexlab{a}})}]{hodge61b}
{Hodge}, P.~W. 1961{\natexlab{a}}, \aj, 66, 384

\bibitem[{{Hodge}(1961{\natexlab{b}})}]{hodge61a}
---. 1961{\natexlab{b}}, \aj, 66, 249

\bibitem[{{Hodge}(1962)}]{hodge62}
---. 1962, \aj, 67, 125

\bibitem[{{Hodge}(1963)}]{hodge63}
---. 1963, \aj, 68, 470

\bibitem[{{Hodge}(1964{\natexlab{a}})}]{hodge64a}
---. 1964{\natexlab{a}}, \aj, 69, 853

\bibitem[{{Hodge}(1966)}]{hodge66}
---. 1966, \apj, 144, 869

\bibitem[{{Hodge} \& {Michie}(1969)}]{hodge69}
{Hodge}, P.~W., \& {Michie}, R.~W. 1969, \aj, 74, 587

\bibitem[{{Hodge}(1964{\natexlab{b}})}]{hodge64b}
{Hodge}, W.~P. 1964{\natexlab{b}}, \aj, 69, 438

\bibitem[{{Hofmann} {et~al.}(2001){Hofmann}, {Schwarz}, \&
  {St{\"o}cker}}]{hofmann01}
{Hofmann}, S., {Schwarz}, D.~J., \& {St{\"o}cker}, H. 2001, \prd, 64, 083507

\bibitem[{{Hooper} \& {Linden}(2011)}]{hooper11}
{Hooper}, D., \& {Linden}, T. 2011, \prd, 83, 083517

\bibitem[{{Hubble}(1930)}]{hubble30}
{Hubble}, E.~P. 1930, \apj, 71, 231

\bibitem[{{Huxor} {et~al.}(2005){Huxor}, {Tanvir}, {Irwin}, {Ibata}, {Collett},
  {Ferguson}, {Bridges}, \& {Lewis}}]{huxor05}
{Huxor}, A.~P., {Tanvir}, N.~R., {Irwin}, M.~J., {Ibata}, R., {Collett}, J.~L.,
  {Ferguson}, A.~M.~N., {Bridges}, T., \& {Lewis}, G.~F. 2005, \mnras, 360,
  1007

\bibitem[{{Ibata} {et~al.}(1994){Ibata}, {Gilmore}, \& {Irwin}}]{ibata94}
{Ibata}, R.~A., {Gilmore}, G., \& {Irwin}, M.~J. 1994, \nat, 370, 194

\bibitem[{{Ibata} {et~al.}(1997){Ibata}, {Wyse}, {Gilmore}, {Irwin}, \&
  {Suntzeff}}]{ibata97b}
{Ibata}, R.~A., {Wyse}, R.~F.~G., {Gilmore}, G., {Irwin}, M.~J., \& {Suntzeff},
  N.~B. 1997, \aj, 113, 634

\bibitem[{{Ibata et al.}(2007)}]{ibata07}
{Ibata et al.} 2007, \apj, 671, 1591

\bibitem[{{Illingworth}(1976)}]{illingworth76}
{Illingworth}, G. 1976, \apj, 204, 73

\bibitem[{{Irwin} \& {Hatzidimitriou}(1995)}]{ih95}
{Irwin}, M., \& {Hatzidimitriou}, D. 1995, \mnras, 277, 1354

\bibitem[{{Irwin} {et~al.}(1990){Irwin}, {Bunclark}, {Bridgeland}, \&
  {McMahon}}]{irwin90}
{Irwin}, M.~J., {Bunclark}, P.~S., {Bridgeland}, M.~T., \& {McMahon}, R.~G.
  1990, \mnras, 244, 16P

\bibitem[{{Irwin} {et~al.}(2008){Irwin}, {Ferguson}, {Huxor}, {Tanvir},
  {Ibata}, \& {Lewis}}]{irwin08}
{Irwin}, M.~J., {Ferguson}, A.~M.~N., {Huxor}, A.~P., {Tanvir}, N.~R., {Ibata},
  R.~A., \& {Lewis}, G.~F. 2008, \apjl, 676, L17

\bibitem[{{Irwin et al.}(2007)}]{irwin07}
{Irwin et al.} 2007, \apjl, 656, L13

\bibitem[{{Jardel} \& {Gebhardt}(2011)}]{jardel11}
{Jardel}, J., \& {Gebhardt}, K. 2011, ArXiv:1112.0319

\bibitem[{{Johnston} {et~al.}(2005){Johnston}, {Law}, \&
  {Majewski}}]{johnston05}
{Johnston}, K.~V., {Law}, D.~R., \& {Majewski}, S.~R. 2005, \apj, 619, 800

\bibitem[{{Jungman} {et~al.}(1996){Jungman}, {Kamionkowski}, \&
  {Griest}}]{jungman96}
{Jungman}, G., {Kamionkowski}, M., \& {Griest}, K. 1996, \physrep, 267, 195

\bibitem[{{Kalirai} {et~al.}(2010){Kalirai}, {Beaton}, {Geha}, {Gilbert},
  {Guhathakurta}, {Kirby}, {Majewski}, {Ostheimer}, {Patterson}, \&
  {Wolf}}]{kalirai10}
{Kalirai}, J.~S., {Beaton}, R.~L., {Geha}, M.~C., {Gilbert}, K.~M.,
  {Guhathakurta}, P., {Kirby}, E.~N., {Majewski}, S.~R., {Ostheimer}, J.~C.,
  {Patterson}, R.~J., \& {Wolf}, J. 2010, \apj, 711, 671

\bibitem[{{Kaplinghat} \& {Strigari}(2008)}]{kaplinghat08}
{Kaplinghat}, M., \& {Strigari}, L.~E. 2008, \apjl, 682, L93

\bibitem[{{King}(1962)}]{king62}
{King}, I. 1962, \aj, 67, 471

\bibitem[{{King}(1966)}]{king66}
{King}, I.~R. 1966, \aj, 71, 64

\bibitem[{{Kirby} {et~al.}(2011){Kirby}, {Lanfranchi}, {Simon}, {Cohen}, \&
  {Guhathakurta}}]{kirby11}
{Kirby}, E.~N., {Lanfranchi}, G.~A., {Simon}, J.~D., {Cohen}, J.~G., \&
  {Guhathakurta}, P. 2011, \apj, 727, 78

\bibitem[{{Kirby} {et~al.}(2008){Kirby}, {Simon}, {Geha}, {Guhathakurta}, \&
  {Frebel}}]{kirby08}
{Kirby}, E.~N., {Simon}, J.~D., {Geha}, M., {Guhathakurta}, P., \& {Frebel}, A.
  2008, \apjl, 685, L43

\bibitem[{{Kleyna} {et~al.}(2002){Kleyna}, {Wilkinson}, {Evans}, {Gilmore}, \&
  {Frayn}}]{kleyna02}
{Kleyna}, J., {Wilkinson}, M.~I., {Evans}, N.~W., {Gilmore}, G., \& {Frayn}, C.
  2002, \mnras, 330, 792

\bibitem[{{Kleyna} {et~al.}(1998){Kleyna}, {Geller}, {Kenyon}, {Kurtz}, \&
  {Thorstensen}}]{kleyna98}
{Kleyna}, J.~T., {Geller}, M.~J., {Kenyon}, S.~J., {Kurtz}, M.~J., \&
  {Thorstensen}, J.~R. 1998, \aj, 115, 2359

\bibitem[{{Kleyna} {et~al.}(2004){Kleyna}, {Wilkinson}, {Evans}, \&
  {Gilmore}}]{kleyna04}
{Kleyna}, J.~T., {Wilkinson}, M.~I., {Evans}, N.~W., \& {Gilmore}, G. 2004,
  \mnras, 354, L66

\bibitem[{{Kleyna} {et~al.}(2005){Kleyna}, {Wilkinson}, {Evans}, \&
  {Gilmore}}]{kleyna05}
---. 2005, \apjl, 630, L141

\bibitem[{{Kleyna} {et~al.}(2003){Kleyna}, {Wilkinson}, {Gilmore}, \&
  {Evans}}]{kleyna03}
{Kleyna}, J.~T., {Wilkinson}, M.~I., {Gilmore}, G., \& {Evans}, N.~W. 2003,
  \apjl, 588, L21

\bibitem[{{Klimentowski} {et~al.}(2007){Klimentowski}, {{\L}okas},
  {Kazantzidis}, {Prada}, {Mayer}, \& {Mamon}}]{klimentowski07}
{Klimentowski}, J., {{\L}okas}, E.~L., {Kazantzidis}, S., {Prada}, F., {Mayer},
  L., \& {Mamon}, G.~A. 2007, \mnras, 378, 353

\bibitem[{{Klypin} {et~al.}(2001){Klypin}, {Kravtsov}, {Bullock}, \&
  {Primack}}]{klypin01}
{Klypin}, A., {Kravtsov}, A.~V., {Bullock}, J.~S., \& {Primack}, J.~R. 2001,
  \apj, 554, 903

\bibitem[{{Klypin} {et~al.}(1999){Klypin}, {Kravtsov}, {Valenzuela}, \&
  {Prada}}]{klypin99}
{Klypin}, A., {Kravtsov}, A.~V., {Valenzuela}, O., \& {Prada}, F. 1999, \apj,
  522, 82

\bibitem[{{Koch} {et~al.}(2007{\natexlab{a}}){Koch}, {Kleyna}, {Wilkinson},
  {Grebel}, {Gilmore}, {Evans}, {Wyse}, \& {Harbeck}}]{koch07b}
{Koch}, A., {Kleyna}, J.~T., {Wilkinson}, M.~I., {Grebel}, E.~K., {Gilmore},
  G.~F., {Evans}, N.~W., {Wyse}, R.~F.~G., \& {Harbeck}, D.~R.
  2007{\natexlab{a}}, \aj, 134, 566

\bibitem[{{Koch} {et~al.}(2007{\natexlab{b}}){Koch}, {Wilkinson}, {Kleyna},
  {Gilmore}, {Grebel}, {Mackey}, {Evans}, \& {Wyse}}]{koch07}
{Koch}, A., {Wilkinson}, M.~I., {Kleyna}, J.~T., {Gilmore}, G.~F., {Grebel},
  E.~K., {Mackey}, A.~D., {Evans}, N.~W., \& {Wyse}, R.~F.~G.
  2007{\natexlab{b}}, \apj, 657, 241

\bibitem[{{Koch} {et~al.}(2009){Koch}, {Wilkinson}, {Kleyna}, {Irwin},
  {Zucker}, {Belokurov}, {Gilmore}, {Fellhauer}, \& {Evans}}]{koch09}
{Koch}, A., {Wilkinson}, M.~I., {Kleyna}, J.~T., {Irwin}, M., {Zucker}, D.~B.,
  {Belokurov}, V., {Gilmore}, G.~F., {Fellhauer}, M., \& {Evans}, N.~W. 2009,
  \apj, 690, 453

\bibitem[{{Koposov} {et~al.}(2011{\natexlab{a}}){Koposov}, {Belokurov},
  {Evans}, {Gilmore}, {Gieles}, {Irwin}, {Lewis}, {Niederste-Ostholt},
  {Pe{\~n}arrubia}, {Smith}, {Bizyaev}, {Malanushenko}, {Malanushenko},
  {Schneider}, \& {Wyse}}]{koposov11b}
{Koposov}, S.~E., {Belokurov}, V., {Evans}, N.~W., {Gilmore}, G., {Gieles}, M.,
  {Irwin}, M.~J., {Lewis}, G.~F., {Niederste-Ostholt}, M., {Pe{\~n}arrubia},
  J., {Smith}, M.~C., {Bizyaev}, D., {Malanushenko}, E., {Malanushenko}, V.,
  {Schneider}, D.~P., \& {Wyse}, R.~F.~G. 2011{\natexlab{a}}, ArXiv:1111.7042

\bibitem[{{Koposov} {et~al.}(2011{\natexlab{b}}){Koposov}, {Gilmore}, {Walker},
  {Belokurov}, {Wyn Evans}, {Fellhauer}, {Gieren}, {Geisler}, {Monaco},
  {Norris}, {Okamoto}, {Pe{\~n}arrubia}, {Wilkinson}, {Wyse}, \&
  {Zucker}}]{koposov11}
{Koposov}, S.~E., {Gilmore}, G., {Walker}, M.~G., {Belokurov}, V., {Wyn Evans},
  N., {Fellhauer}, M., {Gieren}, W., {Geisler}, D., {Monaco}, L., {Norris},
  J.~E., {Okamoto}, S., {Pe{\~n}arrubia}, J., {Wilkinson}, M., {Wyse},
  R.~F.~G., \& {Zucker}, D.~B. 2011{\natexlab{b}}, \apj, 736, 146

\bibitem[{{Koposov} {et~al.}(2009){Koposov}, {Yoo}, {Rix}, {Weinberg},
  {Macci{\`o}}, \& {Escud{\'e}}}]{koposov09}
{Koposov}, S.~E., {Yoo}, J., {Rix}, H.-W., {Weinberg}, D.~H., {Macci{\`o}},
  A.~V., \& {Escud{\'e}}, J.~M. 2009, \apj, 696, 2179

\bibitem[{{Koposov et al.}(2008)}]{koposov08}
{Koposov et al.} 2008, \apj, 686, 279

\bibitem[{{Kormendy}(1985)}]{kormendy85}
{Kormendy}, J. 1985, \apj, 295, 73

\bibitem[{{Kormendy} \& {Freeman}(2004)}]{kormendy04}
{Kormendy}, J., \& {Freeman}, K.~C. 2004, in IAU Symposium, Vol. 220, Dark
  Matter in Galaxies, ed. {S.~Ryder, D.~Pisano, M.~Walker, \& K.~Freeman},
  377--+

\bibitem[{{Kravtsov}(2010)}]{kravtsov10}
{Kravtsov}, A. 2010, Advances in Astronomy, 2010

\bibitem[{{Kroupa}(1997)}]{kroupa97}
{Kroupa}, P. 1997, New Astronomy, 2, 139

\bibitem[{{Kuhlen}(2010)}]{kuhlen10}
{Kuhlen}, M. 2010, Advances in Astronomy, 2010

\bibitem[{{Kuhn}(1993)}]{kuhn93}
{Kuhn}, J.~R. 1993, \apjl, 409, L13

\bibitem[{{Kuhn} \& {Miller}(1989)}]{kuhn89}
{Kuhn}, J.~R., \& {Miller}, R.~H. 1989, \apjl, 341, L41

\bibitem[{{Kusenko}(2006)}]{kusenko06}
{Kusenko}, A. 2006, Physical Review Letters, 97, 241301

\bibitem[{{Kusenko}(2009)}]{kusenko09}
---. 2009, \physrep, 481, 1

\bibitem[{{Kuzio de Naray} {et~al.}(2008){Kuzio de Naray}, {McGaugh}, \& {de
  Blok}}]{kuzio08}
{Kuzio de Naray}, R., {McGaugh}, S.~S., \& {de Blok}, W.~J.~G. 2008, \apj, 676,
  920

\bibitem[{{Kuzio de Naray} {et~al.}(2006){Kuzio de Naray}, {McGaugh}, {de
  Blok}, \& {Bosma}}]{kuzio06}
{Kuzio de Naray}, R., {McGaugh}, S.~S., {de Blok}, W.~J.~G., \& {Bosma}, A.
  2006, \apjs, 165, 461

\bibitem[{{Lake}(1989)}]{lake89}
{Lake}, G. 1989, \aj, 98, 1253

\bibitem[{{Law} \& {Majewski}(2010)}]{law10}
{Law}, D.~R., \& {Majewski}, S.~R. 2010, \apj, 714, 229

\bibitem[{{Lee} {et~al.}(2003){Lee}, {Park}, {Park}, {Sohn}, {Oh}, {Yuk},
  {Rey}, {Lee}, {Lee}, {Kim}, {Han}, {Park}, {Lee}, {Jeon}, \& {Kim}}]{lee03}
{Lee}, M.~G., {Park}, H.~S., {Park}, J.-H., {Sohn}, Y.-J., {Oh}, S.~J., {Yuk},
  I.-S., {Rey}, S.-C., {Lee}, S.-G., {Lee}, Y.-W., {Kim}, H.-I., {Han}, W.,
  {Park}, W.-K., {Lee}, J.~H., {Jeon}, Y.-B., \& {Kim}, S.~C. 2003, \aj, 126,
  2840

\bibitem[{{Li} {et~al.}(2010){Li}, {De Lucia}, \& {Helmi}}]{li10}
{Li}, Y.-S., {De Lucia}, G., \& {Helmi}, A. 2010, \mnras, 401, 2036

\bibitem[{{Lin} \& {Faber}(1983)}]{lin83}
{Lin}, D.~N.~C., \& {Faber}, S.~M. 1983, \apjl, 266, L21

\bibitem[{{Loeb} \& {Weiner}(2011)}]{loeb11}
{Loeb}, A., \& {Weiner}, N. 2011, Physical Review Letters, 106, 171302

\bibitem[{{Loewenstein} \& {Kusenko}(2010)}]{loewenstein10}
{Loewenstein}, M., \& {Kusenko}, A. 2010, \apj, 714, 652

\bibitem[{{Loewenstein} \& {Kusenko}(2012)}]{loewenstein12}
---. 2012, ArXiv:1203.5229

\bibitem[{{Loewenstein} {et~al.}(2009){Loewenstein}, {Kusenko}, \&
  {Biermann}}]{loewenstein09}
{Loewenstein}, M., {Kusenko}, A., \& {Biermann}, P.~L. 2009, \apj, 700, 426

\bibitem[{{{\L}okas}(2009)}]{lokas09}
{{\L}okas}, E.~L. 2009, \mnras, 394, L102

\bibitem[{{{\L}okas} {et~al.}(2005){{\L}okas}, {Mamon}, \& {Prada}}]{lokas05}
{{\L}okas}, E.~L., {Mamon}, G.~A., \& {Prada}, F. 2005, \mnras, 363, 918

\bibitem[{{Lovell} {et~al.}(2012){Lovell}, {Eke}, {Frenk}, {Gao}, {Jenkins},
  {Theuns}, {Wang}, {White}, {Boyarsky}, \& {Ruchayskiy}}]{lovell12}
{Lovell}, M.~R., {Eke}, V., {Frenk}, C.~S., {Gao}, L., {Jenkins}, A., {Theuns},
  T., {Wang}, J., {White}, S.~D.~M., {Boyarsky}, A., \& {Ruchayskiy}, O. 2012,
  \mnras, 420, 2318

\bibitem[{{Lynden-Bell}(1967)}]{lyndenbell67}
{Lynden-Bell}, D. 1967, \mnras, 136, 101

\bibitem[{{Macci{\`o}} \& {Fontanot}(2010)}]{maccio10b}
{Macci{\`o}}, A.~V., \& {Fontanot}, F. 2010, \mnras, 404, L16

\bibitem[{{Macci{\`o}} {et~al.}(2010){Macci{\`o}}, {Kang}, {Fontanot},
  {Somerville}, {Koposov}, \& {Monaco}}]{maccio10}
{Macci{\`o}}, A.~V., {Kang}, X., {Fontanot}, F., {Somerville}, R.~S.,
  {Koposov}, S., \& {Monaco}, P. 2010, \mnras, 402, 1995

\bibitem[{{Macci{\`o}} {et~al.}(2012{\natexlab{a}}){Macci{\`o}}, {Paduroiu},
  {Anderhalden}, {Schneider}, \& {Moore}}]{maccio12}
{Macci{\`o}}, A.~V., {Paduroiu}, S., {Anderhalden}, D., {Schneider}, A., \&
  {Moore}, B. 2012{\natexlab{a}}, ArXiv:1202.1282

\bibitem[{{Macci{\`o}} {et~al.}(2012{\natexlab{b}}){Macci{\`o}}, {Ruchayskiy},
  {Boyarsky}, \& {Munoz-Cuartas}}]{maccio12b}
{Macci{\`o}}, A.~V., {Ruchayskiy}, O., {Boyarsky}, A., \& {Munoz-Cuartas},
  J.~C. 2012{\natexlab{b}}, ArXiv:1202.2858

\bibitem[{{Majewski} {et~al.}(2007){Majewski}, {Beaton}, {Patterson},
  {Kalirai}, {Geha}, {Mu{\~n}oz}, {Seigar}, {Guhathakurta}, {Gilbert}, {Rich},
  {Bullock}, \& {Reitzel}}]{majewski07}
{Majewski}, S.~R., {Beaton}, R.~L., {Patterson}, R.~J., {Kalirai}, J.~S.,
  {Geha}, M.~C., {Mu{\~n}oz}, R.~R., {Seigar}, M.~S., {Guhathakurta}, P.,
  {Gilbert}, K.~M., {Rich}, R.~M., {Bullock}, J.~S., \& {Reitzel}, D.~B. 2007,
  \apjl, 670, L9

\bibitem[{{Majewski} {et~al.}(2005){Majewski}, {Frinchaboy}, {Kunkel}, {Link},
  {Mu{\~n}oz}, {Ostheimer}, {Palma}, {Patterson}, \& {Geisler}}]{majewski05}
{Majewski}, S.~R., {Frinchaboy}, P.~M., {Kunkel}, W.~E., {Link}, R.,
  {Mu{\~n}oz}, R.~R., {Ostheimer}, J.~C., {Palma}, C., {Patterson}, R.~J., \&
  {Geisler}, D. 2005, \aj, 130, 2677

\bibitem[{{Majewski} {et~al.}(2000){Majewski}, {Ostheimer}, {Kunkel}, \&
  {Patterson}}]{majewski00}
{Majewski}, S.~R., {Ostheimer}, J.~C., {Kunkel}, W.~E., \& {Patterson}, R.~J.
  2000, \aj, 120, 2550

\bibitem[{{Majewski} {et~al.}(2003){Majewski}, {Skrutskie}, {Weinberg}, \&
  {Ostheimer}}]{majewski03}
{Majewski}, S.~R., {Skrutskie}, M.~F., {Weinberg}, M.~D., \& {Ostheimer}, J.~C.
  2003, \apj, 599, 1082

\bibitem[{{Mamon} \& {{\L}okas}(2005)}]{mamon05}
{Mamon}, G.~A., \& {{\L}okas}, E.~L. 2005, \mnras, 363, 705

\bibitem[{{Martin} {et~al.}(2008){Martin}, {de Jong}, \& {Rix}}]{martin08}
{Martin}, N.~F., {de Jong}, J.~T.~A., \& {Rix}, H.-W. 2008, \apj, 684, 1075

\bibitem[{{Martin} {et~al.}(2007){Martin}, {Ibata}, {Chapman}, {Irwin}, \&
  {Lewis}}]{martin07}
{Martin}, N.~F., {Ibata}, R.~A., {Chapman}, S.~C., {Irwin}, M., \& {Lewis},
  G.~F. 2007, \mnras, 380, 281

\bibitem[{{Martin} {et~al.}(2006){Martin}, {Irwin}, {Ibata}, {Conn}, {Lewis},
  {Bellazzini}, {Chapman}, \& {Tanvir}}]{martin06}
{Martin}, N.~F., {Irwin}, M.~J., {Ibata}, R.~A., {Conn}, B.~C., {Lewis}, G.~F.,
  {Bellazzini}, M., {Chapman}, S., \& {Tanvir}, N. 2006, \mnras, 367, L69

\bibitem[{{Martin} \& {Jin}(2010)}]{martin10}
{Martin}, N.~F., \& {Jin}, S. 2010, \apj, 721, 1333

\bibitem[{{Martin} {et~al.}(2009){Martin}, {McConnachie}, {Irwin}, {Widrow},
  {Ferguson}, {Ibata}, {Dubinski}, {Babul}, {Chapman}, {Fardal}, {Lewis},
  {Navarro}, \& {Rich}}]{martin09}
{Martin}, N.~F., {McConnachie}, A.~W., {Irwin}, M., {Widrow}, L.~M.,
  {Ferguson}, A.~M.~N., {Ibata}, R.~A., {Dubinski}, J., {Babul}, A., {Chapman},
  S., {Fardal}, M., {Lewis}, G.~F., {Navarro}, J., \& {Rich}, R.~M. 2009, \apj,
  705, 758

\bibitem[{{Martinez} {et~al.}(2009){Martinez}, {Bullock}, {Kaplinghat},
  {Strigari}, \& {Trotta}}]{martinez09}
{Martinez}, G.~D., {Bullock}, J.~S., {Kaplinghat}, M., {Strigari}, L.~E., \&
  {Trotta}, R. 2009, JCAP, 6, 14

\bibitem[{{Martinez} {et~al.}(2011){Martinez}, {Minor}, {Bullock},
  {Kaplinghat}, {Simon}, \& {Geha}}]{martinez11}
{Martinez}, G.~D., {Minor}, Q.~E., {Bullock}, J., {Kaplinghat}, M., {Simon},
  J.~D., \& {Geha}, M. 2011, \apj, 738, 55

\bibitem[{{Mashchenko} {et~al.}(2006){Mashchenko}, {Couchman}, \&
  {Wadsley}}]{mashchenko06}
{Mashchenko}, S., {Couchman}, H.~M.~P., \& {Wadsley}, J. 2006, \nat, 442, 539

\bibitem[{{Mashchenko} {et~al.}(2008){Mashchenko}, {Wadsley}, \&
  {Couchman}}]{mashchenko08}
{Mashchenko}, S., {Wadsley}, J., \& {Couchman}, H.~M.~P. 2008, Science, 319,
  174

\bibitem[{{Mateo} {et~al.}(1996){Mateo}, {Mirabal}, {Udalski}, {Szymanski},
  {Kaluzny}, {Kubiak}, {Krzeminski}, \& {Stanek}}]{mateo96}
{Mateo}, M., {Mirabal}, N., {Udalski}, A., {Szymanski}, M., {Kaluzny}, J.,
  {Kubiak}, M., {Krzeminski}, W., \& {Stanek}, K.~Z. 1996, \apjl, 458, L13+

\bibitem[{{Mateo} {et~al.}(1991){Mateo}, {Olszewski}, {Welch}, {Fischer}, \&
  {Kunkel}}]{mateo91}
{Mateo}, M., {Olszewski}, E., {Welch}, D.~L., {Fischer}, P., \& {Kunkel}, W.
  1991, \aj, 102, 914

\bibitem[{{Mateo} {et~al.}(1993){Mateo}, {Olszewski}, {Pryor}, {Welch}, \&
  {Fischer}}]{mateo93}
{Mateo}, M., {Olszewski}, E.~W., {Pryor}, C., {Welch}, D.~L., \& {Fischer}, P.
  1993, \aj, 105, 510

\bibitem[{{Mateo} {et~al.}(1998){Mateo}, {Olszewski}, {Vogt}, \&
  {Keane}}]{mateo98b}
{Mateo}, M., {Olszewski}, E.~W., {Vogt}, S.~S., \& {Keane}, M.~J. 1998, \aj,
  116, 2315

\bibitem[{{Mateo} {et~al.}(2008){Mateo}, {Olszewski}, \& {Walker}}]{mateo08}
{Mateo}, M., {Olszewski}, E.~W., \& {Walker}, M.~G. 2008, \apj, 675, 201

\bibitem[{{Mateo}(1998)}]{mateo98}
{Mateo}, M.~L. 1998, \araa, 36, 435

\bibitem[{{Mayer} {et~al.}(2001{\natexlab{a}}){Mayer}, {Governato}, {Colpi},
  {Moore}, {Quinn}, {Wadsley}, {Stadel}, \& {Lake}}]{mayer01a}
{Mayer}, L., {Governato}, F., {Colpi}, M., {Moore}, B., {Quinn}, T., {Wadsley},
  J., {Stadel}, J., \& {Lake}, G. 2001{\natexlab{a}}, \apj, 559, 754

\bibitem[{{Mayer} {et~al.}(2001{\natexlab{b}}){Mayer}, {Governato}, {Colpi},
  {Moore}, {Quinn}, {Wadsley}, {Stadel}, \& {Lake}}]{mayer01b}
---. 2001{\natexlab{b}}, \apjl, 547, L123

\bibitem[{{McConnachie} \& {C{\^o}t{\'e}}(2010)}]{mcconnachie10}
{McConnachie}, A.~W., \& {C{\^o}t{\'e}}, P. 2010, \apjl, 722, L209

\bibitem[{{McConnachie} {et~al.}(2008){McConnachie}, {Huxor}, {Martin},
  {Irwin}, {Chapman}, {Fahlman}, {Ferguson}, {Ibata}, {Lewis}, {Richer}, \&
  {Tanvir}}]{mcconnachie08}
{McConnachie}, A.~W., {Huxor}, A., {Martin}, N.~F., {Irwin}, M.~J., {Chapman},
  S.~C., {Fahlman}, G., {Ferguson}, A.~M.~N., {Ibata}, R.~A., {Lewis}, G.~F.,
  {Richer}, H., \& {Tanvir}, N.~R. 2008, \apj, 688, 1009

\bibitem[{{McConnachie} {et~al.}(2009){McConnachie}, {Irwin}, {Ibata},
  {Dubinski}, {Widrow}, {Martin}, {C{\^o}t{\'e}}, {Dotter}, {Navarro},
  {Ferguson}, {Puzia}, {Lewis}, {Babul}, {Barmby}, {Bienaym{\'e}}, {Chapman},
  {Cockcroft}, {Collins}, {Fardal}, {Harris}, {Huxor}, {Mackey},
  {Pe{\~n}arrubia}, {Rich}, {Richer}, {Siebert}, {Tanvir}, {Valls-Gabaud}, \&
  {Venn}}]{mcconnachie09}
{McConnachie}, A.~W., {Irwin}, M.~J., {Ibata}, R.~A., {Dubinski}, J., {Widrow},
  L.~M., {Martin}, N.~F., {C{\^o}t{\'e}}, P., {Dotter}, A.~L., {Navarro},
  J.~F., {Ferguson}, A.~M.~N., {Puzia}, T.~H., {Lewis}, G.~F., {Babul}, A.,
  {Barmby}, P., {Bienaym{\'e}}, O., {Chapman}, S.~C., {Cockcroft}, R.,
  {Collins}, M.~L.~M., {Fardal}, M.~A., {Harris}, W.~E., {Huxor}, A., {Mackey},
  A.~D., {Pe{\~n}arrubia}, J., {Rich}, R.~M., {Richer}, H.~B., {Siebert}, A.,
  {Tanvir}, N., {Valls-Gabaud}, D., \& {Venn}, K.~A. 2009, \nat, 461, 66

\bibitem[{{McGaugh} {et~al.}(2007){McGaugh}, {de Blok}, {Schombert}, {Kuzio de
  Naray}, \& {Kim}}]{mcgaugh07}
{McGaugh}, S.~S., {de Blok}, W.~J.~G., {Schombert}, J.~M., {Kuzio de Naray},
  R., \& {Kim}, J.~H. 2007, \apj, 659, 149

\bibitem[{{McGaugh} {et~al.}(2001){McGaugh}, {Rubin}, \& {de Blok}}]{mcgaugh01}
{McGaugh}, S.~S., {Rubin}, V.~C., \& {de Blok}, W.~J.~G. 2001, \aj, 122, 2381

\bibitem[{{McGaugh} {et~al.}(2000){McGaugh}, {Schombert}, {Bothun}, \& {de
  Blok}}]{mcgaugh00}
{McGaugh}, S.~S., {Schombert}, J.~M., {Bothun}, G.~D., \& {de Blok}, W.~J.~G.
  2000, \apjl, 533, L99

\bibitem[{{McGaugh} \& {Wolf}(2010)}]{mcgaugh10}
{McGaugh}, S.~S., \& {Wolf}, J. 2010, \apj, 722, 248

\bibitem[{{Metz} \& {Kroupa}(2007)}]{metz07}
{Metz}, M., \& {Kroupa}, P. 2007, \mnras, 376, 387

\bibitem[{{Michie}(1963)}]{michie63}
{Michie}, R.~W. 1963, \mnras, 125, 127

\bibitem[{{Milgrom}(1983)}]{milgrom83}
{Milgrom}, M. 1983, \apj, 270, 365

\bibitem[{{Minor} {et~al.}(2010){Minor}, {Martinez}, {Bullock}, {Kaplinghat},
  \& {Trainor}}]{minor10}
{Minor}, Q.~E., {Martinez}, G., {Bullock}, J., {Kaplinghat}, M., \& {Trainor},
  R. 2010, \apj, 721, 1142

\bibitem[{{Moffat}(2006)}]{moffat06}
{Moffat}, J.~W. 2006, JCAP, 3, 4

\bibitem[{{Moore}(1994)}]{moore94}
{Moore}, B. 1994, \nat, 370, 629

\bibitem[{{Moore} {et~al.}(1998){Moore}, {Governato}, {Quinn}, {Stadel}, \&
  {Lake}}]{moore98}
{Moore}, B., {Governato}, F., {Quinn}, T., {Stadel}, J., \& {Lake}, G. 1998,
  \apjl, 499, L5+

\bibitem[{{Mu{\~n}oz} {et~al.}(2006){Mu{\~n}oz}, {Carlin}, {Frinchaboy},
  {Nidever}, {Majewski}, \& {Patterson}}]{munoz06b}
{Mu{\~n}oz}, R.~R., {Carlin}, J.~L., {Frinchaboy}, P.~M., {Nidever}, D.~L.,
  {Majewski}, S.~R., \& {Patterson}, R.~J. 2006, \apjl, 650, L51

\bibitem[{{Mu{\~n}oz} {et~al.}(2010){Mu{\~n}oz}, {Geha}, \&
  {Willman}}]{munoz10}
{Mu{\~n}oz}, R.~R., {Geha}, M., \& {Willman}, B. 2010, \aj, 140, 138

\bibitem[{{Mu{\~n}oz} {et~al.}(2008){Mu{\~n}oz}, {Majewski}, \&
  {Johnston}}]{munoz08}
{Mu{\~n}oz}, R.~R., {Majewski}, S.~R., \& {Johnston}, K.~V. 2008, \apj, 679,
  346

\bibitem[{{Mu\~noz} {et~al.}(2011){Mu\~noz}, {Padmanabhan}, \&
  {Geha}}]{munoz11}
{Mu\~noz}, R.~R., {Padmanabhan}, N., \& {Geha}, M. 2011, ArXiv:1110.1086

\bibitem[{{Mu{\~n}oz et al.}(2005)}]{munoz05}
{Mu{\~n}oz et al.} 2005, \apjl, 631, L137

\bibitem[{{Mu{\~n}oz et al.}(2006)}]{munoz06}
---. 2006, \apj, 649, 201

\bibitem[{{Navarro} {et~al.}(1996){Navarro}, {Eke}, \& {Frenk}}]{navarro96b}
{Navarro}, J.~F., {Eke}, V.~R., \& {Frenk}, C.~S. 1996, \mnras, 283, L72

\bibitem[{{Navarro, Frenk \& White}(1996)}]{navarro96}
{Navarro, Frenk \& White}. 1996, \apj, 462, 563

\bibitem[{{Navarro, Frenk \& White}(1997)}]{navarro97}
---. 1997, \apj, 490, 493

\bibitem[{{Niederste-Ostholt} {et~al.}(2009){Niederste-Ostholt}, {Belokurov},
  {Evans}, {Gilmore}, {Wyse}, \& {Norris}}]{niederste-ostholt09}
{Niederste-Ostholt}, M., {Belokurov}, V., {Evans}, N.~W., {Gilmore}, G.,
  {Wyse}, R.~F.~G., \& {Norris}, J.~E. 2009, \mnras, 398, 1771

\bibitem[{{Norris} {et~al.}(2010){Norris}, {Wyse}, {Gilmore}, {Yong}, {Frebel},
  {Wilkinson}, {Belokurov}, \& {Zucker}}]{norris10}
{Norris}, J.~E., {Wyse}, R.~F.~G., {Gilmore}, G., {Yong}, D., {Frebel}, A.,
  {Wilkinson}, M.~I., {Belokurov}, V., \& {Zucker}, D.~B. 2010, \apj, 723, 1632

\bibitem[{{Odenkirchen} {et~al.}(2001){Odenkirchen}, {Grebel}, {Harbeck},
  {Dehnen}, {Rix}, {Newberg}, {Yanny}, {Holtzman}, {Brinkmann}, {Chen},
  {Csabai}, {Hayes}, {Hennessy}, {Hindsley}, {Ivezi{\'c}}, {Kinney},
  {Kleinman}, {Long}, {Lupton}, {Neilsen}, {Nitta}, {Snedden}, \&
  {York}}]{odenkirchen01}
{Odenkirchen}, M., {Grebel}, E.~K., {Harbeck}, D., {Dehnen}, W., {Rix}, H.-W.,
  {Newberg}, H.~J., {Yanny}, B., {Holtzman}, J., {Brinkmann}, J., {Chen}, B.,
  {Csabai}, I., {Hayes}, J.~J.~E., {Hennessy}, G., {Hindsley}, R.~B.,
  {Ivezi{\'c}}, {\v Z}., {Kinney}, E.~K., {Kleinman}, S.~J., {Long}, D.,
  {Lupton}, R.~H., {Neilsen}, E.~H., {Nitta}, A., {Snedden}, S.~A., \& {York},
  D.~G. 2001, \aj, 122, 2538

\bibitem[{{Oh, Lin \& Aarseth}(1995)}]{oh95}
{Oh, Lin \& Aarseth}. 1995, \apj, 442, 142

\bibitem[{{Okamoto} {et~al.}(2012){Okamoto}, {Arimoto}, {Yamada}, \&
  {Onodera}}]{okamoto12}
{Okamoto}, S., {Arimoto}, N., {Yamada}, Y., \& {Onodera}, M. 2012, \apj, 744,
  96

\bibitem[{{Olszewski} \& {Aaronson}(1985)}]{olszewski85}
{Olszewski}, E.~W., \& {Aaronson}, M. 1985, \aj, 90, 2221

\bibitem[{{Olszewski} {et~al.}(1995){Olszewski}, {Aaronson}, \&
  {Hill}}]{olszewski95}
{Olszewski}, E.~W., {Aaronson}, M., \& {Hill}, J.~M. 1995, \aj, 110, 2120

\bibitem[{{Olszewski} {et~al.}(1996){Olszewski}, {Pryor}, \&
  {Armandroff}}]{edo96}
{Olszewski}, E.~W., {Pryor}, C., \& {Armandroff}, T.~E. 1996, \aj, 111, 750

\bibitem[{{Ostriker} {et~al.}(1974){Ostriker}, {Peebles}, \&
  {Yahil}}]{ostriker74}
{Ostriker}, J.~P., {Peebles}, P.~J.~E., \& {Yahil}, A. 1974, \apjl, 193, L1

\bibitem[{{Pal} \& {Wolfenstein}(1982)}]{pal82}
{Pal}, P.~B., \& {Wolfenstein}, L. 1982, \prd, 25, 766

\bibitem[{{Palma} {et~al.}(2003){Palma}, {Majewski}, {Siegel}, {Patterson},
  {Ostheimer}, \& {Link}}]{palma03}
{Palma}, C., {Majewski}, S.~R., {Siegel}, M.~H., {Patterson}, R.~J.,
  {Ostheimer}, J.~C., \& {Link}, R. 2003, \aj, 125, 1352

\bibitem[{{Parry} {et~al.}(2011){Parry}, {Eke}, {Frenk}, \&
  {Okamoto}}]{parry11}
{Parry}, O.~H., {Eke}, V.~R., {Frenk}, C.~S., \& {Okamoto}, T. 2011,
  ArXiv:1105.3474

\bibitem[{{Pe{\~n}arrubia} {et~al.}(2010){Pe{\~n}arrubia}, {Belokurov},
  {Evans}, {Mart{\'{\i}}nez-Delgado}, {Gilmore}, {Irwin}, {Niederste-Ostholt},
  \& {Zucker}}]{penarrubia10b}
{Pe{\~n}arrubia}, J., {Belokurov}, V., {Evans}, N.~W.,
  {Mart{\'{\i}}nez-Delgado}, D., {Gilmore}, G., {Irwin}, M.,
  {Niederste-Ostholt}, M., \& {Zucker}, D.~B. 2010, \mnras, 408, L26

\bibitem[{{Pe{\~n}arrubia} {et~al.}(2008){Pe{\~n}arrubia}, {Navarro}, \&
  {McConnachie}}]{penarrubia08b}
{Pe{\~n}arrubia}, J., {Navarro}, J.~F., \& {McConnachie}, A.~W. 2008, \apj,
  673, 226

\bibitem[{{Pe{\~n}arrubia} {et~al.}(2009){Pe{\~n}arrubia}, {Navarro},
  {McConnachie}, \& {Martin}}]{penarrubia09}
{Pe{\~n}arrubia}, J., {Navarro}, J.~F., {McConnachie}, A.~W., \& {Martin},
  N.~F. 2009, \apj, 698, 222

\bibitem[{{Piatek} \& {Pryor}(1995)}]{pp95}
{Piatek}, S., \& {Pryor}, C. 1995, \aj, 109, 1071

\bibitem[{{Piatek} {et~al.}(2006){Piatek}, {Pryor}, {Bristow}, {Olszewski},
  {Harris}, {Mateo}, {Minniti}, \& {Tinney}}]{piatek06}
{Piatek}, S., {Pryor}, C., {Bristow}, P., {Olszewski}, E.~W., {Harris}, H.~C.,
  {Mateo}, M., {Minniti}, D., \& {Tinney}, C.~G. 2006, \aj, 131, 1445

\bibitem[{{Piatek} {et~al.}(2007){Piatek}, {Pryor}, {Bristow}, {Olszewski},
  {Harris}, {Mateo}, {Minniti}, \& {Tinney}}]{piatek07}
---. 2007, \aj, 133, 818

\bibitem[{{Piatek} {et~al.}(2002){Piatek}, {Pryor}, {Olszewski}, {Harris},
  {Mateo}, {Minniti}, {Monet}, {Morrison}, \& {Tinney}}]{piatek02}
{Piatek}, S., {Pryor}, C., {Olszewski}, E.~W., {Harris}, H.~C., {Mateo}, M.,
  {Minniti}, D., {Monet}, D.~G., {Morrison}, H., \& {Tinney}, C.~G. 2002, \aj,
  124, 3198

\bibitem[{{Piatek} {et~al.}(2003){Piatek}, {Pryor}, {Olszewski}, {Harris},
  {Mateo}, {Minniti}, \& {Tinney}}]{piatek03}
{Piatek}, S., {Pryor}, C., {Olszewski}, E.~W., {Harris}, H.~C., {Mateo}, M.,
  {Minniti}, D., \& {Tinney}, C.~G. 2003, \aj, 126, 2346

\bibitem[{{Pieri} {et~al.}(2009){Pieri}, {Lattanzi}, \& {Silk}}]{pieri09}
{Pieri}, L., {Lattanzi}, M., \& {Silk}, J. 2009, \mnras, 399, 2033

\bibitem[{{Plummer}(1911)}]{plummer11}
{Plummer}, H.~C. 1911, \mnras, 71, 460

\bibitem[{{Polisensky} \& {Ricotti}(2011)}]{polisensky11}
{Polisensky}, E., \& {Ricotti}, M. 2011, \prd, 83, 043506

\bibitem[{{Pontzen} \& {Governato}(2011)}]{pontzen11}
{Pontzen}, A., \& {Governato}, F. 2011, ArXiv:1106.0499

\bibitem[{{Pryor} \& {Kormendy}(1990)}]{pryor90}
{Pryor}, C., \& {Kormendy}, J. 1990, \aj, 100, 127

\bibitem[{{Queloz} {et~al.}(1995){Queloz}, {Dubath}, \& {Pasquini}}]{queloz95}
{Queloz}, D., {Dubath}, P., \& {Pasquini}, L. 1995, \aap, 300, 31

\bibitem[{{Read} \& {Gilmore}(2005)}]{read05}
{Read}, J.~I., \& {Gilmore}, G. 2005, \mnras, 356, 107

\bibitem[{{Read} {et~al.}(2006{\natexlab{a}}){Read}, {Wilkinson}, {Evans},
  {Gilmore}, \& {Kleyna}}]{read06b}
{Read}, J.~I., {Wilkinson}, M.~I., {Evans}, N.~W., {Gilmore}, G., \& {Kleyna},
  J.~T. 2006{\natexlab{a}}, \mnras, 367, 387

\bibitem[{{Read} {et~al.}(2006{\natexlab{b}}){Read}, {Wilkinson}, {Evans},
  {Gilmore}, \& {Kleyna}}]{read06}
---. 2006{\natexlab{b}}, \mnras, 366, 429

\bibitem[{{Richardson} {et~al.}(2011){Richardson}, {Irwin}, {McConnachie},
  {Martin}, {Dotter}, {Ferguson}, {Ibata}, {Chapman}, {Lewis}, {Tanvir}, \&
  {Rich}}]{richardson11}
{Richardson}, J.~C., {Irwin}, M.~J., {McConnachie}, A.~W., {Martin}, N.~F.,
  {Dotter}, A.~L., {Ferguson}, A.~M.~N., {Ibata}, R.~A., {Chapman}, S.~C.,
  {Lewis}, G.~F., {Tanvir}, N.~R., \& {Rich}, R.~M. 2011, \apj, 732, 76

\bibitem[{{Richstone} \& {Tremaine}(1986)}]{richstone86}
{Richstone}, D.~O., \& {Tremaine}, S. 1986, \aj, 92, 72

\bibitem[{{Riemer-S{\o}rensen} \& {Hansen}(2009)}]{riemer09}
{Riemer-S{\o}rensen}, S., \& {Hansen}, S.~H. 2009, \aap, 500, L37

\bibitem[{{Romano-D{\'{\i}}az} {et~al.}(2009){Romano-D{\'{\i}}az}, {Shlosman},
  {Heller}, \& {Hoffman}}]{romanodiaz09}
{Romano-D{\'{\i}}az}, E., {Shlosman}, I., {Heller}, C., \& {Hoffman}, Y. 2009,
  \apj, 702, 1250

\bibitem[{{Salucci} \& {Burkert}(2000)}]{salucci00}
{Salucci}, P., \& {Burkert}, A. 2000, \apjl, 537, L9

\bibitem[{{Salucci} {et~al.}(2011){Salucci}, {Wilkinson}, {Walker}, {Gilmore},
  {Grebel}, {Koch}, {Frigerio Martins}, \& {Wyse}}]{salucci11}
{Salucci}, P., {Wilkinson}, M.~I., {Walker}, M.~G., {Gilmore}, G.~F., {Grebel},
  E.~K., {Koch}, A., {Frigerio Martins}, C., \& {Wyse}, R.~F.~G. 2011,
  ArXiv:1111.1165

\bibitem[{{S{\'a}nchez-Salcedo} {et~al.}(2006){S{\'a}nchez-Salcedo},
  {Reyes-Iturbide}, \& {Hernandez}}]{sanchez06}
{S{\'a}nchez-Salcedo}, F.~J., {Reyes-Iturbide}, J., \& {Hernandez}, X. 2006,
  \mnras, 370, 1829

\bibitem[{{Sand} {et~al.}(2010){Sand}, {Seth}, {Olszewski}, {Willman},
  {Zaritsky}, \& {Kallivayalil}}]{sand10}
{Sand}, D.~J., {Seth}, A., {Olszewski}, E.~W., {Willman}, B., {Zaritsky}, D.,
  \& {Kallivayalil}, N. 2010, \apj, 718, 530

\bibitem[{{Sand} {et~al.}(2011){Sand}, {Strader}, {Willman}, {Zaritsky},
  {McLeod}, {Caldwell}, {Seth}, \& {Olszewski}}]{sand11}
{Sand}, D.~J., {Strader}, J., {Willman}, B., {Zaritsky}, D., {McLeod}, B.,
  {Caldwell}, N., {Seth}, A., \& {Olszewski}, E. 2011, ArXiv:1111.6608

\bibitem[{{Saviane} {et~al.}(2000){Saviane}, {Held}, \& {Bertelli}}]{saviane00}
{Saviane}, I., {Held}, E.~V., \& {Bertelli}, G. 2000, \aap, 355, 56

\bibitem[{{Sawala} {et~al.}(2010){Sawala}, {Scannapieco}, {Maio}, \&
  {White}}]{sawala10}
{Sawala}, T., {Scannapieco}, C., {Maio}, U., \& {White}, S. 2010, \mnras, 402,
  1599

\bibitem[{{Schwarzschild}(1979)}]{schwarzschild79}
{Schwarzschild}, M. 1979, \apj, 232, 236

\bibitem[{{Schweitzer} {et~al.}(1995){Schweitzer}, {Cudworth}, {Majewski}, \&
  {Suntzeff}}]{schweitzer95}
{Schweitzer}, A.~E., {Cudworth}, K.~M., {Majewski}, S.~R., \& {Suntzeff}, N.~B.
  1995, \aj, 110, 2747

\bibitem[{{Scott} {et~al.}(2010){Scott}, {Conrad}, {Edsj{\"o}},
  {Bergstr{\"o}m}, {Farnier}, \& {Akrami}}]{scott10}
{Scott}, P., {Conrad}, J., {Edsj{\"o}}, J., {Bergstr{\"o}m}, L., {Farnier}, C.,
  \& {Akrami}, Y. 2010, JCAP, 1, 31

\bibitem[{{Seitzer} \& {Frogel}(1985)}]{seitzer85}
{Seitzer}, P., \& {Frogel}, J.~A. 1985, \aj, 90, 1796

\bibitem[{{Shapley}(1938{\natexlab{a}})}]{shapley38}
{Shapley}, H. 1938{\natexlab{a}}, Harvard College Observatory Bulletin, 908, 1

\bibitem[{{Shapley}(1938{\natexlab{b}})}]{shapley38b}
---. 1938{\natexlab{b}}, \nat, 142, 715

\bibitem[{{Simon} {et~al.}(2005){Simon}, {Bolatto}, {Leroy}, {Blitz}, \&
  {Gates}}]{simon05}
{Simon}, J.~D., {Bolatto}, A.~D., {Leroy}, A., {Blitz}, L., \& {Gates}, E.~L.
  2005, \apj, 621, 757

\bibitem[{{Simon} \& {Geha}(2007)}]{simon07}
{Simon}, J.~D., \& {Geha}, M. 2007, \apj, 670, 313

\bibitem[{{Simon} {et~al.}(2011){Simon}, {Geha}, {Minor}, {Martinez}, {Kirby},
  {Bullock}, {Kaplinghat}, {Strigari}, {Willman}, {Choi}, {Tollerud}, \&
  {Wolf}}]{simon11}
{Simon}, J.~D., {Geha}, M., {Minor}, Q.~E., {Martinez}, G.~D., {Kirby}, E.~N.,
  {Bullock}, J.~S., {Kaplinghat}, M., {Strigari}, L.~E., {Willman}, B., {Choi},
  P.~I., {Tollerud}, E.~J., \& {Wolf}, J. 2011, \apj, 733, 46

\bibitem[{{Slater} {et~al.}(2011){Slater}, {Bell}, \& {Martin}}]{slater11}
{Slater}, C.~T., {Bell}, E.~F., \& {Martin}, N.~F. 2011, \apjl, 742, L14

\bibitem[{{Sohn} {et~al.}(2007){Sohn}, {Majewski}, {Mu{\~n}oz}, {Kunkel},
  {Johnston}, {Ostheimer}, {Guhathakurta}, {Patterson}, {Siegel}, \&
  {Cooper}}]{sohn07}
{Sohn}, S.~T., {Majewski}, S.~R., {Mu{\~n}oz}, R.~R., {Kunkel}, W.~E.,
  {Johnston}, K.~V., {Ostheimer}, J.~C., {Guhathakurta}, P., {Patterson},
  R.~J., {Siegel}, M.~H., \& {Cooper}, M.~C. 2007, \apj, 663, 960

\bibitem[{{Spergel} \& {Steinhardt}(2000)}]{spergel00}
{Spergel}, D.~N., \& {Steinhardt}, P.~J. 2000, Physical Review Letters, 84,
  3760

\bibitem[{{Spergel} {et~al.}(2003){Spergel}, {Verde}, {Peiris}, {Komatsu},
  {Nolta}, {Bennett}, {Halpern}, {Hinshaw}, {Jarosik}, {Kogut}, {Limon},
  {Meyer}, {Page}, {Tucker}, {Weiland}, {Wollack}, \& {Wright}}]{spergel03}
{Spergel}, D.~N., {Verde}, L., {Peiris}, H.~V., {Komatsu}, E., {Nolta}, M.~R.,
  {Bennett}, C.~L., {Halpern}, M., {Hinshaw}, G., {Jarosik}, N., {Kogut}, A.,
  {Limon}, M., {Meyer}, S.~S., {Page}, L., {Tucker}, G.~S., {Weiland}, J.~L.,
  {Wollack}, E., \& {Wright}, E.~L. 2003, \apjs, 148, 175

\bibitem[{{Springel} {et~al.}(2008){Springel}, {Wang}, {Vogelsberger},
  {Ludlow}, {Jenkins}, {Helmi}, {Navarro}, {Frenk}, \& {White}}]{springel08}
{Springel}, V., {Wang}, J., {Vogelsberger}, M., {Ludlow}, A., {Jenkins}, A.,
  {Helmi}, A., {Navarro}, J.~F., {Frenk}, C.~S., \& {White}, S.~D.~M. 2008,
  \mnras, 391, 1685

\bibitem[{{Stecker}(1978)}]{stecker78}
{Stecker}, F.~W. 1978, \apj, 223, 1032

\bibitem[{{Stetson} {et~al.}(1998){Stetson}, {Hesser}, \&
  {Smecker-Hane}}]{stetson98}
{Stetson}, P.~B., {Hesser}, J.~E., \& {Smecker-Hane}, T.~A. 1998, \pasp, 110,
  533

\bibitem[{{Strigari}(2010)}]{strigari10}
{Strigari}, L.~E. 2010, Advances in Astronomy, 2010

\bibitem[{{Strigari} {et~al.}(2007{\natexlab{a}}){Strigari}, {Bullock},
  {Kaplinghat}, {Diemand}, {Kuhlen}, \& {Madau}}]{strigari07}
{Strigari}, L.~E., {Bullock}, J.~S., {Kaplinghat}, M., {Diemand}, J., {Kuhlen},
  M., \& {Madau}, P. 2007{\natexlab{a}}, \apj, 669, 676

\bibitem[{{Strigari} {et~al.}(2006){Strigari}, {Bullock}, {Kaplinghat},
  {Kravtsov}, {Gnedin}, {Abazajian}, \& {Klypin}}]{strigari06}
{Strigari}, L.~E., {Bullock}, J.~S., {Kaplinghat}, M., {Kravtsov}, A.~V.,
  {Gnedin}, O.~Y., {Abazajian}, K., \& {Klypin}, A.~A. 2006, \apj, 652, 306

\bibitem[{{Strigari} {et~al.}(2008{\natexlab{a}}){Strigari}, {Bullock},
  {Kaplinghat}, {Simon}, {Geha}, {Willman}, \& {Walker}}]{strigari08}
{Strigari}, L.~E., {Bullock}, J.~S., {Kaplinghat}, M., {Simon}, J.~D., {Geha},
  M., {Willman}, B., \& {Walker}, M.~G. 2008{\natexlab{a}}, \nat, 454, 1096

\bibitem[{{Strigari} {et~al.}(2007{\natexlab{b}}){Strigari}, {Koushiappas},
  {Bullock}, \& {Kaplinghat}}]{strigari07b}
{Strigari}, L.~E., {Koushiappas}, S.~M., {Bullock}, J.~S., \& {Kaplinghat}, M.
  2007{\natexlab{b}}, \prd, 75, 083526

\bibitem[{{Strigari} {et~al.}(2008{\natexlab{b}}){Strigari}, {Koushiappas},
  {Bullock}, {Kaplinghat}, {Simon}, {Geha}, \& {Willman}}]{strigari08b}
{Strigari}, L.~E., {Koushiappas}, S.~M., {Bullock}, J.~S., {Kaplinghat}, M.,
  {Simon}, J.~D., {Geha}, M., \& {Willman}, B. 2008{\natexlab{b}}, \apj, 678,
  614

\bibitem[{{Suntzeff} {et~al.}(1986){Suntzeff}, {Aaronson}, {Olszewski}, \&
  {Cook}}]{suntzeff86}
{Suntzeff}, N.~B., {Aaronson}, M., {Olszewski}, E.~W., \& {Cook}, K.~H. 1986,
  \aj, 91, 1091

\bibitem[{{Suntzeff} {et~al.}(1993){Suntzeff}, {Mateo}, {Terndrup},
  {Olszewski}, {Geisler}, \& {Weller}}]{suntzeff93}
{Suntzeff}, N.~B., {Mateo}, M., {Terndrup}, D.~M., {Olszewski}, E.~W.,
  {Geisler}, D., \& {Weller}, W. 1993, \apj, 418, 208

\bibitem[{{the VERITAS collaboration: Vivier et al.}(2011)}]{vivier11}
{the VERITAS collaboration: Vivier et al.} 2011, ArXiv:1110.6615

\bibitem[{{Tollerud} {et~al.}(2011){Tollerud}, {Beaton}, {Geha}, {Bullock},
  {Guhathakurta}, {Kalirai}, {Majewski}, {Kirby}, {Gilbert}, {Yniguez},
  {Patterson}, {Ostheimer}, \& {Choudhury}}]{tollerud11}
{Tollerud}, E.~J., {Beaton}, R.~L., {Geha}, M.~C., {Bullock}, J.~S.,
  {Guhathakurta}, P., {Kalirai}, J.~S., {Majewski}, S.~R., {Kirby}, E.~N.,
  {Gilbert}, K.~M., {Yniguez}, B., {Patterson}, R.~J., {Ostheimer}, J.~C., \&
  {Choudhury}, A. 2011, ArXiv:1112.1067

\bibitem[{{Tollerud} {et~al.}(2008){Tollerud}, {Bullock}, {Strigari}, \&
  {Willman}}]{tollerud08}
{Tollerud}, E.~J., {Bullock}, J.~S., {Strigari}, L.~E., \& {Willman}, B. 2008,
  \apj, 688, 277

\bibitem[{{Tolstoy} {et~al.}(2009){Tolstoy}, {Hill}, \& {Tosi}}]{tolstoy09}
{Tolstoy}, E., {Hill}, V., \& {Tosi}, M. 2009, \araa, 47, 371

\bibitem[{{Tolstoy et al.}(2004)}]{tolstoy04}
{Tolstoy et al.} 2004, \apjl, 617, L119

\bibitem[{{Tonini} {et~al.}(2006){Tonini}, {Lapi}, \& {Salucci}}]{tonini06}
{Tonini}, C., {Lapi}, A., \& {Salucci}, P. 2006, \apj, 649, 591

\bibitem[{{Tremaine} \& {Gunn}(1979)}]{tremaine79}
{Tremaine}, S., \& {Gunn}, J.~E. 1979, Physical Review Letters, 42, 407

\bibitem[{{van den Bergh}(1972)}]{vandenbergh72}
{van den Bergh}, S. 1972, \apjl, 171, L31

\bibitem[{{van der Marel}(1994)}]{vandermarel94}
{van der Marel}, R.~P. 1994, \mnras, 270, 271

\bibitem[{{Vogelsberger} {et~al.}(2012){Vogelsberger}, {Zavala}, \&
  {Loeb}}]{vogelsberger12}
{Vogelsberger}, M., {Zavala}, J., \& {Loeb}, A. 2012, ArXiv:1201.5892

\bibitem[{{Vogt} {et~al.}(1995){Vogt}, {Mateo}, {Olszewski}, \&
  {Keane}}]{vogt95}
{Vogt}, S.~S., {Mateo}, M., {Olszewski}, E.~W., \& {Keane}, M.~J. 1995, \aj,
  109, 151

\bibitem[{{von Hoerner}(1957)}]{vonhoerner57}
{von Hoerner}, S. 1957, \apj, 125, 451

\bibitem[{{Walcher} {et~al.}(2003){Walcher}, {Fried}, {Burkert}, \&
  {Klessen}}]{walcher03}
{Walcher}, C.~J., {Fried}, J.~W., {Burkert}, A., \& {Klessen}, R.~S. 2003,
  \aap, 406, 847

\bibitem[{{Walker} {et~al.}(2009{\natexlab{a}}){Walker}, {Belokurov}, {Evans},
  {Irwin}, {Mateo}, {Olszewski}, \& {Gilmore}}]{walker09c}
{Walker}, M.~G., {Belokurov}, V., {Evans}, N.~W., {Irwin}, M.~J., {Mateo}, M.,
  {Olszewski}, E.~W., \& {Gilmore}, G. 2009{\natexlab{a}}, \apjl, 694, L144

\bibitem[{{Walker} {et~al.}(2008){Walker}, {Mateo}, \& {Olszewski}}]{walker08}
{Walker}, M.~G., {Mateo}, M., \& {Olszewski}, E.~W. 2008, \apjl, 688, L75

\bibitem[{{Walker} {et~al.}(2007{\natexlab{a}}){Walker}, {Mateo}, {Olszewski},
  {Bernstein}, {Sen}, \& {Woodroofe}}]{walker07}
{Walker}, M.~G., {Mateo}, M., {Olszewski}, E.~W., {Bernstein}, R., {Sen}, B.,
  \& {Woodroofe}, M. 2007{\natexlab{a}}, \apjs, 171, 389

\bibitem[{{Walker} {et~al.}(2006){Walker}, {Mateo}, {Olszewski}, {Bernstein},
  {Wang}, \& {Woodroofe}}]{walker06a}
{Walker}, M.~G., {Mateo}, M., {Olszewski}, E.~W., {Bernstein}, R., {Wang}, X.,
  \& {Woodroofe}, M. 2006, \aj, 131, 2114

\bibitem[{{Walker} {et~al.}(2007{\natexlab{b}}){Walker}, {Mateo}, {Olszewski},
  {Gnedin}, {Wang}, {Sen}, \& {Woodroofe}}]{walker07b}
{Walker}, M.~G., {Mateo}, M., {Olszewski}, E.~W., {Gnedin}, O.~Y., {Wang}, X.,
  {Sen}, B., \& {Woodroofe}, M. 2007{\natexlab{b}}, \apjl, 667, L53

\bibitem[{{Walker} {et~al.}(2009{\natexlab{b}}){Walker}, {Mateo}, {Olszewski},
  {Pe{\~n}arrubia}, {Wyn Evans}, \& {Gilmore}}]{walker09d}
{Walker}, M.~G., {Mateo}, M., {Olszewski}, E.~W., {Pe{\~n}arrubia}, J., {Wyn
  Evans}, N., \& {Gilmore}, G. 2009{\natexlab{b}}, \apj, 704, 1274

\bibitem[{{Walker} {et~al.}(2010){Walker}, {McGaugh}, {Mateo}, {Olszewski}, \&
  {Kuzio de Naray}}]{walker10}
{Walker}, M.~G., {McGaugh}, S.~S., {Mateo}, M., {Olszewski}, E.~W., \& {Kuzio
  de Naray}, R. 2010, \apjl, 717, L87

\bibitem[{{Walker} \& {Pe{\~n}arrubia}(2011)}]{wp11}
{Walker}, M.~G., \& {Pe{\~n}arrubia}, J. 2011, \apj, 742, 20

\bibitem[{{Walker, Mateo \& Olszewski}(2009)}]{walker09a}
{Walker, Mateo \& Olszewski}. 2009, \aj, 137, 3100

\bibitem[{{Walsh} {et~al.}(2007){Walsh}, {Jerjen}, \& {Willman}}]{walsh07}
{Walsh}, S.~M., {Jerjen}, H., \& {Willman}, B. 2007, \apjl, 662, L83

\bibitem[{{Walsh} {et~al.}(2009){Walsh}, {Willman}, \& {Jerjen}}]{walsh09}
{Walsh}, S.~M., {Willman}, B., \& {Jerjen}, H. 2009, \aj, 137, 450

\bibitem[{{Wang} {et~al.}(2005){Wang}, {Woodroofe}, {Walker}, {Mateo}, \&
  {Olszewski}}]{wang05}
{Wang}, X., {Woodroofe}, M., {Walker}, M.~G., {Mateo}, M., \& {Olszewski}, E.
  2005, \apj, 626, 145

\bibitem[{{Watkins} {et~al.}(2009){Watkins}, {Evans}, {Belokurov}, {Smith},
  {Hewett}, {Bramich}, {Gilmore}, {Irwin}, {Vidrih}, {Wyrzykowski}, \&
  {Zucker}}]{watkins09}
{Watkins}, L.~L., {Evans}, N.~W., {Belokurov}, V., {Smith}, M.~C., {Hewett},
  P.~C., {Bramich}, D.~M., {Gilmore}, G.~F., {Irwin}, M.~J., {Vidrih}, S.,
  {Wyrzykowski}, {\L}., \& {Zucker}, D.~B. 2009, \mnras, 398, 1757

\bibitem[{{Westfall} {et~al.}(2006){Westfall}, {Majewski}, {Ostheimer},
  {Frinchaboy}, {Kunkel}, {Patterson}, \& {Link}}]{westfall06}
{Westfall}, K.~B., {Majewski}, S.~R., {Ostheimer}, J.~C., {Frinchaboy}, P.~M.,
  {Kunkel}, W.~E., {Patterson}, R.~J., \& {Link}, R. 2006, \aj, 131, 375

\bibitem[{{Wilkinson} {et~al.}(2002){Wilkinson}, {Kleyna}, {Evans}, \&
  {Gilmore}}]{wilkinson02}
{Wilkinson}, M.~I., {Kleyna}, J., {Evans}, N.~W., \& {Gilmore}, G. 2002,
  \mnras, 330, 778

\bibitem[{{Wilkinson} {et~al.}(2004){Wilkinson}, {Kleyna}, {Evans}, {Gilmore},
  {Irwin}, \& {Grebel}}]{wilkinson04}
{Wilkinson}, M.~I., {Kleyna}, J.~T., {Evans}, N.~W., {Gilmore}, G.~F., {Irwin},
  M.~J., \& {Grebel}, E.~K. 2004, \apjl, 611, L21

\bibitem[{{Willman} {et~al.}(2011){Willman}, {Geha}, {Strader}, {Strigari},
  {Simon}, {Kirby}, {Ho}, \& {Warres}}]{willman11}
{Willman}, B., {Geha}, M., {Strader}, J., {Strigari}, L.~E., {Simon}, J.~D.,
  {Kirby}, E., {Ho}, N., \& {Warres}, A. 2011, \aj, 142, 128

\bibitem[{{Willman et al.}(2005{\natexlab{a}})}]{willman05b}
{Willman et al.} 2005{\natexlab{a}}, \aj, 129, 2692

\bibitem[{{Willman et al.}(2005{\natexlab{b}})}]{willman05a}
---. 2005{\natexlab{b}}, \apjl, 626, L85

\bibitem[{{Wilson}(1955)}]{wilson55}
{Wilson}, A.~G. 1955, \pasp, 67, 27

\bibitem[{{Wolf} \& {Bullock}(2012)}]{wolf12}
{Wolf}, J., \& {Bullock}, J.~S. 2012, ArXiv:1203.4240

\bibitem[{{Wolf} {et~al.}(2010){Wolf}, {Martinez}, {Bullock}, {Kaplinghat},
  {Geha}, {Mu{\~n}oz}, {Simon}, \& {Avedo}}]{wolf10}
{Wolf}, J., {Martinez}, G.~D., {Bullock}, J.~S., {Kaplinghat}, M., {Geha}, M.,
  {Mu{\~n}oz}, R.~R., {Simon}, J.~D., \& {Avedo}, F.~F. 2010, \mnras, 406, 1220

\bibitem[{{Zhao}(1996)}]{zhao96}
{Zhao}, H. 1996, \mnras, 278, 488

\bibitem[{{Zucker} {et~al.}(2004){Zucker}, {Kniazev}, {Bell},
  {Mart{\'{\i}}nez-Delgado}, {Grebel}, {Rix}, {Rockosi}, {Holtzman},
  {Walterbos}, {Annis}, {York}, {Ivezi{\'c}}, {Brinkmann}, {Brewington},
  {Harvanek}, {Hennessy}, {Kleinman}, {Krzesinski}, {Long}, {Newman}, {Nitta},
  \& {Snedden}}]{zucker04}
{Zucker}, D.~B., {Kniazev}, A.~Y., {Bell}, E.~F., {Mart{\'{\i}}nez-Delgado},
  D., {Grebel}, E.~K., {Rix}, H.-W., {Rockosi}, C.~M., {Holtzman}, J.~A.,
  {Walterbos}, R.~A.~M., {Annis}, J., {York}, D.~G., {Ivezi{\'c}}, {\v Z}.,
  {Brinkmann}, J., {Brewington}, H., {Harvanek}, M., {Hennessy}, G.,
  {Kleinman}, S.~J., {Krzesinski}, J., {Long}, D., {Newman}, P.~R., {Nitta},
  A., \& {Snedden}, S.~A. 2004, \apjl, 612, L121

\bibitem[{{Zucker} {et~al.}(2007){Zucker}, {Kniazev},
  {Mart{\'{\i}}nez-Delgado}, {Bell}, {Rix}, {Grebel}, {Holtzman}, {Walterbos},
  {Rockosi}, {York}, {Barentine}, {Brewington}, {Brinkmann}, {Harvanek},
  {Kleinman}, {Krzesinski}, {Long}, {Neilsen}, {Nitta}, \&
  {Snedden}}]{zucker07}
{Zucker}, D.~B., {Kniazev}, A.~Y., {Mart{\'{\i}}nez-Delgado}, D., {Bell},
  E.~F., {Rix}, H.-W., {Grebel}, E.~K., {Holtzman}, J.~A., {Walterbos},
  R.~A.~M., {Rockosi}, C.~M., {York}, D.~G., {Barentine}, J.~C., {Brewington},
  H., {Brinkmann}, J., {Harvanek}, M., {Kleinman}, S.~J., {Krzesinski}, J.,
  {Long}, D., {Neilsen}, Jr., E.~H., {Nitta}, A., \& {Snedden}, S.~A. 2007,
  \apjl, 659, L21

\bibitem[{{Zucker et al.}(2006{\natexlab{a}})}]{zucker06b}
{Zucker et al.} 2006{\natexlab{a}}, \apjl, 650, L41

\bibitem[{{Zucker et al.}(2006{\natexlab{b}})}]{zucker06a}
---. 2006{\natexlab{b}}, \apjl, 643, L103

\end{thebibliography}

\end{document}